\documentclass[12pt]{article}

\usepackage{url}
\usepackage{graphicx}
\usepackage{fullpage}
\usepackage{amsfonts}
\usepackage{textcomp}
\usepackage{gensymb}
\usepackage{mathtools}
\usepackage{amsmath,amssymb,amsfonts,amsthm}
\usepackage{dsfont}
\usepackage{graphicx, nicefrac}
\theoremstyle{plain}
\newtheorem{theorem}{Theorem}
\newtheorem{corollary}[theorem]{Corollary}
\newtheorem{lemma}[theorem]{Lemma}

\newtheorem{proposition}[theorem]{Proposition}
\theoremstyle{definition}
\newtheorem{definition}[theorem]{Definition}

\theoremstyle{remark}

\usepackage[titletoc,toc,page,title,header]{appendix}
\usepackage{fullpage}
\usepackage{amsfonts}
\usepackage{tikz}
\usepackage{graphicx}
\usepackage{amsmath}
\usepackage{nicefrac}
\usepackage{array, makecell}
\usepackage{longtable}
\usepackage{blindtext}
\usepackage{graphicx}
\usepackage{subcaption}
\usepackage{xcolor}
\usepackage{xparse}
\usepackage{longtable}
\usepackage{pdfpages}
\usepackage{comment}
\usepackage{algorithm,algpseudocode}% http://ctan.org/pkg/{algorithms,algorithmx}
\algnewcommand{\Input}[1]{%
  \State \textbf{Input:}
   \hspace*{\algorithmicindent}\parbox[t]{.8\linewidth}{\raggedright #1}
}
\algnewcommand{\Output}[1]{%
  \State \textbf{Output:}
  \hspace*{\algorithmicindent}\parbox[t]{.8\linewidth}{\raggedright #1}
}
\algnewcommand{\Initialize}[1]{%
  \State \textbf{Initialize:}
  \hspace*{\algorithmicindent}\parbox[t]{.8\linewidth}{\raggedright #1}
}

\begin{document}
\newcommand{\Depth}{1.3}
\newcommand{\Height}{1.3}
\newcommand{\Width}{1.3}

% ===============================================================
% Title page
% ===============================================================

\title{A common lines approach for ab-initio modeling of molecules with tetrahedral and octahedral symmetry}

\author{Adi Shasha Geva and Yoel Shkolnisky}

%\maketitle

\maketitle

\bigskip

\noindent Adi Shasha Geva \\
Department of Applied Mathematics, School of Mathematical Sciences \\
Tel-Aviv University \\
{\tt shashaadi@gmail.com}

\bigskip
\bigskip

\noindent Yoel Shkolnisky \\
Department of Applied Mathematics, School of Mathematical Sciences \\
Tel-Aviv University \\
{\tt yoelsh@tauex.tau.ac.il}

\bigskip
\bigskip

\begin{center}
Please address manuscript correspondence to Adi Shasha Geva,
{\tt shashaadi@gmail.com}
\end{center}
\newpage

% ===============================================================
% Abstract
% ===============================================================
\newpage
\begin{abstract}

A main task in cryo-electron microscopy single particle reconstruction is to find a three-dimensional model of a molecule given a set of its randomly oriented and positioned noisy projection-images. In this work, we propose an algorithm for ab-initio reconstruction for molecules with tetrahedral or octahedral symmetry. The algorithm exploits the multiple common lines between each pair of projection-images as well as self common lines within each image. It is robust to noise in the input images as it integrates the information from all images at once. The applicability of the proposed algorithm is demonstrated using experimental cryo-electron microscopy data.

\end{abstract}
% ===============================================================
% Listing
% ===============================================================

% ===============================================================
% Chapter 1: Introduction
% ===============================================================

\section{Introduction} \label{chapter:Intro}

%%% cryo-EM %%%

Cryo-electron microscopy (cryo-EM) is a method for determining the high-resolution three-dimensional structure of biomolecules~\cite{Frank2006}.
The method involves imaging frozen copies of the investigated molecule by an electron-microscope, with each copy assuming some unknown random orientation fixed at the moment of freezing. 
Due to the low electron dose that can be applied to the imaged molecules, the projection-images produced by cryo-EM are very noisy. Once the orientation of each of the imaged copies of the molecule has been determined, a low-resolution ab-initio model of the molecule may be recovered from the acquired projection-images by tomographic reconstruction algorithms.
An accurate ab-initio model is crucial for obtaining a high-resolution model, which is  determined by iterative procedures from the set of raw input projection-images.
The task of finding the orientation of the molecule giving rise to each projection-image is known as the
``orientation assignment problem'', and is the main objective of this work.

%%% orientation assignment problem %%%

Formally,
if we denote the electrostatic potential of the molecule by $\psi: \mathbb{R}^3 \rightarrow \mathbb{R}$, 
and consider a set of $N$ rotation matrices 
\begin{equation} \label{eq:rotation matrix}
	R_{i} =
	\begin{pmatrix}
    	\vert & \vert & \vert\\
    	R^{1}_i   & R^{2}_i   & R^{3}_i\\
    	\vert & \vert & \vert
	\end{pmatrix}
	\in SO(3), \quad i \in [N]=\{1,\dots,N\},
\end{equation}
where $SO(3)$ is the group of all rotations in $\mathbb{R}^3$,
then the projection-image $P_{R_i}$, $i \in [N]$, that was generated by imaging $\psi$ rotated by $R_i$, is given by the line integrals of $\psi$ along the lines parallel to $R^{3}_i$ (the third column of $R_i$), namely
\begin{equation} \label{eq:line integrals}
P_{R_{i}}(x,y)=\int_{-\infty}^{\infty} \psi(R_{i}r) dz = 
\int_{-\infty}^{\infty} \psi(xR^{1}_{i}+yR^{2}_{i}+zR^{3}_{i}) dz, \quad r=(x,y,z)^{T}.
\end{equation}
The ``orientation assignment problem'' is defined as finding a set of $N$ rotation matrices $\{R_i\}_{i=1}^N$ such that~\eqref{eq:line integrals} holds for all $i \in [N]$, given only the set of projection-images $\{P_{R_i}\}_{i=1}^N$.

%%% handedness ambiguity %%%
 
An inherent ambiguity in cryo-electron microscopy stems from the fact that the handedness (chirality) of the molecule cannot be resolved from its projection-images. This ambiguity is referred to as the handedness ambiguity. 
Consequently, any projection-image is compatible with two distinct orientations as follows.
We denote by $J=\operatorname{diag}(1,1,-1)$ the reflection matrix through the $xy$-plane, and define by $\tilde \psi(r) = \psi(Jr)$ the mirror image of the molecule $\psi(r)$, $r=(x, y, z)^T$. Since $J^2 = I$, $\psi(r) =\psi(J^2r) =\tilde \psi(Jr)$, and along with~\eqref{eq:line integrals} we have
$$
P_{R_{i}}(x,y)=
\int_{-\infty}^{\infty} \psi(R_{i}r) dz = 
\int_{-\infty}^{\infty} \tilde \psi(JR_{i}r) dz = 
\int_{-\infty}^{\infty} \tilde \psi(JR_{i}JJr) dz.
$$
By noting that $Jr= (x, y, -z)^T$ and using the change of variables $z \rightarrow z'=-z$ we have
\begin{equation} \label{eq:handedness ambiguity}
\begin{split}
P_{R_{i}}(x,y)&=
\int_{-\infty}^{\infty} \tilde \psi((JR_{i}J)Jr) dz=
\int_{-\infty}^{\infty} \tilde \psi((JR_{i}J)(x,y,-z)^T) dz\\&=
\int_{-\infty}^{\infty} \tilde \psi((JR_{i}J)(x,y,z')^T) dz'=
\tilde P_{JR_{i}J}(x,y),
\end{split}
\end{equation}
where $\tilde{P}$ is a projection-image generated from $\tilde{\psi}$.
Equation~\eqref{eq:handedness ambiguity} shows that a projection-image of the molecule $\psi$ at orientation $R_{i}$ is identical to a projection-image of its mirror image molecule 
$\tilde \psi$ at orientation $J R_{i} J$.
Thus, both sets of orientations assignments $\{R_{i}\}^{N}_{i=1}$ and $\{JR_{i}J\}^{N}_{i=1}$ are consistent with the same set of projection-images $\{P_{R_{i}}\}^{N}_{i=1}$.
Biologically, only the model reconstructed using the orientations $\{ R_{i} \}_{i=1}^{N}$ is valid, yet distinguishing whether a reconstruction corresponds to $\{ R_{i} \}_{i=1}^{N}$ or $\{ J R_{i} J\}_{i=1}^{N}$ is impossible without utilizing other structural information.

%%% symmetry ambiguity %%%

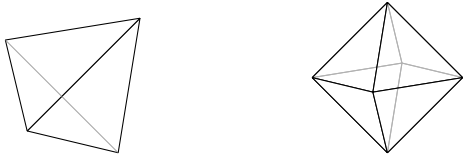
\begin{figure}
\centering
\begin{tikzpicture}[line join=bevel,z=-5.5]
\coordinate (O) at (0,0,0);
\coordinate (B) at (0,1.5,1.5);
\coordinate (E) at (1.5,1.5,0);
\coordinate (G) at (1.5,0,1.5);
\draw[black,fill=white] (O) -- (E) -- (G) -- cycle;% right Face
\draw[black,fill=white] (O) -- (B) -- (E) -- cycle;% Left Face
\draw[black!50,opacity=0.7] (B) -- (G);% Bottom Face
\end{tikzpicture}
\hspace{2cm}
\begin{tikzpicture}[line join=bevel,z=-5.5]
\coordinate (A1) at (0,0,-1);
\coordinate (A2) at (-1,0,0);
\coordinate (A3) at (0,0,1);
\coordinate (A4) at (1,0,0);
\coordinate (B1) at (0,1,0);
\coordinate (C1) at (0,-1,0);
\draw (A1) -- (A2) -- (B1) -- cycle;
\draw (A4) -- (A1) -- (B1) -- cycle;
\draw (A1) -- (A2) -- (C1) -- cycle;
\draw (A4) -- (A1) -- (C1) -- cycle;
\draw [fill opacity=0.7,fill=white] (A2) -- (A3) -- (B1) -- cycle;
\draw [fill opacity=0.7,fill=white] (A3) -- (A4) -- (B1) -- cycle;
\draw [fill opacity=0.7,fill=white] (A2) -- (A3) -- (C1) -- cycle;
\draw [fill opacity=0.7,fill=white] (A3) -- (A4) -- (C1) -- cycle;
\end{tikzpicture}
\if
\quad\quad
\begin{tikzpicture}[line join=bevel]
\coordinate (O) at (0,0,0);
\coordinate (A) at (0,\Width,0);
\coordinate (B) at (0,\Width,\Height);
\coordinate (C) at (0,0,\Height);
\coordinate (D) at (\Depth,0,0);
\coordinate (E) at (\Depth,\Width,0);
\coordinate (F) at (\Depth,\Width,\Height);
\coordinate (G) at (\Depth,0,\Height);
\draw[black,fill=white] (O) -- (C) -- (G) -- (D) -- cycle;% Bottom Face
\draw[black,fill=white] (O) -- (A) -- (E) -- (D) -- cycle;% Back Face
\draw[black,fill=white] (O) -- (A) -- (B) -- (C) -- cycle;% Left Face
\draw[black,fill=white,opacity=0.7] (D) -- (E) -- (F) -- (G) -- cycle;% Right Face
\draw[black,fill=white,opacity=0.7] (C) -- (B) -- (F) -- (G) -- cycle;% Front Face
\draw[black,fill=white,opacity=0.7] (A) -- (B) -- (F) -- (E) -- cycle;% Top Face
\end{tikzpicture}
\fi
\caption{A regular tetrahedron and a regular octahedron.} \label{fig: T and O}
\end{figure}

In this work, we propose an algorithm for solving the ``orientation assignment problem'' for molecules with tetrahedral or octahedral symmetry~\cite{golubitsky1988singularities}, denoted by $\mathbb T \subset SO(3)$ and $\mathbb O \subset SO(3)$, respectively. To present these symmetries, we denote by $C_{n}$ the group of all rotations by $2\pi/n$ radians around some fixed axis (rotational symmetry of order $n$). Then, the elements of the tetrahedral symmetry group~$\mathbb{T}$ are the identity, the elements of 
4 $C_3$ rotation groups whose axes pass through each vertex of the regular tetrahedron (see Fig.~\ref{fig: T and O}) and the corresponding midpoint of
the opposite face, and the elements of 3 $C_2$ rotation groups whose axes pass through the midpoints of two of its opposite edges. In total, the tetrahedral group $\mathbb{T}$ has 12 elements. The elements of the octahedral symmetry group $\mathbb{O}$ are the identity, the elements of 
3 $C_4$ rotation groups whose axes pass through two opposite vertices of the regular octahedron (see Fig.~\ref{fig: T and O}),
4 $C_3$ rotation groups whose axes pass through the midpoints of two of its opposite faces,
and 6 $C_2$ rotation groups whose axe pass through the midpoints of two of its opposite edges. In total, the octahedral group $\mathbb{O}$ has 24 elements. 

%%% choosing the symmetry axes %%%

Since the structure of a molecule is independent of its coordinate system, we choose without loss of generality a coordinate system in which the rotational axes mentioned above coincide with the axes listed in Table~\ref{tab:symmetry-group-elements}.
In this coordinate system, 
the symmetry group elements of a molecule with tetrahedral symmetry are given in Appendix~\ref{tab: T symmetry group elements representations} 
and the symmetry group elements of a molecule with octahedral symmetry are given in Appendix~\ref{tab: O symmetry group elements representations}. 

\begin{table}
\centering
\begin{tabular}{ c c c }
 symmetry & axes & angles \\ 
 \hline 
 $\mathbb T$ &  [1,1,1], [-1,-1,1], [-1,1,-1], [1,-1,-1] &  $\nicefrac{2\pi}{3}$, $\nicefrac{4\pi}{3}$ \\  
 &  [1,0,0], [0,1,0], [0,0,1] & $\pi$ \\
 \hline
 &  [1,0,0], [0,1,0], [0,0,1] & $\nicefrac{\pi}{2}$, $\pi$, $\nicefrac{3\pi}{2}$ \\  
 $\mathbb O$ &  [1,1,1], [-1,1,1], [1,-1,1], [1,1,-1] & $\nicefrac{2\pi}{3}$, $\nicefrac{4\pi}{3}$ \\
 &  [1,1,0], [-1,1,0], [1,0,1], [-1,0,1], [0,1,1], [0,-1,1]& $\pi$\\ \hline
\end{tabular}
\caption{\label{tab:symmetry-group-elements}The nontrivial elements of the tetrahedral and octahedral symmetries.}
\end{table}

To see the effect of symmetry on the orientation assigment problem, we denote by $g^{(k)}$ the $k$-th symmetry group element of the symmetry group~$\mathbb T$ or~$\mathbb O$, $ k \in [n]$, 
where $n$ is the number of elements in the symmetry group.
Mathematically, a molecule $\psi$ has symmetry $G$ ($G=\mathbb{T}$ or $G=\mathbb{O}$) if
\begin{equation} \label{eq:molecule symmetry}
\psi(r) = \psi(g^{(k)}r), \  k \in [n],
\end{equation}
for any $r=(x,y,z)^{T}$. 
Together with~\eqref{eq:line integrals}, it holds that for any $R_i \in SO(3)$ and any $r=(x,y,z)^{T}$,
\begin{equation} \label{eq:images symmetry}
P_{R_{i}}(x,y)=\int_{-\infty}^{\infty} \psi(R_{i}r) dz = 
\int_{-\infty}^{\infty} \psi(g^{(k)}R_{i}r) dz = 
P_{g^{(k)}R_{i}}(x,y),
\end{equation} 
for all $k \in [n]$, implying that the $n$ projection-images $\{P_{g^{(k)}R_i}\}_{k=1}^n$ are identical.
Hence, equation~\eqref{eq:images symmetry} reveals another ambiguity of the set of projection-images $\{P_{R_{i}}\}_{i=1}^{N}$, referred to as the symmetry ambiguity, in which all orientation assignments of the form $\{g_i R_i\}_{i=1}^N$, where ${g_i} \in G$ is an arbitrary symmetry group element, are consistent with the same set of images $\{P_{R_i}\}_{i=1}^N$.

%%% orientation assignment problem %%%

Combining the symmetry ambiguity with the handedness ambiguity described in~\eqref{eq:handedness ambiguity}, the orientation assignment problem can be stated as finding either one of the sets of orientations  $\{R_{i}\}^{N}_{i=1}$ or $\{JR_{i}J\}^{N}_{i=1}$, where each $R_{i}$ may be replaced by $g_iR_{i}$, with $g_i \in G$ being an arbitrary symmetry group element, independently for each $i$ (that is independently for each rotation).

%%% common lines %%%

Solving the orientation assignment problem, in its broadest sense, amounts to relating between images and rotation matrices.
% In the following, we introduce the concept of common lines as our method for solving is based on.
% and 
% show how it can be used to relate between two images and their relative rotation.
% the describe the relation between 
In this work, we introduce a common lines based method for solving the orientation assignment problem for molecules having either tetrahedral or octahedral symmetry.
The common lines, defined in the following and discussed in detail in subsequent sections, 
reveal useful relations 
between images and rotation matrices. To define common lines, we recall the  Fourier projection slice theorem~\cite{Natr2001a}, which provides an important relation between the Fourier transform of $\psi$  and the Fourier transform of $P_{R_{i}}$ (see~\eqref{eq:line integrals}). 
We define the two-dimensional Fourier transform of a projection-image~\eqref{eq:line integrals} by
\begin{equation*}
\hat{P}_{R_i} (\omega_{x},\omega_{y}) = \iint_{\mathbb{R}^{2}} P_{R_i}(x,y)
e^{-\imath \left ( x \omega_{x} + y \omega_{y} \right )}\, dx dy,
\end{equation*}
and the three-dimensional Fourier transform of the molecule by
\begin{equation*}
\hat{\psi} (\omega_{x},\omega_{y},\omega_{z}) = \iiint_{\mathbb{R}^{3}}
\psi(x,y,z) e^{-\imath \left ( x \omega_{x} + y \omega_{y} + z
\omega_{z}\right )}\, dx dy dz.
\end{equation*}
Using this notation, the Fourier projection slice theorem states that 
\begin{equation} \label{eq:Projection slice theorem}
\hat{P}_{R_i}(\omega_{x},\omega_{y})=
\hat{\psi}(\omega_xR_i^{1}+\omega_yR_i^{2}), 
\quad (\omega_{x},\omega_{y})\in\mathbb{R}^{2},
\end{equation}
where $R_i^{1}$ and $R_{i}^{2}$ are the first and second columns of $R_{i}$, respectively. In words, the two-dimensional Fourier transform of any projection-image $P_{R_i}$ is equal to the restriction of the three-dimensional Fourier transform of the molecule $\psi$ to the plane through the origin spanned by~$R_i^{1}$ and~$R_i^{2}$, or equivalently, to the central plane whose normal coincides with~$R_i^{3}$. 
As any two central planes intersect along a single line through the origin (as long as the central planes do not coincide), the central planes corresponding to any pair of Fourier-transformed projection-images $\hat{P}_{R_{i}}$ and $\hat{P}_{R_{j}}$ intersect along such a line, and therefore, both (Fourier transformed) images share a pair of lines on which their Fourier transforms coincide, thus referred to as common lines.
Given that $\hat{P}_{R_{i}}$ and $\hat{P}_{R_{j}}$ are images of a molecule with tetrahedral or octahedral symmetry, each $\hat{P}_{{g^{(k)}}R_{j}}$, $k \in [n]$, is identical to $\hat{P}_{R_{j}}$ (see~\eqref{eq:images symmetry}).
In addition, each $\hat{P}_{{g^{(k)}}R_{j}}$, $k \in [n]$, also shares a common line with~$\hat{P}_{R_{i}}$. Since the rotations $g^{(k)}R_{j}$, $k \in [n]$, are in general different from each other, the planes spanned by their first two columns are also different. Thus, $\hat{P}_{R_{i}}$ and $\hat{P}_{R_{j}}$ have $n$ common lines altogether.

% ===============================================================
% Chapter 2: Related work
% ===============================================================

\section{Related work}\label{chapter: Related work}

%%% The angular reconstitution method %%%

Common lines methods for ab-initio reconstruction of macromolecules have originated with the angular reconstitution method by Van Heel~\cite{VanHeel1987}. It is a  sequential method
in which given a triplet of projection-images $\{ P_{R_i}, P_{R_j}, P_{R_k}\}$,
the set of relative rotations 
$\{ R_i^TR_j, R_i^TR_k, R_j^TR_k\}$
is first estimated by detecting common lines between $ P_{R_i}, P_{R_j},$ and $P_{R_k}$.
Then, setting $R_i=I$ without loss of generality, determines $R_j$ and $R_k$ from 
$R_i^TR_j$ and $R_i^TR_k$. 
By applying this method sequentially to each triplet $\{ R_i^TR_j, R_i^TR_l, R_j^TR_l\}$ where $l \neq i,j$, the orientation $R_l$ of the image $P_{R_l}$ is determined from $R_i^TR_l$ simply by $R_l=R_i^TR_l$.

%%% The angular reconstitution method is non robust %%%

Detecting common lines between a pair of images is typically done by finding the pair of central lines in the Fourier transforms of the images that have the highest correlation~\cite{voting}.
In cryo-EM, the images are contaminated with high levels of noise, thus making the detection of common lines error prone. 
Consequently, the relative rotations in the angular reconstitution method are estimated with errors, which render the method not robust to noise.

%%% synchronization method %%%

A common lines based approach for molecules without symmetry that is robust to noisy input images is the synchronization method~\cite{asymsync,pragier2016graph}. In this approach, 
all relative rotations $\{R_i^TR_j\}_{i<j \in [N]} $ are first estimated using common lines (robust estimation of common lines is described in~\cite{voting,greenberg2017common}). Then,
the rotations 
$\{R_i\}_{i \in [N]} $ are estimated simultaneously,
by constructing a $3N\times 3N$ matrix whose $(i,j)$ block of size $3\times 3$ contains the estimate for $R_i^TR_j$, and factorizing this matrix using SVD.  However, this method is not applicable to symmetric molecules due to the symmetry ambiguity described by~\eqref{eq:images symmetry}. 
Specifically, consider a pair of images $P_{R_i}$ and $P_{R_j}$, $i<j \in [N]$, of a molecule with tetrahedral or octahedral symmetry.
By the discussion above, there are $n$ pairs of common lines between the images, corresponding to $n$  pairs of projection planes, 
but it is unknown which pair of common lines corresponds to which pair of projection planes. As a result, the best one can estimate from a single pair of common lines between the images is the relative rotation $R_i^Tg_{ij}R_j$, where 
 $g_{ij}$ is an unknown arbitrary symmetry group element. In such a case, factorizing the  $3N \times 3N$ matrix whose blocks are $R_i^Tg_{ij}R_j$ does not give $R_{i}$ (more precisely $g_{i}R_{i}$ for some arbitrary~$g_{i}$), unless we are able to carefully choose $g_{ij}$.
 
%%% Cn and D2 methods %%%

Two robust common lines based methods which are applicable to symmetric molecules are described in~\cite{pragier2019common} for molecules with~$C_{n}$ symmetry and in~\cite{syncD2} for molecules with~$D_{2}$ symmetry. In both methods, all common lines between each pair of images are utilized to estimate a set
$\{R_i^Tg_{ij}R_j \}_{i<j \in [N]}$, with $g_{ij}$ being an unknown symmetry group element of the~$C_{n}$ or~$D_{2}$ symmetry groups. Once the set $\{R_i^Tg_{ij}R_j \}_{i<j \in [N]} $ has been estimated, the methods exploit the symmetry group properties to obtain the set of rotation matrices $\{ R_i\}_{i \in [N]}$, with each rotation matrix satisfying $ R_i \in \{ g_k R_i\}_{k=1}^n$. 
%In both methods, all common lines between each pair of images are utilized to estimate a set $\{\bar R_i^T \bar R_j\}_{i<j \in [N]} $ which satisfies $\bar R_i^T \bar R_j = $ $R_i^Tg_{ij}R_j$ for all ${i<j \in [N]}$, with $g_{ij}$ being an unknown symmetry group element of the $C_{n}$ or $D_{2}$ symmetry groups. Once the set $\{\tilde R_i^T \tilde R_j\}_{i<j \in [N]} $ has been estimated, the methods exploit the symmetry group properties to obtain the set of rotation matrices $\{\tilde R_i\}_{i \in [N]}$, with each rotation matrix satisfying $\tilde R_i \in \{ g_kR_i\}_{k=1}^n$. 
Unfortunately, the methods in~\cite{pragier2019common,syncD2} are not applicable to molecules with~$\mathbb{T}$ or~$\mathbb{O}$ symmetries. Specifically, the  method in~\cite{pragier2019common} for molecules with $C_n$ symmetry uses
the property that the average of the group elements of $C_{n}$ is the matrix $\operatorname{diag}(0,0,1)$. 
This property doesn't hold for molecules with~$\mathbb{T}$ or $\mathbb{O}$ symmetry,
as the average over all group elements of the groups $\mathbb{T}$ and $\mathbb{O}$ is the zero matrix.
As for the method in~\cite{syncD2} for molecules with $D_2$ symmetry, this method assumes that the rotational symmetry axes of the molecule coincide with the $x,y$ and $z$ axes, which does not hold for~$\mathbb{T}$ and~$\mathbb{O}$ symmetries.

While we propose a method that is based on common lines, there exist algorithms for finding ab-initio models that are based on casting the reconstruction problem as an optimization problem~\cite{cryosparc, relion3, EMAN,xmipp}. Such algorithms use some general-purpose optimization algorithm (such as stochastic gradient descent, stochastic hill climbing, projection-matching, and simulated annealing, to name a few), on a subset of the data or its class averages. All these methods boil down to a non-convex optimization, which is susceptible to getting stuck in a local minimum corresponding to a structure that is inconsistent with the investigated molecule. Yet, these methods are widely used in practice and many times produce satisfactory initial models. In Section~\ref{sec:results} we show an example of both a success as well as a failure of such a method. 

%%% work organization %%%

The paper is organized as follows. 
In Section~\ref{chapter:Common lines}, we formally define the common lines, introduce the notion of self common lines and derive basic properties of common lines and self common lines.
In Section~\ref{capter:solving the orientation assignment problem}, we describe our algorithm for estimating the orientations of a given set of projection-images. 
Then, in Section~\ref{sec:results}, we report some numerical experiments we conducted using simulated and experimental data sets, demonstrating the robustness and effectiveness of our proposed method.
Finally, in Section~\ref{Future work}, we discuss possible future work.

% ===============================================================
% Chpter 3: Common lines and self common lines
% ===============================================================

\section{Common lines and self common lines} \label{chapter:Common lines}

Formally, for each $k \in [n]$, the unit vector
\begin{equation} \label{eq:cl vec k}
q_{R_i,R_j}^k = \frac{R^{3}_{i} \times g^{(k)}R^{3}_{j}}
{\vert\vert{R^{3}_{i} \times g^{(k)}R^{3}_{j}}\vert\vert}
\end{equation}
gives the direction of the common line between the central planes of $\hat P_{R_i}$ and $\hat P_{g^{(k)}R_j}$, since it is perpendicular to the normal vectors of both of them.
We can express $q_{R_i,R_j}^k$ using its local coordinates on both central planes by
\begin{equation} \label{eq:q sin cos}
q_{R_i,R_j}^k = 
\cos(\alpha^{k,1}_{R_i,R_j})R^{1}_{i}+\sin(\alpha^{k,1}_{R_i,R_j})R^{2}_{i} =
\cos(\alpha^{k,2}_{R_i,R_j})g^{(k)} R^{1}_{j}+\sin(\alpha^{k,2}_{R_i,R_j})g^{(k)} R^{2}_{j},
\end{equation}
where $\alpha^{k,1}_{R_i,R_j}$ and  $\alpha^{k,2}_{R_i,R_j}$ are the angles between $q_{R_i,R_j}^k$ and the local $x$-axes of the planes.
Using this notation along with~\eqref{eq:Projection slice theorem}, we have that
for any $\xi \in \mathbb{R}$ and $k\in[n]$,
\begin{equation} \label{eq:PRi=PRj}
\begin{split}
\hat{P}_{R_{i}}(\xi\cos(\alpha^{k,1}_{R_i,R_j}),\xi\sin(\alpha^{k,1}_{R_i,R_j}))
 = &
 \hat{\psi}(\xi\cos(\alpha^{k,1}_{R_i,R_j}) R_i^{1}+\xi\sin(\alpha^{k,1}_{R_i,R_j})R_i^{2})\\ = & 
 \hat{\psi}(\xi q_{R_i,R_j}^k)\\
 = &
\hat{\psi}( \cos(\alpha^{k,2}_{R_i,R_j})g^{(k)} R^{1}_{j}+\sin(\alpha^{k,2}_{R_i,R_j})g^{(k)} R^{2}_{j} )\\
 = &
\hat{P}_{g^{(k)}R_{j}}(\xi\cos(\alpha^{k,2}_{R_i,R_j}),\xi\sin(\alpha^{k,2}_{R_i,R_j}))
\\
 = &
\hat{P}_{R_{j}}(\xi\cos(\alpha^{k,2}_{R_i,R_j}),\xi\sin(\alpha^{k,2}_{R_i,R_j})).
\end{split}
\end{equation}
Following~\eqref{eq:PRi=PRj}, we express the set of common lines between the pair of images 
$\hat{P}_{R_{i}}$ and $\hat{P}_{R_{j}}$ 
by the set of local coordinates 
$\{(\alpha^{k,1}_{R_i,R_j}, \alpha^{k,2}_{R_i,R_j})\}_{k \in [n]}$.
In particular, 
% by~\cite{viewang}, 
$\alpha^{k,1}_{R_i,R_j}$ and  $\alpha^{k,2}_{R_i,R_j}$ 
may be recovered from the entries of $R_i^T g^{(k)} R_j$ using
\begin{equation} \label{eq:recover angles}
\begin{split}
\alpha^{k,1}_{R_i,R_j} = \text{atan2}
\left( (R_i^T g^{(k)} R_j)_{1,3},-(R_i^T g^{(k)} R_j)_{2,3} \right),
\\
\alpha^{k,2}_{R_i,R_j} = \text{atan2} 
\left(-(R_i^T g^{(k)} R_j)_{3,1},(R_i^T g^{(k)} R_j)_{3,2} \right).    
\end{split}
\end{equation}

Note that in~\eqref{eq:cl vec k}, the vector $R_j^3$ is multiplied by the symmetry group element $g^{(k)}$, $k \in [n]$, while $R_i^3$ is not. Since this choice is arbitrary, we  show in the following that we get the same common lines if we multiply $R_i^3$ by $g^{(k)}$ instead. In other words, we show in the following that the set of local coordinates $\{(\alpha^{k,1}_{R_i,R_j}, \alpha^{k,2}_{R_i,R_j})\}_{k \in [n]}$ for the common lines between the pair of images 
$\hat{P}_{R_{i}}$ and $\hat{P}_{R_{j}}$ is well defined.

Similarly to~\eqref{eq:cl vec k}, for each $l \in [n]$, the unit vector
\begin{equation} \label{eq:cl vec k 1}
\tilde q_{R_i,R_j}^l = \frac{g^{(l)}R^{3}_{i} \times R^{3}_{j}}
{\vert\vert{g^{(l)}R^{3}_{i} \times R^{3}_{j}}\vert\vert}
\end{equation}
gives the direction of the common line between the central planes of the Fourier transformed images $\hat P_{g^{(l)}R_i}$ and $\hat P_{R_j}$.
As $G$ is a group, for each $l \in [n]$ there is $k \in [n]$ such that $ (g^{(k)})^T=g^{(l)}$.
Then, using~\eqref{eq:cl vec k}, it holds that
\begin{equation} \label{eq: equivalent definitions}
(g^{(k)})^T q_{R_i,R_j}^k 
= (g^{(k)})^T \frac{R^{3}_{i} \times g^{(k)}R^{3}_{j}}
{\vert\vert{R^{3}_{i} \times g^{(k)}R^{3}_{j}}\vert\vert}
= \frac{g^{(l)}R^{3}_{i} \times R^{3}_{j}}
{\vert\vert{R^{3}_{i} \times g^{(k)}R^{3}_{j}}\vert\vert}
= \frac{g^{(l)}R^{3}_{i} \times R^{3}_{j}}
{\vert\vert{g^{(l)}R^{3}_{i} \times R^{3}_{j}}\vert\vert}
= \tilde q_{R_i,R_j}^l,
\end{equation}
where the second equality follows since for any rotation $R$ it holds that $R(a\times b) = (Ra)\times (Rb)$,
and the third equality follows from the latter property along with the anti-commutative property of the cross product, i.e., $a\times b = -(b\times a)$, and the invariance of the 2-norm to orthogonal transformations.
By multiplying~\eqref{eq:q sin cos} by $(g^{(k)})^T$ from the left we get using~\eqref{eq: equivalent definitions}
\begin{equation} \label{eq:q sin cos 1}
\tilde q_{R_j,R_i}^l = 
\cos(\alpha^{k,1}_{R_i,R_j})g^{(l)}R^{1}_{i}+\sin(\alpha^{k,1}_{R_i,R_j})g^{(l)}R^{2}_{i} =
\cos(\alpha^{k,2}_{R_i,R_j}) R^{1}_{j}+\sin(\alpha^{k,2}_{R_i,R_j}) R^{2}_{j}.
\end{equation}
Equation~\eqref{eq:q sin cos 1} implies that $\alpha^{k,1}_{R_i,R_j}$ and  $\alpha^{k,2}_{R_i,R_j}$ are also the angles between $\tilde q_{R_i,R_j}^l$ and the local $x$-axes of the planes of the Fourier transformed images $\hat P_{g^{(l)}R_i}$ and $\hat P_{R_j}$.
Then, similarly to~\eqref{eq:PRi=PRj},
\begin{equation} \label{eq:PRi=PRj 2}
\begin{split}
\hat{P}_{R_{i}}(\xi\cos(\alpha^{k,1}_{R_i,R_j}),\xi\sin(\alpha^{k,1}_{R_i,R_j}))
 = &
 \hat{P}_{g^{(l)}R_{i}}(\xi\cos(\alpha^{k,1}_{R_i,R_j}),\xi\sin(\alpha^{k,1}_{R_i,R_j}))\\ = & 
 \hat{\psi}(\xi \tilde q_{R_i,R_j}^l)\\
 = &
\hat{P}_{R_{j}}(\xi\cos(\alpha^{k,2}_{R_i,R_j}),\xi\sin(\alpha^{k,2}_{R_i,R_j})).
\end{split}
\end{equation}
Thus, the set of local coordinates
for the common lines is well defined, as the same set is obtained from the two equivalent definitions~\eqref{eq:cl vec k} and~\eqref{eq:cl vec k 1}.

%%% self common lines %%%

Another important property of projection-images of symmetric molecules, and in particular of molecules with tetrahedral or octahedral symmetry, is the existence of self common lines, which 
 are common lines between any two (identical) images $\hat{P}_{R_{i}}$ and $\hat{P}_{g^{(k)} R_{i}}$, $k \in \{2, \dots, n\}$.
The direction vector of the self common line between $\hat P_{R_i}$ and $\hat P_{g^{(k)}R_i}$ is
\begin{equation} \label{eq:scl vec k}
q_{R_i,R_i}^k = \frac{R^{3}_{i} \times g^{(k)}R^{3}_{i}}
{\vert\vert{R^{3}_{i} \times g^{(k)}R^{3}_{i}}\vert\vert}.
\end{equation}
When expressing $q_{R_i,R_i}^k$ by the local coordinates
 $(\alpha^{k,1}_{R_i,R_i}, \alpha^{k,2}_{R_i,R_i})$, we get similarly to~\eqref{eq:PRi=PRj} that
\begin{equation} \label{eq:PRi=PRi}
\begin{split}
\hat{P}_{R_{i}}(\xi\cos(\alpha^{k,1}_{R_i,R_i}),\xi\sin(\alpha^{k,1}_{R_i,R_i}))
 =
\hat{P}_{R_{i}}(\xi\cos(\alpha^{k,2}_{R_i,R_i}),\xi\sin(\alpha^{k,2}_{R_i,R_i})),
\end{split}
\end{equation}
% where by~\cite{viewang}, 
and $\alpha^{k,1}_{R_i,R_i}$ and  $\alpha^{k,2}_{R_i,R_i}$ 
may be recovered from the entries of $R_i^T g^{(k)} R_i$ using
\begin{equation} \label{eq:recover angles scl}
\begin{split}
\alpha^{k,1}_{R_i,R_i} = \text{atan2}
\left( (R_i^T g^{(k)} R_i)_{1,3},-(R_i^T g^{(k)} R_i)_{2,3} \right),
\\
\alpha^{k,2}_{R_i,R_i} = \text{atan2} 
\left(-(R_i^T g^{(k)} R_i)_{3,1},(R_i^T g^{(k)} R_i)_{3,2} \right).    
\end{split}
\end{equation}
Thus, the set of self common lines of the image $\hat{P}_{R_{i}}$
is expressed by the set of local coordinates 
$\{(\alpha^{k,1}_{R_i,R_i}, \alpha^{k,2}_{R_i,R_i})\}_{k \in \{2, \dots, n\} }$.

% %%% shifts %%%

% Lastly, the Fourier projection-slice theorem~\eqref{eq:Projection slice theorem} relies on the stipulation that the centers of all projection-images coincide with the center of the three-dimensional molecule.
% In practice, it is unlikely that all projections are simultaneously aligned with respect to a common three-dimensional origin, making the Fourier projection-slice theorem, as stated in~\eqref{eq:Projection slice theorem}, not applicable to pairs of experimental projection-images.
% Thus, the procedure for detecting common lines between projection-images,
% which is given in Section~\ref{capter:solving the orientation assignment problem} below as $\pi_{ij}$,
% needs to be modified along the lines of~\cite{shifts} in order to handle the presence of unknown shifts.

% ===============================================================
% Chpter 5: Relative rotations estimation
% ===============================================================

\section{Algorithm}\label{capter:solving the orientation assignment problem}

%%% intro %%%

In this section, we derive our method for solving the orientation assignment problem for molecules with tetrahedral or octahedral symmetry. Throughout this section, we denote by~$G$ either the group $\mathbb{T}$ or the group $\mathbb{O}$.
Our method consists of two steps; 
first, we assign to each pair of projection-images  
$P_{R_i}$ and $P_{R_j}$ (see~\eqref{eq:line integrals}) , $i<j \in [N]$, 
of a molecule whose symmetry group is $G$,
a pair of rotation matrices $(\tilde R_{ij}, \tilde{R}_{ji})$
which is an estimate to a pair of rotation matrices $( R_{ij}, {R}_{ji})$
which satisfies
\begin{equation} \label{eq:common line approach}
\{ R^T_{ij} g^{(k)}  R_{ji} \}_{k=1}^n = \{ R^T_{i} g^{(k)} R_{j} \}_{k=1}^n, \quad g^{(k)} \in G.
\end{equation}
Then, we estimate the orientations of all projection-images 
$\{P_{R_i}\}_{i \in [N]} $ 
from the set of rotation matrices 
$\{ (\tilde R_{ij}, \tilde{R}_{ji}) \}_{i<j \in [N]}$.

%%% score function %%%

To find the rotations $\tilde R_{ij}$ and $\tilde{R}_{ji}$ which estimate $ R_{ij}$ and ${R}_{ji}$ of~\eqref{eq:common line approach},
% To estimate the rotations $\tilde R_{ij}$ and $\tilde{R}_{ji}$ which satisfy~\eqref{eq:common line approach}, 
we follow the maximum likelihood approach described in~\cite{pragier2019common,syncD2} as follows. First, we construct a function 
 $\pi_{ij}(Q_r,Q_s)$, which for any two rotations $Q_r,Q_s \in SO(3)$ computes  a score that indicates how well $\{ Q_r^T g^{(k)}{Q_s}\}_{k=1}^n $ approximates $\{ R_i^T g^{(k)}{R_j}\}_{k=1}^n$. Since it is impossible to  find efficiently the optimum of~$\pi_{ij}$ over $SO(3) \times SO(3)$, we show in Appendix~\ref{Constructing $SO_G(3)$} how to construct a finite subset $SO_G(3) \subset SO(3)$ on which we search for the optimum of~$\pi_{ij}$. 
 In particular, the subset $SO_G(3)$ takes advantage of the fact that the underlying molecule is symmetric, to reduce the number of rotations in $SO_G(3)$, while maintaining high accuracy of our algorithm.
We use the pair $(Q_r,Q_s) \in SO_{G}(3) \times SO_{G}(3)$ that attains the highest score $\pi_{ij}$ as our 
estimate $(\tilde R_{ij},\tilde{R}_{ji})$ for $(R_{ij},{R}_{ji})$ of~\eqref{eq:common line approach}. 

We next describe the construction of the function
$\pi_{ij}: SO_G(3) \times SO_G(3) \to [0,1]$ 
for each pair of images $P_{R_i}$ and $P_{R_j}$.
We denote by
\begin{equation} \label{eq:fourier ray}
\nu_{i, \theta}(\xi) = \hat P_{R_i}(\xi\cos\theta,\xi\sin\theta), \ \ \xi \in (0,\infty)
\end{equation}
the half line (known as a Fourier ray) in the direction which forms an angle $\theta$ with the $x$-axis of the Fourier transformed image $\hat P_{R_i}$, $i \in [N]$, 
and by
\begin{equation} \label{eq:NNC}
 \rho_{ij}(\theta, \phi) = \Re \ 
\int_{0}^{\infty}\frac{(\nu_{i, \theta}(\xi))^{*}\nu_{j, \phi}(\xi)d\xi}
{||\nu_{i, \theta}(\xi)||_{L_2}||\nu_{j, \phi}(\xi)||_{L_2}}
\end{equation}
the real part of the normalized cross correlations between $\nu_{i, \theta}(\xi)$ and $\nu_{j, \phi}(\xi)$.
Note that due to~\eqref{eq:PRi=PRj}, it holds that
$ \rho_{ij}(\alpha^{k,1}_{R_i,R_j}, \alpha _{R_i,R_j}^{k,2}) = 1$ 
for all $k \in [n]$, where $\{(\alpha^{k,1}_{R_i,R_j}, \alpha _{R_i,R_j}^{k,2})\}_{k \in [n]}$ is the set of common lines between $\hat P_{R_i}$ and $\hat P_{R_j}$.

Now, consider a pair of rotations $Q_r, Q_s \in SO_G(3)$.
Analogously to~\eqref{eq:recover angles},
we compute the set of local coordinates
$\{(\alpha^{k,1}_{Q_r,Q_s}, \alpha _{Q_r,Q_s}^{k,2})\}_{k \in [n]}$
from the set $\{ Q_r^T g^{(k)}{Q_s}\}_{k=1}^n$ using
\begin{equation} \label{eq:recover angles cands}
\begin{split}
\alpha^{k,1}_{Q_r,Q_s} = \text{atan2}
\left( (Q_r^T g^{(k)} Q_s)_{1,3},-(Q_r^T g^{(k)} Q_s)_{2,3} \right),
\\ 
\alpha^{k,2}_{Q_r,Q_s} = \text{atan2}
\left(-(Q_r^T g^{(k)} Q_s)_{3,1},(Q_r^T g^{(k)} Q_s)_{3,2} \right).
\end{split}
\end{equation}
If 
$\{ Q_r^T g^{(k)}{Q_s}\}_{k=1}^n = \{ R_i^T g^{(k)}{R_j}\}_{k=1}^n$,
then~\eqref{eq:recover angles cands} along with~\eqref{eq:recover angles} imply that the set
of local coordinates $\{(\alpha^{k,1}_{Q_r,Q_s}, \alpha _{Q_r,Q_s}^{k,2})\}_{k \in [n]}$ 
is 
% \rvc{\rvg{contained in}}
% {I know the reviewer ask for that, but I think he is wrong. We do not claim this in the case of $\rho=1$, but rather in the case $\{ Q_r^T g^{(k)}{Q_s}\}_{k=1}^n = \{ R_i^T g^{(k)}{R_j}\}_{k=1}^n$, in which equality of the common lines follows from the equations.}
the set of common lines of the pair of images $P_{R_i}$ and~$P_{R_j}$,
i.e, it is equal to $\{(\alpha^{k,1}_{R_i,R_j}, \alpha _{R_i,R_j}^{k,2})\}_{k \in [n]}$.
The score function $\pi_{ij}$ is thus defined as
\begin{equation} \label{eq:score cl}
\pi_{ij}(Q_r,Q_s) = 
\prod_{k \in [n]}  \rho_{ij}(\alpha^{k,1}_{Q_r,Q_s}, \alpha _{Q_r,Q_s}^{k,2}),
\end{equation}
satisfying $\pi_{ij}(Q_r,Q_s) = 1$
% whenever
if
$\{ Q_r^T g^{(k)}{Q_s}\}_{k=1}^n = \{ R_i^T g^{(k)}{R_j}\}_{k=1}^n$.
We note that we define~$\pi_{ij}$ in~\eqref{eq:score cl} as a product to enforce that all~$n$ correlations are large simultaneously. Each $\rho_{ij}(\alpha^{k,1}_{Q_r,Q_s}, \alpha _{Q_r,Q_s}^{k,2})$ is a proxy to the  probability that  $(\alpha^{k,1}_{Q_r,Q_s}, \alpha _{Q_r,Q_s}^{k,2})$ is a common line between $P_{R_i}$ and $P_{R_j}$, and we want all these probabilities to be large simultaneously.

%%% it is only an estimation %%%

In practice, since $\pi_{ij}$ is computed using noisy images, and since $SO_G(3)$ is only a finite subset of $SO(3)$, $\pi_{ij}(Q_r,Q_s)$ is unlikely to be exactly 1. Thus, the pair of candidates $Q_{r},Q_{s} \in SO_{G}(3)$ with the highest score $\pi_{ij}(Q_{r},Q_{s})$ is used to construct $\{ Q_{r}^T g^{(k)}{Q_{s}}\}_{k=1}^n$, which serves as an estimate for $\{ R_i^T g^{(k)}{R_j}\}_{k=1}^n$.

%%% combine self common lines %%%

In order to achieve a more robust estimate of $\{ R_i^T g^{(k)}{R_j}\}_{k=1}^n$, 
we also combine the set of self common lines into the score function $\pi_{ij}$ of~\eqref{eq:score cl}.
Specifically, as $\{(\alpha^{k,1}_{R_i,R_i}, \alpha _{R_i,R_i}^{k,2})\}_{k \in \{2, \dots, n\}}$ and 
$\{(\alpha^{k,1}_{R_j,R_j}, \alpha _{R_j,R_j}^{k,2})\}_{k \in\{2, \dots, n\}}$ are the sets of self common lines of $P_{R_i}$ and $P_{R_j}$ respectively (see~\eqref{eq:PRi=PRi} and its following paragraph), we define
\begin{equation} \label{eq:score cl scl}
\pi_{ij}(Q_r,Q_s) = 
\prod_{k \in [n]} 
 \rho_{ij}(\alpha^{k,1}_{Q_r,Q_s}, \alpha _{Q_r,Q_s}^{k,2})
\prod_{k \in \{2, \dots, n\}} 
\rho_{ii}(\alpha^{k,1}_{Q_r,Q_r}, \alpha _{Q_r,Q_r}^{k,2})
 \rho_{jj}(\alpha^{k,1}_{Q_s,Q_s}, \alpha _{Q_s,Q_s}^{k,2}).
\end{equation}
%In Appendix \ref{sec:SG},
%we show how to replace the set $\{2, \dots, n\}$ in~\eqref{eq:score cl scl not optimal} with the subset $S_G \subset \{2, \dots, n\}$ defined in~\eqref{eq:scl inds}, thereby allowing a faster evaluation of~\eqref{eq:score cl scl not optimal}.
%The subset~$S_G$ is constructed by eliminating redundant self common lines, in the sense that the same optimum of $\tilde \pi_{ij}$ is attained either with or without them. 
%Thus we define (with a slight abuse of notation compared with~\eqref{eq:score cl})
%\begin{equation} \label{eq:score cl scl}
%\pi_{ij}(Q_r,Q_s) = 
%\prod_{k \in [n]} 
%\rho_{ij}(\alpha^{k,1}_{Q_r,Q_s}, \alpha _{Q_r,Q_s}^{k,2})
%\prod_{k \in S_G} 
%\rho_{ii}(\alpha^{k,1}_{Q_r,Q_r}, \alpha _{Q_r,Q_r}^{k,2})
%\rho_{jj}(\alpha^{k,1}_{Q_s,Q_s}, \alpha _{Q_s,Q_s}^{k,2}).
%\end{equation} 

%%% get highest score pair of candidates %%%

Using the score function $\pi_{ij}$ of~\eqref{eq:score cl scl}, we choose for each $i<j\in[N]$ the pair $(\tilde{R}_{ij},\tilde{R}_{ji})$ that satisfies 
\begin{equation} \label{eq:ml score}
(\tilde R_{ij}, \tilde R_{ji}) = 
\underset{\substack{(Q_r, Q_s) \in SO_G(3) \times SO_G(3)}}{\operatorname{\arg max}} \pi_{ij} (Q_r, Q_s),
\end{equation}
and use it as an estimate for $( R_{ij},  R_{ji})$.
% construct the set $\{ \tilde R^T_{ij} g^{(k)} \tilde R_{ij} \}_{k=1}^n$, and use it as an estimate for $\{ R_i^T g^{(k)}{R_j}\}_{k=1}^n$.
% We point out that for each $i<j\in[N]$ the pair $(\tilde{R}_{ij},\tilde{R}_{ji})$ which satisfies~\eqref{eq:ml score} is not unique.
The procedure for computing the set 
$\{ (\tilde R_{ij}, \tilde{R}_{ji}) \}_{i<j \in [N]}$
is summarized in Algorithm~\ref{alg:relative}.

%%% Algorithm 1 %%%
\begin{algorithm}
  \caption{Computing $\{ (\tilde R_{ij}, \tilde{R}_{ji}) \}_{i<j \in [N]}$ for molecules with $\mathbb T$ or $\mathbb O$ symmetry}\label{alg:relative} 
  \begin{algorithmic}[1]
    \Input{ (i) $SO_G(3)$
    \quad
    (ii) Images $\hat P_{R_i}$, $i \in [N]$}
    \For{$i<j = 1,\ldots,N$}
      \State $(\tilde R_{ij},\tilde R_{ji}) \gets \underset{\substack{(Q_r, Q_s) \in SO_G(3) \times SO_G(3)}}{\operatorname{\arg max}} \pi_{ij} (Q_r, Q_s)$
    \Comment See~\eqref{eq:ml score}
    \EndFor
    \Output{$\{ (\tilde R_{ij}, \tilde{R}_{ji}) \}_{i<j \in [N]}$}
  \end{algorithmic}
\end{algorithm}

%%%% shifts %%%%

% \rvg{
% In order to handle the presence of shifts between the Fourier-transformed projection-images $\hat P_{R_i}$ and $\hat P_{R_j}$, 
% we first select a maximal allowed shift $\Delta s_{max}$, a shift resolution $\delta_s$ and set $M = \lceil\Delta s_{max}/\delta_s\rceil$.
% For all shifts $\Delta s = m\delta_s$, $m=-M,\dots,M$, the shifted projection-image $\hat P_{R_j}^m$ is where each of its Fourier rays has been shifted by a one-dimensional shift $\Delta s$, and is denoted
% by $\nu_{i, \theta}^m(\xi)$.
% Then~\eqref{eq:NNC} is replaced by
% \begin{equation} \label{eq:NNC shifts}
%  \rho_{ij}(\theta, \phi) = \underset{m=-M,\dots,M}{\operatorname{ max}} \Re \ 
% \int_{0}^{\infty}\frac{(\nu^m_{i, \theta}(\xi))^{*}\nu_{j, \phi}(\xi)d\xi}
% {||\nu^m_{i, \theta}(\xi)||_{L_2}||\nu_{j, \phi}(\xi)||_{L_2}}.
% \end{equation}
% }

The Fourier projection-slice theorem~\eqref{eq:Projection slice theorem} relies on the stipulation that the centers of all projection-images coincide with the center of the three-dimensional molecule.
In practice, it is unlikely that all projection-images are simultaneously aligned with respect to a common three-dimensional origin, making the Fourier projection-slice theorem, as stated in~\eqref{eq:Projection slice theorem}, not applicable to pairs of experimental projection-images.
Thus, the procedure for detecting common lines between projection-images
% which is given in Section~\ref{capter:solving the orientation assignment problem} below as $\pi_{ij}$,
needs to be modified 
% along the lines of~\cite{shifts} 
in order to handle the presence of unknown shifts, as we now describe.
% XXX Here is my suggestion: The Fourier slice theorem~\eqref{eq:PRi=PRj} assumes that all projection images are properly centered (that is, were generated according to~\eqref{eq:line integrals}). 
% However, in practice, 
Since
each input image is centered differently, 
% namely, 
we do not observe $P_{R_{i}}(x,y)$ that follows~\eqref{eq:line integrals}, but rather $P_{R_{i}}(x-\Delta x_{i},y-\Delta y_{i})$ for some unknown $(\Delta x_{i}, \Delta y_{i})$. This results in the Fourier ray~\eqref{eq:fourier ray} in the direction of a common line being multiplied by some phases that correspond to a one-dimensional shift along the common line, though the directions of the common lines~\eqref{eq:recover angles} do not change (see~\cite{springer2012} for a detailed derivation). Thus, we replace in~\eqref{eq:score cl scl} each $\rho_{ij}$ (see~\eqref{eq:NNC}) by
\begin{equation*}
\rho_{ij}(\theta, \phi) = \max_{s} \Re \ 
\int_{0}^{\infty}\frac{(\nu^s_{i, \theta}(\xi))^{*}\nu_{j, \phi}(\xi)d\xi}
{||\nu^s_{i, \theta}(\xi)||_{L_2}||\nu_{j, \phi}(\xi)||_{L_2}},
\end{equation*}
where $\nu^s_{i, \theta}(\xi))$ is the Fourier ray $\nu_{i, \theta}(\xi)$  (see~\eqref{eq:fourier ray}) multiplied by the phases that correspond to a shift of the common line by $s$. In practice, we maximize over a finite set of values of~$s$ (for example, up to 10\% of the size of the image in steps of 1~pixel). We make a similar change in $\rho_{ii}$ and $\rho_{jj}$ in~\eqref{eq:score cl scl}.

%%% handedness ambiguity %%%

Due to the inherent handedness ambiguity of~\eqref{eq:handedness ambiguity},
the images $\hat P_{R_i} $ and $\hat{\tilde P}_{JR_iJ} $ are identical. 
Thus, the common line between each pair of projection-images $\hat P_{R_i} $ and $\hat P_{g^{(k)}R_j} $, $i<j \in [N]$, $k \in [n]$, is identical to the common line between the pair of projection-images $\hat{\tilde P}_{JR_iJ} $ and $\hat{\tilde P}_{Jg^{(k)}R_jJ} $, 
and so the set of common lines between $\hat P_{R_i} $ and $\hat P_{R_j} $ is identical to the set of common lines between $\hat{\tilde P}_{JR_iJ} $ and $\hat{\tilde P}_{JR_jJ} $.
Similarly, the self common lines of each projection-image $\hat P_{R_i} $ are identical to the self common lines of the projection-image $\hat{\tilde P}_{JR_iJ} $. 
Since by a direct calculation it can be shown for the symmetry group $G$ that $\{Jg^{(k)}J\}_{k=1}^n = \{g^{(k)}\}_{k=1}^n$,
% {Say the this is for $G$ equal to $\mathcal{T}$ or $\mathcal{O}$?} 
it holds that $$\{ (JR_i J)^Tg^{(k)}J{R_j}J\}_{k=1}^n=
\{ JR_i^T g^{(k)}{R_j}J\}_{k=1}^n.$$
According to~\eqref{eq:recover angles} and~\eqref{eq:recover angles scl}, we note that the set
$\{ JR_i^T g^{(k)}{R_j}J\}_{k=1}^n$
produces the same set of local coordinates as the set $\{ R_i^T g^{(k)}{R_j}\}_{k=1}^n$,
and thus also maximizes $\pi_{ij}$ of~\eqref{eq:score cl scl}.
As a result, 
the pair $(\tilde R_{ij}, \tilde R_{ji})$ estimates either the pair $( R_{ij},  R_{ji})$ or the pair $( JR_{ij}J, JR_{ji}J)$,
% the set 
% $\{ \tilde R^T_{ij} g^{(k)} \tilde R_{ij} \}_{k=1}^n$
% estimates either the set  $\{ R_i^T g^{(k)}{R_j}\}_{k=1}^n$
% \rv{
% (in case the pair $(\tilde R_{ij}, \tilde R_{ji})$ estimates the pair $( R_{ij},  R_{ji})$)
% }
% or the set  $\{ JR_i^T g^{(k)}{R_j}J\}_{k=1}^n$ 
% \rv{
% (in case the pair $(\tilde R_{ij}, \tilde R_{ji})$ estimates the pair $( JR_{ij}J, JR_{ji}J)$)
% }
yet it is impossible to distinguish between the two. %sets.
Moreover, the estimate for each pair of indices ($i$,$j$) is independent from other pairs of indices. 
% Therefore, 
% we apply the handedness synchronization procedure~\cite{pragier2016graph} in order to approximate one of the sets 
% \begin{equation} \label{eq:J ync to one set}
% \{\{ R^T_{i} g^{(k)} R_{j} \}_{k=1}^n\}_{i<j \in [N]} 
% \quad \text{or} \quad  
% \{\{ J R^T_{i} g^{(k)}  R_{j}J \}_{k=1}^n \}_{i<j \in [N]}.
% \end{equation}
% From now on, we assume without loss of generality that
% each pair 
% $ (\tilde R_{ij}, \tilde{R}_{ji}) $
% computed by Algorithm~\ref{alg:relative}, $i<j \in [N]$, satisfies that 
% $\{ \tilde R^T_{ij} g^{(k)} \tilde R_{ij} \}_{k=1}^n$ is an estimate of
% $ \{ R^T_{i} g^{(k)} R_{j} \}_{k=1}^n. $
Therefore, 
we apply the handedness synchronization procedure~\cite{pragier2016graph} in order to partition the set $\{ (\tilde R_{ij}, \tilde R_{ji}) \ | \ i<j \in [N]\}$ into two subsets, given by
\begin{equation}
    \begin{split}
        & A = \{ (\tilde R_{ij}, \tilde R_{ji}) \ | \ (\tilde R_{ij}, \tilde R_{ji}) \text{ estimates } ( R_{ij},  R_{ji})\},  \\ &
        B = \{ (\tilde R_{ij}, \tilde R_{ji}) \ | \ (\tilde R_{ij}, \tilde R_{ji}) \text{ estimates } ( JR_{ij}J,  JR_{ji}J)\}.
    \end{split}
\end{equation}
Then we choose either one of the subsets, and replace each estimate $(\tilde R_{ij}, \tilde R_{ji})$ in it with $(J\tilde R_{ij}J, J\tilde R_{ji}J)$. Since $J^2 = I$, all estimates are now consistent with the same hand.
From now on, we assume without loss of generality that
each pair 
$ (\tilde R_{ij}, \tilde{R}_{ji}) $
computed by Algorithm~\ref{alg:relative}, $i<j \in [N]$, estimates $ ( R_{ij}, {R}_{ji}) $.

% ===============================================================
% Chpter 6: Rotations estimation
% ===============================================================

Once we have computed  $ ( \tilde R_{ij}, \tilde {R}_{ji})$ for all $ i<j \in [N]$ (using Algorithm~\ref{alg:relative}), which estimate $ ( R_{ij}, {R}_{ji})$ satisfying~\eqref{eq:common line approach}, in the second step of
the proposed method, we estimate the orientations of all projection-images 
$\{P_{R_i}\}_{i \in [N]} $.
% namely the set of rotations
% $\{g_iR_i\}_{i=1}^N$, 
% where each $g_i \in G$ is an arbitrary symmetry group element.
%%% estimation to Ri and Rij : proposition and corollary %%%
This step relies on the following two propositions and corollary,
showing that $ R_{ij}$ and ${R}_{ji}$ which satisfy~\eqref{eq:common line approach} are not unique, and may be expressed by  $R_{i}$ and ${R}_{j}$ of~\eqref{eq:common line approach} up to a symmetry group element in~$\mathbb{O}$. As a result, the pair $ ( \tilde R_{ij}, \tilde {R}_{ji})$ of~\eqref{eq:ml score} is also not unique, in the sense that it may estimate any pair $ (  R_{ij},  {R}_{ji})$ which satisfies~\eqref{eq:common line approach}.

% \rv{
% We first show using the following propositions that $ R_{ij}$ and ${R}_{ji}$ of~\eqref{eq:common line approach}, which are estimated by $\tilde R_{ij}$ and $\tilde{R}_{ji}$ 
% (where $(\tilde R_{ij}, \tilde{R}_{ji})$ computed by Algorithm~\ref{alg:relative}), respectively,
% may be expressed by  $R_{i}$ and ${R}_{j}$ (the ground truth rotations by which the projection-images $P_{R_i}$ and $P_{R_j}$ were formed) up to a symmetry group element in~$\mathbb{O}$.
% The relation between the two pairs is important for the second step of our proposed method.
% }
% We first show using the following propositions that $\tilde R_{ij}$ and $\tilde{R}_{ji}$ of each pair
% $(\tilde R_{ij}, \tilde{R}_{ji})$
% computed by Algorithm~\ref{alg:relative}
% may be expressed by
%  $R_{i}$ and ${R}_{j}$ up to a symmetry group element in~$\mathbb{O}$.
We start by recalling that the normalizer of a subgroup $\tilde G$ in a group $\tilde H$ $(\tilde G \subseteq \tilde H)$ is given by
$$N_{\tilde H}(\tilde G) = \{\tilde h \in \tilde H: \tilde h^T\tilde G \tilde h = \tilde G\}.$$
\begin{proposition} \label{prop: unique relation}
Let $  R_{ij}, {R}_{ji}$ and $ R_{i},{R}_{j}$ be two pairs of rotations
satisfying~\eqref{eq:common line approach}, $i<j \in [N]$.
Then $  R_{ij}, {R}_{ji}$ and  $ R_{i},{R}_{j}$ satisfy
\begin{equation} \label{eq:relation to ground truth}
 R_{ij} = h_{ij}R_i \textit{ and }  R_{ji} = h_{ji}R_j,
\quad h_{ij},h_{ji}\in N_{SO(3)}(G).                                    
\end{equation}
\end{proposition} 
\begin{proof}
Since  $ R_{ij}, {R}_{ji}, R_{i}, {R}_{j} \in SO(3)$, there exist $h_{ij},h_{ji}\in SO(3)$ such that
\begin{equation} \label{eq:relation to ground truth 2}
 R_{ij} = h_{ij}R_i \textit{ and } R_{ji} = h_{ji}R_j.
\end{equation}
Substituting~\eqref{eq:relation to ground truth 2}
into~\eqref{eq:common line approach}
and multiplying both sides of the resulting equation by $R_i$ from the left and by $R_j^T$ from the right results in
\begin{equation} \label{eq:H is normalizer 1}
\{ h_{ij}^Tg^{(k)} h_{ji}\}_{k=1}^n = \{ g^{(k)} \}_{k=1}^n.
\end{equation}
Since $I \in G$ (the identity element of $G$), we deduce from~\eqref{eq:H is normalizer 1} that there exists $ g_{ji} \in G$
such that $$h_{ij}^T I h_{ji} = h_{ij}^T h_{ji} = g_{ji}$$
and thus 
\begin{subequations}
\begin{equation} \label{eq:H is normalizer 3}
h_{ji}= h_{ij}g_{ji}, 
\end{equation} 
\begin{equation} \label{eq:H is normalizer 5}
 h_{ij}^T = g_{ji}h_{ji}^T.
\end{equation}
\end{subequations}
Plugging~\eqref{eq:H is normalizer 3} and~\eqref{eq:H is normalizer 5}
into~\eqref{eq:H is normalizer 1} results in
\begin{subequations}
\begin{equation} \label{eq:H is normalizer 2}
\{  h_{ij}^Tg^{(k)} h_{ij}g_{ji} \}_{k=1}^n =
\{ g^{(k)} \}_{k=1}^n,
\end{equation}
\begin{equation} \label{eq:H is normalizer 4}
\{  g_{ji}h_{ji}^Tg^{(k)} h_{ji} \}_{k=1}^n =
\{ g^{(k)} \}_{k=1}^n.
\end{equation}
\end{subequations}
Since $G$ is a finite group, it holds that 
$\{ g^{(k)} g_{ji}^T \}_{k=1}^n = \{ g_{ji}^T g^{(k)} \}_{k=1}^n = \{ g^{(k)} \}_{k=1}^n$.
Thus, by multiplying~\eqref{eq:H is normalizer 2}
 by $g_{ji}^T$ from the right and by multiplying~\eqref{eq:H is normalizer 4}
 by $g_{ji}^T$ from the left, we get
\begin{equation} \label{eq:H is normalizer}
\begin{split}
\{  h_{ij}^Tg^{(k)} h_{ij} \}_{k=1}^n = \{ g^{(k)} \}_{k=1}^n,
\\
\{  h_{ji}^Tg^{(k)} h_{ji} \}_{k=1}^n = \{ g^{(k)} \}_{k=1}^n.
\end{split}
\end{equation}
Equation~\eqref{eq:H is normalizer} implies that $h_{ij}$ and $h_{ji}$ belong to the normalizer of the group $G$ in $SO(3)$, i.e, $h_{ij}, h_{ji}\in N_{SO(3)}(G)$.
\end{proof}
\begin{proposition} \label{prop: T O normalizer}
$N_{SO(3)}(\mathbb T) = \mathbb O \text{ and } N_{SO(3)}(\mathbb O) = \mathbb O.$
\end{proposition} 
The proof of Proposition~\ref{prop: T O normalizer} is given in Appendix~\ref{prop: group normalizers}.
%%% estimation to Ri and Rij : corollary %%%
\begin{corollary}\label{corollary: unique relation}
Let $  R_{ij}, {R}_{ji}$ and $ R_{i},{R}_{j}$ be two pairs of rotations
satisfying~\eqref{eq:common line approach}, $i<j \in [N]$.
For the symmetry group $\mathbb T$ it holds that
\begin{equation} \label{eq:relation to ground truth T}
 R_{ij} = h_{ij}g_{ij}R_i 
\textit{ and } 
 R_{ji} = h_{ij}g_{ji}R_j,
\quad g_{ij}, g_{ji}\in \mathbb T, \quad h_{ij}\in \mathbb O.
\end{equation}
For the symmetry group $\mathbb O$ it holds that
\begin{equation} \label{eq:relation to ground truth O}
 R_{ij} = g_{ij}R_i \textit{ and }  R_{ji} = g_{ji}R_j,
\quad g_{ij},g_{ji}\in \mathbb O.                                               
\end{equation}
\end{corollary}
\begin{proof}
For the symmetry group $\mathbb T$,
by~\eqref{eq:H is normalizer 3} and Proposition~\ref{prop: T O normalizer}, there exist $g_{ji}\in \mathbb T$ and $h_{ji}, h_{ij} \in \mathbb O $ such that $h_{ji} = h_{ij} g_{ji}$. 
Thus by~\eqref{eq:relation to ground truth},
$ R_{ji} = h_{ji}R_j = h_{ij}g_{ji}R_j$
and $ R_{ij} = h_{ij}R_i = h_{ij}g_{ij}R_i$ with $g_{ij}=I\in \mathbb T$, implying~\eqref{eq:relation to ground truth T}.
For the symmetry group $\mathbb O$,~\eqref{eq:relation to ground truth O} follows directly from~\eqref{eq:relation to ground truth} and Proposition~\ref{prop: T O normalizer}.
\end{proof}

% We next show how to estimate the 
% % set of rotations $\{g_iR_i\}_{i=1}^N$ 
% \rv{orientations}
% of the projection-images 
% $\{P_{R_i}\}_{i \in [N]} $.

We now describe the second step of our proposed method, i.e. how to estimate the orientations of the projection-images 
$\{P_{R_i}\}_{i \in [N]} $ from the set of rotations $\{(\tilde R_{ij}, \tilde R_{ji}) \}_{i<j \in [N]}$ computed by Algorithm~\ref{alg:relative}. This step fundamentally relies on the choice of axes in Table~\ref{tab:symmetry-group-elements}, which implies that the matrices corresponding to the group elements of~$\mathbb T$ (Appendix~\ref{tab: T symmetry group elements representations}) and~$\mathbb O$ (Appendix~\ref{tab: O symmetry group elements representations}) all have exactly one nonzero entry in each row and each column which is equal to either $1$ or $-1$.
A key property of these symmetry group elements is that they may be represented uniquely using addition and subtraction of single entry matrices, defined as follows.

\begin{definition}\label{def:single entry matrix}
A single-entry matrix, denoted by $e_{ij} \in \mathbb{R}^{3\times3}$, is a matrix whose $(i,j)$ element is one and the rest of its elements are zero.
Moreover, we define 
$$
e_{(-i)(-j)} = e_{ij},\quad e_{(-i)j} = e_{i(-j)} = -e_{ij}.
$$
\end{definition}

%%% one line notation %%%

\begin{definition}\label{def:one line notation} 
Given a symmetry group element $g$ from Appendix~\ref{tab: T symmetry group elements representations} for~$\mathbb T$ or Appendix~\ref{tab: O symmetry group elements representations} for~$\mathbb O$,
we define the one-line notation of $g$ by the vector
$\sigma = (\sigma(1)\ \sigma(2)\ \sigma(3))$
given by 
$$
\sigma^T = g
\begin{pmatrix}
1\\
2\\
3
\end{pmatrix}.
$$
\end{definition}
In words, we multiply the matrix that corresponds to a group element by the vector $(1,2,3)^{T}$.
 
%%% symmetry group elements representation %%%

\begin{lemma}\label{lema:group rep}
Each symmetry group element $g$ from Appendix~\ref{tab: T symmetry group elements representations} for~$\mathbb T$ or Appendix~\ref{tab: O symmetry group elements representations} for~$\mathbb O$ may be represented uniquely by
the sum 
$$g = e_{1\sigma(1)}+ e_{2\sigma(2)}+ e_{3\sigma(3)},$$
where $\sigma$ is the one-line notation of $g$, and $e_{m\sigma(m)}$, $m=1,2,3$, are given in Definition~\ref{def:single entry matrix}.
\end{lemma}

\begin{proof}
By a direct calculation using the representation given in Appendix~\ref{tab: T symmetry group elements representations} for $\mathbb T$ and Appendix~\ref{tab: O symmetry group elements representations} for $\mathbb O$.
\end{proof}

Recall that due to the symmetry ambiguity (discussed in Section~\ref{chapter:Intro}), all orientations assignments of the form $\{g_iR_i\}_{i=1}^N$ , where $g_i \in G$ is an arbitrary symmetry group element, are consistent with the same set of images $\{P_{R_i}\}_{i \in [N]} $.
Hence, there are $n^N$ valid assignments while only one of them is required.
Therefore, our method is designed to obtain one arbitrary valid assignment $\{g_iR_i\}_{i=1}^N$ from this set of $n^{N}$ valid assignments.

% \rvc{Therefore, our method was designed to obtain one (unknown) valid assignment.}{Therefore, our method is designed to obtain one arbitrary valid assignment from this set of $n^{N}$ assignments.}

The key idea of our method for obtaining a valid assignment $\{g_iR_i\}_{i=1}^N$ to the set of projection-images $\{P_{R_i}\}_{i \in [N]} $, 
% \rvc{is to retrieve the three rows of each $g_iR_i$, $i \in [N]$, as described in the following.}{How about something like ``
is to estimate one of the rows of all matrices $g_{i}R_{i}$ simultaneously, then another row, and finally the last row, and then assemble the matrices $g_{i}R_{i}$ from these estimations. Note that we do not know which row of the matrices (first, second, or third) we estimate at each step, as explained below.
% ''}

Let $\{g_iR_i\}_{i=1}^N$ be any valid assignment to the set of projection-images 
$\{P_{R_i}\}_{i \in [N]} $, and let $m=1,2,3$.
We denote the one-line notation of each $g_i \in G$, $i \in [N]$, from $\{g_iR_i\}_{i=1}^N$ by $\sigma_i$
(see Definition~\ref{def:one line notation}),  
and define 
$\sigma = (\sigma_1,\sigma_2,\dots,\sigma_N)$.
We also denote the $m$th row of $R_i$ by $v_i^{(m)}$, 
and for simplicity, we denote by $v_i^{(-m)}$ the vector $-v_i^{(m)}$, i.e., the minus of the $m$th row of $R_i$.
In particular, $v_i^{(\sigma_i(m))}$ is the $m$th row of $g_iR_i$ since 
$$
g_iR_i = (e_{1\sigma_i(1)} + e_{2\sigma_i(2)} + e_{3\sigma_i(3)})
\begin{pmatrix}
-\  v_i^{(1)}\ -\\
-\  v_i^{(2)}\ -\\
-\  v_i^{(3)}\ -
\end{pmatrix}
=
\begin{pmatrix}
-\  v_i^{(\sigma_i(1))}\ -\\
-\  v_i^{(\sigma_i(2))}\ -\\
-\  v_i^{(\sigma_i(3))}\ -
\end{pmatrix},
$$
where the first equality follows by expressing $g_i$ using Lemma~\ref{lema:group rep} and the second equality follows by a direct calculation.
Moreover, let $\tau$ denote a one-line notation corresponding to an arbitrary symmetry group element $ g_{\tau} \in \mathbb{O}$. Then $v_i^{(\sigma_i(\tau(m)))}$ is the $\tau(m)$th row of $g_iR_i$.
Finally, denote by $v_{\sigma,\tau(m)}$ the concatenation (as a row vector of length $3N$) of the $\tau(m)$th row of each $g_iR_i$, $i \in [N]$, i.e.
\begin{equation} \label{def:v_sig,m} 
v_{\sigma,\tau(m)} = (v_1^{(\sigma_{1}(\tau(m)))}, \dots , v_N^{(\sigma_{N}(\tau(m)))}),
\end{equation}
and define
\begin{equation} \label{H factorization}
H_{\sigma,m} = v_{\sigma,\tau(m)}^Tv_{\sigma,\tau(m)}.
\end{equation}
Then $H_{\sigma,m}$ is a rank-1 $3N\times 3N$ block matrix 
 whose $(i,j)$ $3\times3$ block is given by the rank-1 matrix $v_i^{(\sigma_i(\tau(m)))^T}v_j^{(\sigma_{j}(\tau(m)))}$.

Once we construct the three matrices $H_{\sigma,m}$, $m=1,2,3$, then
factorizing each matrix $H_{\sigma,m}$ using SVD results in 
either the vector $v_{\sigma,\tau(m)}$ or the vector $-v_{\sigma,\tau(m)}$, which we denote by $s_m \cdot v_{\sigma,\tau(m)}$, $s_m \in \{-1,1\}$.
% the vector $v_{\sigma,r}$
% \rv{a permutation matrix which is identified with the permutation $\tau$ or $-\tau$}. 
%with $\pm 1$ on its diagonal.
Note that
\begin{align*}
\begin{pmatrix}
-\ s_1 \cdot v_{\sigma,\tau(1)}\ -\\
-\ s_2 \cdot v_{\sigma,\tau(2)}\ -\\
-\ s_3 \cdot v_{\sigma,\tau(3)}\ -
\end{pmatrix} 
& =
S \cdot
\begin{pmatrix}
-\ v_{\sigma,\tau(1)}\ -\\
-\ v_{\sigma,\tau(2)}\ -\\
-\ v_{\sigma,\tau(3)}\ -
\end{pmatrix} 
=
S \cdot
g_{\tau} \cdot
\begin{pmatrix}
-\ v_{\sigma,1}\ -\\
-\ v_{\sigma,2}\ -\\
-\ v_{\sigma,3}\ -
\end{pmatrix} 
\\ & =
S \cdot
g_{\tau} \cdot
\begin{pmatrix}
v_1^{(\sigma_{1}(1))}, \dots , v_N^{(\sigma_{N}(1))}\\
v_1^{(\sigma_{1}(2))}, \dots , v_N^{(\sigma_{N}(2))}\\
v_1^{(\sigma_{1}(3))}, \dots , v_N^{(\sigma_{N}(3))}
\end{pmatrix},
\end{align*}
where $S = \operatorname{diag}(s_1,s_2,s_3)$.
% \rvc{where $S = \operatorname{diag}(s_1,s_2,s_3)$.}{Maybe add something like ``
The latter equation together with~\eqref{def:v_sig,m} and~\eqref{H factorization} means that if we are able to construct the matrices $H_{\sigma,m}$, then factorizing each of these matrices gives us simultaneously the rows of all matrices $g_{i}R_{i}$, up to multiplication by the matrix $O=S\cdot g_{\tau}$, $O \in O(3)$.
If $\det (Og_iR_i) = -1$, we simply multiply $Og_iR_i$ by $-1$, and thus, we may assume without loss of generality that $ O$ is a rotation. 
The matrix~$O$ is an inherent degree of freedom of the orientation assignment problem, with $\{Og_iR_i\}_{i=1}^N$ being a valid solution.

Next, we describe how to construct
three rank-1 matrices $H_{\sigma,m}$, $m=1,2,3$,
given a set of matrices $\{(R_{ij}, R_{ji})\}_{i<j \in [N]}$ where each $(R_{ij}, R_{ji})$  satisfies~\eqref{eq:common line approach}. Since by Proposition~\ref{prop: unique relation} there are many sets of matrices $\{(R_{ij}, R_{ji})\}_{i<j \in [N]}$ satisfying~\eqref{eq:common line approach}, there are also many possible triplets of matrices $H_{\sigma,m}$, $m=1,2,3$, and our algorithm will return one of these triplets.
Overall, we will obtain a valid assignment to the set of projection-images $\{P_{R_i}\}_{i \in [N]} $.

% so that $H_{\sigma,m}$ encodes the $\tau(m)$th row of all matrices $g_{i}R_{i}$, $i \in [N]$,
% where each $g_{i} \in G$ is an arbitrary symmetry group element, and $\tau$ is a one-line notation corresponding to an arbitrary symmetry group element in $\mathbb{O}$. 
% Since each $g_{i}$ is arbitrary, there are many such triplets of matrices, and our algorithm will return one of these triplets.

% a set of matrices $H_{\sigma,m}$, $m=1,2,3$, from which we obtain the valid assignment $\{g_iR_i\}_{i=1}^N$ to the set of projection-images $\{P_{R_i}\}_{i \in [N]} $.

We start by showing how to recover 
from the pair $(R_{ij}, R_{ji})$ of~\eqref{eq:common line approach}
the rank-1 $3 \times 3$ matrices 
$(v_i^{(\sigma_{ij}(\tau_{ij}(m)))})^Tv_j^{(\sigma_{ji}(\tau_{ij}(m)))}$, $m=1,2,3$, $i<j\in[N]$,
% from $(R_{ij}, R_{ji})$ of~\eqref{eq:common line approach} for all $i<j\in[N]$, 
where $v_i^{(\sigma_{ij}(\tau_{ij}(m)))}$
and $v_j^{(\sigma_{ji}(\tau_{ij}(m)))}$ are the $\tau_{ij}(m)$th rows of $g_{ij} R_i$ and $g_{ji} R_j$, with $g_{ij},g_{ji} \in G$, whose one-line notations are $\sigma_{ij}$, $\sigma_{ji}$ respectively, and $\tau_{ij}$ is a one-line notation corresponding to a symmetry group element in $\mathbb{O}$.
These matrices are the building blocks 
from which the three rank-1 $3N \times 3N$ block matrices $H_{\sigma,m}$, $m=1,2,3$, will be constructed. We will use the following lemma, whose proof is given in Appendix~\ref{Proof of Lemma sym prop}.

%%% symmetry group elements properties %%%

\begin{lemma}\label{lema:group properties}
Let $g_1, g_2$
be any two symmetry group elements from Appendix~\ref{tab: T symmetry group elements representations} for $\mathbb T$ or Appendix~\ref{tab: O symmetry group elements representations} for $\mathbb O$,
with $\sigma_1,\sigma_2 $ being their one-line notations, respectively.
Then,
\begin{align}
g_1^Te_{mm}g_2 &= e_{\sigma_1(m)\sigma_2(m)}, \quad m=1,2,3, \label{eq:property 1} \\
g_1^Te_{mm}g_1 &= e_{\sigma_1(m)\sigma_1(m)}, \quad m=1,2,3, \label{eq:property 2} \\
\{\pm g_1^Te_{mr}g_2\}_{m,r=1}^3 &=  \{\pm e_{mr}\}_{m,r=1}^3. \label{eq:property 3}
\end{align}
\end{lemma}
Using Lemma~\ref{lema:group properties}, we prove the following proposition, which relates $(R_{ij}, R_{ji})$ of~\eqref{eq:common line approach} with 
the rank-1 $3 \times 3$ matrices 
$(v_i^{(\sigma_{ij}(\tau_{ij}(m)))})^Tv_j^{(\sigma_{ji}(\tau_{ij}(m)))}$, $m=1,2,3$.
\begin{proposition} \label{prop:relative rows}
Let $  R_{ij}, {R}_{ji}$ and $ R_{i},{R}_{j}$ be two pairs of rotations
satisfying~\eqref{eq:common line approach}, $i<j \in [N]$.
Then, for $m=1,2,3$,
\begin{equation} \label{eq:rows products}
 R_{ij}^Te_{mm} R_{ji}=
(v_i^{(\sigma_{ij}(\tau_{ij}(m)))})^Tv_j^{(\sigma_{ji}(\tau_{ij}(m)))},
\end{equation}
where the matrices $e_{mm}$, $m=1,2,3$, are single entry matrices defined in Definition~\ref{def:single entry matrix},
$v_i^{(\sigma_{ij}(\tau_{ij}(m)))}$
and $v_j^{(\sigma_{ji}(\tau_{ij}(m)))}$ are the $\tau_{ij}(m)$th rows of $g_{ij} R_i$ and $g_{ji} R_j$, with $g_{ij},g_{ji} \in G$,  whose one-line notations are $\sigma_{ij}$ and $\sigma_{ji}$ respectively, and $\tau_{ij}$ is a one-line notation corresponding to a symmetry group element in $\mathbb{O}$ (see Definition~\ref{def:one line notation} for the definition of a one-line notation).
\end{proposition}
\begin{proof}
For any two symmetry group elements $g_{ij}, g_{ji} \in G$,
% $g_{ij}, g_{ji} \in \mathbb{T}$ or $g_{ij}, g_{ji} \in \mathbb{O}$
it holds by~\eqref{eq:property 1} that for $m=1,2,3$
\begin{equation}  \label{eq:rows products 1}
g_{ij}^{T}e_{mm}g_{ji}
=e_{\sigma_{ij}(m)\sigma_{ji}(m)},
\end{equation}
where $\sigma_{ij},\sigma_{ji}$ are the one-line notations of $g_{ij}, g_{ji} \in G$, respectively.
For any $h_{ij} \in \mathbb{O}$, it holds by~\eqref{eq:property 2} that for $m=1,2,3$
\begin{equation}  \label{eq:rows products 2}
h_{ij}^{T}e_{mm}h_{ij}
=  e_{\tau_{ij}(m)\tau_{ij}(m)},
\end{equation}
where $\tau_{ij}$ is the one-line notation of $h_{ij}$.
% Since by Definition~\ref{def:one line notation} $e_{\tau_{ij}(m)\tau_{ij}(m)}=e_{|\tau_{ij}(m)||\tau_{ij}(m)|}$, we may assume that $\tau_{ij}(m)=|\tau_{ij}(m)|$, which implies together with table~\ref{tab: O symmetry group elements representations} that $\tau_{ij}$ is a permutation on the wet $\{1,2,3\}$.

For the symmetry group $\mathbb T$, we get by~\eqref{eq:relation to ground truth T} that for $m=1,2,3$
% \begin{equation} \label{eq: rows product T}
\begin{align*}
 R_{ij}^Te_{mm} R_{ji} &=
R_i^{T}g_{ij}^Th_{ij}^Te_{mm}h_{ij}g_{ji}R_j = 
R_i^{T}g_{ij}^Te_{\tau_{ij}(m)\tau_{ij}(m)}g_{ji}R_j      \\ & =
R_i^{T}e_{\sigma_{ij}(\tau_{ij}(m))\sigma_{ji}(\tau_{ij}(m))}R_j  =
(v_i^{(\sigma_{ij}(\tau_{ij}(m)))})^Tv_j^{(\sigma_{ji}(\tau_{ij}(m)))},
\end{align*}
% \end{equation}
where the second equality follows from~\eqref{eq:rows products 2}, the third equality follows from~\eqref{eq:rows products 1}, and
the last equality follows by a direct calculation.
For the symmetry group $\mathbb O$, we get by~\eqref{eq:relation to ground truth O} that for $m=1,2,3$
\begin{align*}
 R_{ij}^Te_{mm} R_{ji} &= 
R_i^{T}g_{ij}^{T}e_{mm}g_{ji}R_j = 
R_i^{T}e_{\sigma_{ij}(m)\sigma_{ji}(m)}R_j  =
(v_i^{(\sigma_{ij}(m))})^Tv_j^{(\sigma_{ji}(m))},
\end{align*}
where the second equality follows from~\eqref{eq:rows products 1}, and the last equality follows by a direct calculation.
For convenience only, we write for the symmetry group $\mathbb O$
\begin{align*}
 R_{ij}^Te_{mm} R_{ji} &=
% \rvc{\tilde R_{ij}^Te_{mm}\tilde R_{ji}}{Why do you have a tilde?} &=
(v_i^{(\sigma_{ij}(m))})^Tv_j^{(\sigma_{ji}(m))} =
(v_i^{(\sigma_{ij}(\tau_{ij}(m)))})^Tv_j^{(\sigma_{ji}(\tau_{ij}(m)))},
\end{align*}
where $\tau_{ij}(m) = m$, so we use consistent notation for both symmetry groups $\mathbb{T}$ and $\mathbb{O}$.
% For convenience only, we write for the symmetry group $\mathbb T$ (see~\eqref{eq: rows product T})
% \begin{align*}
% \{\tilde R_{ij}^Te_{mm}\tilde R_{ji}\}_{m=1}^3 &=
% \{(v_i^{(m)})^Tv_j^{(\sigma_{ji}(m))}\}_{m=1}^3 =
% \{(v_i^{(\sigma_{ij}(m))})^Tv_j^{(\sigma_{ji}(m))}\}_{m=1}^3,
% \end{align*}
% where $\sigma_{ij}(m) = m$, so we use consistent notation for both symmetry groups $\mathbb{T}$ and $\mathbb{O}$.
\end{proof}

We next construct the $3N \times 3N$ matrices $H_{\sigma,m}$, $m=1,2,3$, by setting their $3 \times 3$ blocks  one by one using Proposition~\ref{prop:relative rows}, making sure at each step  
that each $H_{\sigma,m}$ is a rank-1 matrix satisfying  $H_{\sigma,m} = v_{\sigma,\tau(m)}^Tv_{\sigma,\tau(m)}$, where $v_{\sigma,\tau(m)}$ is the concatenation of the $\tau(m)$th row of  $g_iR_i$, $i \in [N]$, each $g_{i} \in G$ is an arbitrary symmetry group element, and $\tau$ is a one-line notation corresponding to an arbitrary symmetry group element in~$\mathbb{O}$.

We start by setting the $(1,2)$ blocks of the matrices $H_{\sigma,m}$, $m=1,2,3$, though the method may be adjusted to start with any other block $(i,j)$, $i<j \in [N]$. %or $i>j \in [N]$.
% i.e., selecting the $(1,2)$ block to start with is an arbitrary choice.
% \rvc{By Proposition~\ref{prop:relative rows}, we compute from $R_{12},R_{21}$ which satisfy~\eqref{eq:common line approach} the three rank-1 $3\times 3$ matrices
% $\{(v_1^{(\sigma_{12}(\tau_{12}(m)))})^Tv_2^{(\sigma_{21}(\tau_{12}(m)))}\}_{m=1,2,3}$,
% where $v_1^{(\sigma_{12}(\tau_{12}(m)))}$
% and $v_2^{(\sigma_{21}(\tau_{12}(m)))}$ are the $\tau_{12}(m)$th rows of $g_{12} R_1$ and $g_{21} R_2$, respectively,
% with $\sigma_{12},\sigma_{21}$ being the one-line notations corresponding to $g_{12},g_{21} \in G$, and $\tau_{12}$ is a one-line notation corresponding to a symmetry group element in~$\mathbb{O}$.
% We then set
% the $(1,2)$ block of the matrix $H_{\sigma,m}$ by 
% \begin{equation} \label{block 12}
%  H_{\sigma,m}^{(1,2)} :=  R_{12}^Te_{mm}  R_{21} = 
%  (v_1^{(\sigma_{12}(\tau_{12}(m)))})^Tv_2^{(\sigma_{21}(\tau_{12}(m)))}.
% \end{equation}}
We set
the $(1,2)$ block of the matrix $H_{\sigma,m}$ to be
\begin{equation} \label{block 12}
 H_{\sigma,m}^{(1,2)} :=  R_{12}^Te_{mm}  R_{21}.
\end{equation}
By Proposition~\ref{prop:relative rows}, it holds that $H_{\sigma,m}^{(1,2)}=  (v_1^{(\sigma_{12}(\tau_{12}(m)))})^Tv_2^{(\sigma_{21}(\tau_{12}(m)))}$
where $v_1^{(\sigma_{12}(\tau_{12}(m)))}$
and $v_2^{(\sigma_{21}(\tau_{12}(m)))}$ are the $\tau_{12}(m)$th rows of $g_{12} R_1$ and $g_{21} R_2$, respectively,
with $\sigma_{12},\sigma_{21}$ being the one-line notations corresponding to $g_{12},g_{21} \in G$, and $\tau_{12}$ is a one-line notation corresponding to a symmetry group element in~$\mathbb{O}$.
% where the matrices $R_{12},R_{21}$ satisfy~\eqref{eq:common line approach}, 
% where each $e_{mm}$ is a single entry matrix (see Definition~\ref{def:single entry matrix}). 
% By Proposition~\ref{prop:relative rows} it holds for each $m$ that 
% $R_{12}^Te_{mm}  R_{21} = (v_1^{(\sigma_{12}(\tau_{12}(m)))})^Tv_2^{(\sigma_{21}(\tau_{12}(m)))}$.
% $v_1^{(\sigma_{12}(\tau_{12}(m)))}$
% and $v_2^{(\sigma_{21}(\tau_{21}(m)))}$ are the $\tau_{12}(m)$th rows of $g_{12} R_1$ and $g_{21} R_2$, where
% $\sigma_{12},\sigma_{21}$ are the one-line notations corresponding to $g_{12},g_{21} \in G$, and $\tau_{12}$ is a one-line notation corresponding to a symmetry group element in $\mathbb{O}$.
Thus, the $(1,2)$ block of $H_{\sigma,m}$ is a rank-1 $3 \times 3$ matrix which encodes the $\tau_{12}(m)$th row of $g_{12}R_1$ and $g_{21}R_2$.

Once the $(1,2)$ blocks of the matrices $H_{\sigma,m}$, $m=1,2,3$ have been set, to ensure that the matrices $H_{\sigma,m}$ are of rank-1,
the $(1,i)$ and  $(2,i)$ blocks of each matrix $H_{\sigma,m}$, $i = 3, \dots ,N$, must be of the form 
$(v_1^{(\sigma_{12}(\tau_{12}(m)))})^Tv_i^{(\sigma_{i}(\tau_{12}(m)))}$ and
$(v_2^{(\sigma_{21}(\tau_{12}(m)))})^Tv_i^{(\sigma_{i}(\tau_{12}(m)))}$, respectively, where $\sigma_{i}$ is a one-line notation corresponding to a symmetry group element $g_{i} \in G$.
Without loss of generality, we continue by setting the $(1,i)$ blocks of the matrices $H_{\sigma,m}$, $i = 3, \dots ,N$ (i.e., we could also continue by setting the  $(2,i)$ blocks instead).

% To set the $(1,i)$ blocks of the matrices $H_{\sigma,m}$, $m=1,2,3$, we first note that by
By Proposition~\ref{prop:relative rows},
we can compute from $R_{1i},R_{i1}$ which satisfy~\eqref{eq:common line approach}, $i = 3, \dots ,N$, the three rank-1 $3\times 3$ matrices
$(v_1^{(\sigma_{1i}(\tau_{1i}(r)))})^Tv_i^{(\sigma_{i1}(\tau_{1i}(r)))}$, $r=1,2,3,$
where $v_1^{(\sigma_{1i}(\tau_{1i}(r)))}$
and $v_i^{(\sigma_{i1}(\tau_{1i}(r)))}$ are the $\tau_{1i}(r)$th rows of $g_{1i} R_1$ and $g_{i1} R_i$, respectively,
with $\sigma_{1i},\sigma_{i1}$ being the one-line notations corresponding to $g_{1i},g_{i1} \in G$,
and $\tau_{1i}$ is a one-line notation corresponding to a symmetry group element in $\mathbb{O}$.
Since each product $(v_1^{(\sigma_{1i}(\tau_{1i}(r)))})^Tv_i^{(\sigma_{i1}(\tau_{1i}(r)))}$
is some row of $R_{1}$ (or minus some row of $R_{1}$) times some row of $R_{i}$ (or minus some row of $R_{i}$), it must belong to one of the $H_{\sigma,m}$.
We therefore
% Note that while $v_1^{(\sigma_{12}(\tau_{12}(m)))}$ is the $\tau_{12}(m)$th row of $g_{12} R_1$, it is also the $\sigma_{12}(\tau_{12}(m))$th row of $ R_1$. 
% Similarly, $v_1^{(\sigma_{1i}(\tau_{1i}(r)))}$ is the $\tau_{1i}(r)$th row of $g_{1i} R_1$ but also the $\sigma_{1i}(\tau_{1i}(r)))$th row of $ R_1$. 
% Therefore, the $(1,2)$ block of each $H_{\sigma,m}$, $m=1,2,3$, which encodes the $\tau_{12}(m)$th row of $g_{12} R_1$, also encodes the $\sigma_{12}(\tau_{12}(m))$th row of $ R_1$.
% Since all the $(1,i)$ blocks of $H_{\sigma,m}$, $i \in [N]$, must encode the same row of $R_i$ as in the $(1,2)$ block, 
% must encode the same row of $R_1$ in all the $(1,i)$ blocks, $i \in [N]$, we 
show how to find which
matrix $(v_1^{(\sigma_{1i}(\tau_{1i}(r)))})^Tv_i^{(\sigma_{i1}(\tau_{1i}(r)))}$, $r=1,2,3,$ belongs to which $H_{\sigma,m}$, $m=1,2,3$.
% as the $(1,2)$ block has already been determined.
% \rvc{We therefore show how to find which
	% matrix $(v_1^{(\sigma_{1i}(\tau_{1i}(r)))})^Tv_i^{(\sigma_{i1}(\tau_{1i}(r)))}$, $r=1,2,3,$ belongs to which $H_{\sigma,m}$, $m=1,2,3$.}{Why are these matrices belong to any $H_{\sigma,m}$ at all? Maybe remind the reader that this product is just some row of $R_{i}$ times some row of $R_{j}$? Something like "\ldots (since the latter product is  some row of $R_{i}$ times some row of $R_{j}$, it must below to one of the $H_{\sigma,m}$)''}

% Thus, for each matrix $H_{\sigma,m}$, $m=1,2,3$, we search for $r \in \{1,2,3\}$ such that 
% \rvc{$$
%  R_{1i}^Te_{rr}  R_{i1} = 
% (v_1^{(\sigma_{1i}(\tau_{1i}(r)))})^Tv_i^{(\sigma_{i1}(\tau_{1i}(r)))} =
% (v_1^{(\sigma_{12}(\tau_{12}(m)))})^Tv_i^{(\sigma_{i}(\tau_{12}(m)))}.
% $$}{I did not understand the rightmost term of this equation. What is $\sigma_{i}$?}
% To find such $r$ for each $m$, we note that as
By noting that
 $v_1^{(s)}$, $s=1,2,3,$ are the rows of the orthogonal matrix $R_1$, we have that for $m,r=1,2,3$
\begin{equation} \label{block 1i prod}
\begin{split}
H_{\sigma,m}^{(1,2)^T}R_{1i}^Te_{rr}  R_{i1}   & =
{(v_1^{(\sigma_{12}(\tau_{12}(m)))^T}v_2^{(\sigma_{21}(\tau_{12}(m)))})}^T 
(v_1^{(\sigma_{1i}(\tau_{1i}(r)))^T}v_i^{(\sigma_{i1}(\tau_{1i}(r)))})  \\ & =
v_2^{(\sigma_{21}(\tau_{12}(m)))^T}v_1^{(\sigma_{12}(\tau_{12}(m)))}v_1^{(\sigma_{1i}(\tau_{1i}(r)))^T}v_i^{(\sigma_{i1}(\tau_{1i}(r)))} \\ &=
\begin{cases}
    \pm v_2^{(\sigma_{21}(\tau_{12}(m)))^T}v_i^{(\sigma_{i1}(\tau_{1i}(r)))} & \text{if } \sigma_{12}(\tau_{12}(m)) = \pm \sigma_{1i}(\tau_{1i}(r))\\
%   - v_2^{(\sigma_{21}(\tau_{12}(m)))^T}v_i^{(\sigma_{i1}(\tau_{1i}(r)))} & \text{if } \sigma_{12}(\tau_{12}(m)) = -\sigma_{1i}(\tau_{1i}(r))\\
  \quad \quad \quad  0_{3\times3} & \text{else}  
\end{cases}
\\ &=
\begin{cases}
    v_2^{(\sigma_{21}(\tau_{12}(m)))^T}v_i^{(\pm \sigma_{i1}(\tau_{1i}(r)))} & \text{if } \sigma_{12}(\tau_{12}(m)) = \pm \sigma_{1i}(\tau_{1i}(r))\\
%   v_2^{(\sigma_{21}(\tau_{12}(m)))^T}v_i^{(-\sigma_{i1}(\tau_{1i}(r)))} & \text{if } \sigma_{12}(\tau_{12}(m)) = -\sigma_{1i}(\tau_{1i}(r))\\
  \quad \quad \quad  0_{3\times3} & \text{else}  
\end{cases}
\end{split}
\end{equation}
where $ H_{\sigma,m}^{(1,2)}$ has already been set by~\eqref{block 12},
and thus
\begin{equation} \label{block 1i cond}
    \begin{split}
        \lVert
H_{\sigma,m}^{(1,2)^T}R_{1i}^Te_{rr}  R_{i1}
\rVert_F \neq 0
& \iff 
v_1^{(\sigma_{12}(\tau_{12}(m)))}v_1^{(\sigma_{1i}(\tau_{1i}(r)))^T} = \pm 1
\\ & \iff 
\sigma_{12}(\tau_{12}(m)) = \pm \sigma_{1i}(\tau_{1i}(r)).
    \end{split}
\end{equation}
Take $r \in \{1,2,3\}$ for which $\lVert
H_{\sigma,m}^{(1,2)^T}R_{1i}^Te_{rr}  R_{i1}
\rVert_F \neq 0$.
% $ H_{\sigma,m}^{(1,i)} =  R_{1i}^Te_{rr}  R_{i1}$ according to~\eqref{block 1i} ($r$ that maximizes~\eqref{block 1i}). 
Then 
\begin{align*}
    % H_{\sigma,m}^{(1,i)} = 
    R_{1i}^Te_{rr}  R_{i1}  & = v_1^{(\sigma_{1i}(\tau_{1i}(r)))^T}v_i^{(\sigma_{i1}(\tau_{1i}(r)))} =
v_1^{(\pm \sigma_{12}(\tau_{12}(m)))^T}v_i^{(\sigma_{i1}(\tau_{1i}(r)))} \\&=
\pm v_1^{( \sigma_{12}(\tau_{12}(m)))^T}v_i^{(\sigma_{i1}(\tau_{1i}(r)))} =
v_1^{( \sigma_{12}(\tau_{12}(m)))^T}v_i^{(\pm \sigma_{i1}(\tau_{1i}(r)))},
\end{align*}
where 
% the first equality follows by~\eqref{block 1i}, 
the first equality follows by Proposition~\ref{prop:relative rows}
and
the second equality follows by~\eqref{block 1i cond}.
% \rvc{To express $v_i^{(\pm \sigma_{i1}(\tau_{1i}(r)))}$ as $v_i^{(\sigma_{i}(\tau_{12}(m)))}$}{Why should we to do that?} for some symmetry group element $g_i \in G$ whose one-line notation is $\sigma_{i}$,
% we note that}

Next, by~\eqref{block 1i cond}, $\tau_{1i}(r) = \pm \sigma_{1i}^{-1}(\sigma_{12}(\tau_{12}(m)))$, where $\sigma_{1i}^{-1}$ denotes the one-line notation corresponding to $g_{1i}^{-1} \in G$, and thus, $\sigma_{1i}^{-1}(\sigma_{12})$ is the one-line notation corresponding to $g_{1i}^{-1}g_{12} \in G$.
% {You are trying to eliminate the $\tau_{1i}$ in the second term, right?} 
Then
\begin{align*}
    % H_{\sigma,m}^{(1,i)} = 
    R_{1i}^Te_{rr}  R_{i1}   =
v_1^{( \sigma_{12}(\tau_{12}(m)))^T}v_i^{(\pm \sigma_{i1}(\tau_{1i}(r)))} &=
v_1^{( \sigma_{12}(\tau_{12}(m)))^T}v_i^{(\pm \sigma_{i1}( \pm \sigma_{1i}^{-1}(\sigma_{12}(\tau_{12}(m)))))}  \\ & = 
v_1^{( \sigma_{12}(\tau_{12}(m)))^T}v_i^{( \sigma_{i1}(\sigma_{1i}^{-1}(\sigma_{12}(\tau_{12}(m)))))},
% {Why can you eliminate the $\pm$? Are the two $\pm$ in the previous line ``synchronized''? I think it does not matter since you introduce the notation $\sigma^{*}$ anyway.}
\end{align*}
where $\sigma_{i1}(\sigma_{1i}^{-1}(\sigma_{12}))$ is the one-line notation corresponding to $g_{i1}g_{1i}^{-1}g_{12} \in G$.
In order to simplify the notation, we define for $i=3,\dots,N$, $g_i^* = g_{i1}g_{1i}^{-1}g_{12} \in G$,
% {For what $i$? Including 1 and 2? You say that below, but it is confusing here.}
% {You already used the notation $g_{i}$, no? Where do you use it?} 
and denote by $\sigma_i^*$ its one-line notation. 
Thus
\begin{align*} 
    R_{1i}^Te_{rr}  R_{i1}  & =
v_1^{( \sigma_{12}(\tau_{12}(m)))^T}v_i^{( \sigma_{i1}(\sigma_{1i}^{-1}(\sigma_{12}(\tau_{12}(m)))))} = v_1^{( \sigma_{12}(\tau_{12}(m)))^T}v_i^{(\sigma_{i}^*(\tau_{12}(m)))},
\end{align*}
implying that for $r\in \{1,2,3\}$ which satisfies $\lVert
H_{\sigma,m}^{(1,2)^T}R_{1i}^Te_{rr}  R_{i1}
\rVert_F \neq 0$,  the matrix $R_{1i}^Te_{rr}  R_{i1}$ is the $3 \times 3$ rank-1 matrix which is the $(1,i)$ block of the matrix $H_{\sigma,m}$.

Hence, we set the $(1,i)$ blocks of the matrices $H_{\sigma,m}$, $m=1,2,3$, $i=3,\dots,N$,
as
\begin{equation} \label{block 1i}
% v_1^{(\sigma_{1}(m))^T}v_i^{(\sigma_{i}(m))} := 
 H_{\sigma,m}^{(1,i)} :=
\underset{\substack M \in 
\{  R_{1i}^Te_{rr}  R_{i1} \}_{r=1,2,3}}
{\operatorname{\arg max}} \quad  
\lVert 
{H}_{\sigma,m}^{(1,2)^T}
M
\rVert_F.
\end{equation}
Thus, for $i=2,\dots,N$,
each $(1,i)$ block of $H_{\sigma,m}$ is given by the rank-1 $3 \times 3$ matrix 
$v_1^{( \sigma_{1}^*(\tau_{12}(m)))^T}v_i^{(\sigma_{i}^*(\tau_{12}(m)))}$ (for notation consistency, we also denote by $\sigma_1^*$ and $\sigma_2^*$ the one-line notations $\sigma_{12}$ and $\sigma_{21}$, respectively).

At this point, we note that as each $H_{\sigma,m}$, $m=1,2,3$, must be of rank-1, the $(i,j)$ block of $H_{\sigma,m}$, $i,j \in [N]$, 
% $i\neq j$, 
must be equal to the  $3 \times 3$ rank-1 matrix
$v_i^{( \sigma_{i}^*(\tau_{12}(m)))^T}
 v_j^{( \sigma_{j}^*(\tau_{12}(m)))}$. 
%  and 
% the $(i,i)$ block of $H_{\sigma,m}$, $i = 1,\dots,N$, must be equal to the rank-1 $3 \times 3$ matrix
% $v_i^{( \sigma_{i}(\tau_{12}(m)))^T}
%  v_i^{( \sigma_{i}(\tau_{12}(m)))}$.
This implies that
 the $(i,1)$  block of $H_{\sigma,m}$, $i = 2,\dots,N$, is defined as
 \begin{equation} \label{block i1}
 H_{\sigma,m}^{(i,1)} :=
 H_{\sigma,m}^{(1,i)^T},
\end{equation}
 the $(1,1)$  block of $H_{\sigma,m}$ is defined as
 \begin{equation} \label{block 11}
 H_{\sigma,m}^{(1,1)} :=
 H_{\sigma,m}^{(1,2)} H_{\sigma,m}^{(1,2)^T},
\end{equation}
and the $(i,j)$ block of $H_{\sigma,m}$, $i,j = 2,\dots,N$, % $i\neq j$ 
is defined as
\begin{equation} \label{block ij}
 H_{\sigma,m}^{(i,j)} :=
 H_{\sigma,m}^{(1,i)^T} H_{\sigma,m}^{(1,j)}.
\end{equation}
% {It must be clear that you are not really doing this. This was one of the issues the reviewer raised.}
%  as $H_{\sigma,m}^{(1,i)^T} H_{\sigma,m}^{(1,j)} = 
%  v_i^{( \sigma_{i}(\tau_{12}(m)))^T}
%  v_j^{( \sigma_{j}(\tau_{12}(m)))}$.
%  and the $(i,i)$ block of $H_{\sigma,m}$, $i = 2,\dots,N$, by
%  \begin{equation} \label{block ii}
%  H_{\sigma,m}^{(i,i)} :=
%  H_{\sigma,m}^{(1,i)^T} H_{\sigma,m}^{(1,i)}
% \end{equation}
%  as $H_{\sigma,m}^{(1,i)^T} H_{\sigma,m}^{(1,i)} = v_i^{( \sigma_{i}(\tau_{12}(m)))^T}
%  v_i^{( \sigma_{i}(\tau_{12}(m)))}$.
Overall, we showed that given the set $\{(R_{ij}, R_{ji})\}_{i<j \in [N]}$, we can construct
three rank-1 $3N \times 3N$ block matrices $H_{\sigma,m}$, $m=1,2,3$,
with each $H_{\sigma,m}$ satisfying $H_{\sigma,m} = v_{\sigma,\tau_{12}(m)}^Tv_{\sigma,\tau_{12}(m)}$ where 
$$v_{\sigma,\tau_{12}(m)} = (
v_1^{(\sigma_{1}^*(\tau_{12}(m)))}
, \dots,
v_i^{( \sigma_{i}^*(\tau_{12}(m)))}
, \dots,
v_N^{( \sigma_{N}^*(\tau_{12}(m)))}
).$$

Note that the construction of $H_{\sigma,m}$, $m=1,2,3$, described above uses only part of the data $\{( R_{ij}, {R}_{ji})\}_{i<j \in [N]}$, namely, it uses only the pairs $\{( R_{1i}, {R}_{i1})\}_{i= 2}^N$ to construct the matrices $ H_{\sigma,m}$, $m=1,2,3$.
Since in practice 
the set $\{(\tilde R_{ij},\tilde {R}_{ji})\}_{i<j \in [N]}$ computed by
Algorithm~\ref{alg:relative} is only an estimate of the set $\{( R_{ij}, {R}_{ji})\}_{i<j \in [N]}$,
we only obtain an estimate to each matrix $H_{\sigma,m}$, $m=1,2,3$, which we denote by $\tilde H_{\sigma,m}$.
Thus, we next show how to modify our construction such that all pairs $\{( R_{ij}, {R}_{ji})\}_{i<j \in [N]}$ are considered.
This way, constructing  $\tilde H_{\sigma,m}$, $m=1,2,3$, using $\{(\tilde R_{ij},\tilde {R}_{ji})\}_{i<j \in [N]}$ would result in more robust estimates to the orientations of the projection-images $\{P_{R_i} \}_{i=1}^N$.
The main idea of the modification is to 
compute the $(i,j)$ blocks of $H_{\sigma,m}$, $m=1,2,3$
from 
% \rvc{their corresponding estimation}{Remove?} 
$( R_{ij}, {R}_{ji})$, $i,j = 2, \dots, N$, $i \neq j$, as follows. 

First, for $i,j = 2, \dots, N$, $i \neq j$,
we have that $ H_{\sigma,m}^{(i,j)}$ as defined in~\eqref{block ij} is equal to the $3 \times 3$ rank-1 matrix
$v_i^{( \sigma_{i}^*(\tau_{12}(m)))^T}
 v_j^{( \sigma_{j}^*(\tau_{12}(m)))}$.
Note that
$ v_i^{( \sigma_{i}^*(\tau_{12}(m)))^T}
 v_j^{( \sigma_{j}^*(\tau_{12}(m)))} \in 
\{\pm (v_i^{(k)})^Tv_j^{(l)}\}_{k,l=1}^3,
$
% {With or without tilde? Now that you have two different notations, this may be confusing.},
as the set of $3 \times 3$ rank-1 matrices $\{\pm (v_i^{(k)})^Tv_j^{(l)}\}_{k,l=1}^3$ consists of all possible products between the rows of $R_i$ and the rows of $R_j$. 
In the following proposition, we show how to obtain the set $\{\pm (v_i^{(k)})^Tv_j^{(l)}\}_{k,l=1}^3$ from $(R_{ij}, R_{ji})$.

\begin{proposition}
Let $  R_{ij}, {R}_{ji}$ and $ R_{i},{R}_{j}$ be two pairs of rotations
satisfying~\eqref{eq:common line approach}, $i<j \in [N]$.
Then
\begin{align*}
\{\pm  R_{ij}^Te_{kl} R_{ji}\}_{k,l=1}^3 &= 
\{\pm (v_i^{(k)})^Tv_j^{(l)}\}_{k,l=1}^3,
\end{align*}
where the matrices $e_{kl}$, $k,l=1,2,3$, are single entry matrices defined in Definition~\ref{def:single entry matrix},
$v_i^{(k)}$ is the $k$th row of $R_i$ and $v_j^{(l)}$ is the $l$th row of $R_j$.
\end{proposition}
\begin{proof}
\begin{align*}
\{\pm  R_{ij}^Te_{kl} R_{ji}\}_{k,l=1}^3 &= 
\{\pm R_i^{T}h_{ij}^{T}e_{kl}h_{ji}R_j\}_{k,l=1}^3 = 
\{\pm R_i^{T}e_{kl}R_j\}_{k,l=1}^3 = 
\{\pm (v_i^{(k)})^Tv_j^{(l)}\}_{k,l=1}^3,
\end{align*}
where the first equality is due to~\eqref{eq:relation to ground truth}, the second equality is due to~\eqref{eq:property 3}, and the last equality follows by a direct calculation.
\end{proof}
Thus there exist $k,l \in \{1,2,3\}$ such that either $ R_{ij}^Te_{kl}  R_{ji}$ or $- R_{ij}^Te_{kl}  R_{ji}$ is equal to
 $v_i^{( \sigma_{i}^*(\tau_{12}(m)))^T}
 v_j^{( \sigma_{j}^*(\tau_{12}(m)))}$. 
 As the current $(i,j)$ blocks of $H_{\sigma,m}$, $m=1,2,3,$ are given by $ H_{\sigma,m}^{(1,i)^T}  H_{\sigma,m}^{(1,j)}$ (equation~\eqref{block ij}),
we replace them with
\begin{equation} \label{block ij 2} 
 H_{\sigma,m}^{(i,j)}
:= \underset{\substack{M \in 
\{{\pm  R_{ij}^Te_{kl}  R_{ji}\}_{k,l=1}^3}}}{\operatorname{\arg min}} \quad  
\lVert
M
- 
 H_{\sigma,m}^{(1,i)^T}  H_{\sigma,m}^{(1,j)}
\rVert_F,
\end{equation}
where $ H_{\sigma,m}^{(1,i)^T}, H_{\sigma,m}^{(1,j)}$ are defined in~\eqref{block 1i}.

Finally, we describe how to construct the three matrices $\tilde H_{\sigma,m}$, $m=1,2,3$, from all estimates $\{(\tilde R_{ij},\tilde {R}_{ji})\}_{i<j \in [N]}$ computed by Algorithm~\ref{alg:relative} (replacing $\{(R_{ij},R_{ji})\}_{i<j \in [N]}$ used above).
Denoting by $\tilde H_{\sigma,m}^{(i,j)}$ the estimate of the $3\times 3$ block
$H_{\sigma,m}^{(i,j)}$, $m=1,2,3$, $i,j \in [N]$,
we have by~\eqref{block 12}
\begin{equation} \label{block 12 est}
 \tilde H_{\sigma,m}^{(1,2)} :=  \tilde R_{12}^Te_{mm} \tilde R_{21},
\end{equation}
by~\eqref{block 1i}
\begin{equation} \label{block 1i est}
\tilde H_{\sigma,m}^{(1,i)} :=
\underset{\substack M \in 
\{ \tilde R_{1i}^Te_{rr} \tilde R_{i1} \}_{r=1,2,3}}
{\operatorname{\arg max}} \quad  
\lVert 
\tilde{H}_{\sigma,m}^{(1,2)^T}
M
\rVert_F, \quad i =3,\dots, N,
\end{equation}
by~\eqref{block i1}
\begin{equation} \label{block i1 est}
 \tilde H_{\sigma,m}^{(i,1)} :=
 \tilde H_{\sigma,m}^{(1,i)^T},  \quad i =2,\dots, N,
\end{equation}
% by~\eqref{block 11}
% \begin{equation} \label{block 11 est}
%  \tilde H_{\sigma,m}^{(1,1)} :=
%  \tilde H_{\sigma,m}^{(1,2)} \tilde H_{\sigma,m}^{(1,2)^T}, 
% \end{equation}
and by~\eqref{block ij 2}
\begin{equation} \label{block ij est}
 \tilde H_{\sigma,m}^{(i,j)} :=
 \underset{\substack{M \in 
\{{\pm \tilde R_{ij}^Te_{kl} \tilde R_{ji}\}_{k,l=1}^3}}}{\operatorname{\arg min}} \quad  
\lVert
M
- 
\tilde H_{\sigma,m}^{(1,i)^T} \tilde H_{\sigma,m}^{(1,j)}
\rVert_F, \quad i,j =2,\dots, N, \quad i \neq j.
\end{equation}
Lastly, we note that for $i \in [N]$, each $\tilde H_{\sigma,m}^{(j,i)^T} \tilde H_{\sigma,m}^{(j,i)}$, where $\tilde H_{\sigma,m}^{(j,i)}$ was computed by~\eqref{block ij est}, is an estimate to $H_{\sigma,m}^{(i,i)}$, $j \in [N] \setminus \{i\}$.
% Finally, to estimate the $(i,i)$ block of $H_{\sigma^*,m}$, $i \in [N]$, we first note that by property~\eqref{eq:v_m condition}
% $$
% H_{\sigma^*,m}^{(i,i)}=
% H_{\sigma^*,m}^{(j,i)^T}H_{\sigma^*,m}^{(j,i)} = 
% {(v_j^{(\sigma^*_j(r))^T}v_i^{(\sigma^*_{i}(r))})}^T
% (v_j^{(\sigma^*_j(r))^T}v_i^{(\sigma^*_{i}(r))})
%  = 
% v_i^{(\sigma^*_i(r))^T}v_i^{(\sigma^*_{i}(r))}
% $$
% for all $j \in [N] \setminus \{i\}$.
% Therefore, each $\tilde H_{\sigma^*,m}^{(j,i)^T} \tilde H_{\sigma^*,m}^{(j,i)}$ is an estimate to $H_{\sigma^*,m}^{(i,i)}$.
% % Since the products $v_i^{(\sigma^*_{i}(m))^T}v_j^{(\sigma^*_{j}(m))}$ are computed from noisy images,
Thus,
we get a more robust
estimate for the $(i,i)$ block of $H_{\sigma,m}$ by computing the average
\begin{equation} \label{block ii} 
\tilde H_{\sigma,m}^{(i,i)}: = \frac{1}{N-1} \sum_{\substack{ j=1 \\ j\neq i}}^{N} \tilde H_{\sigma,m}^{(j,i)^T}\tilde H_{\sigma,m}^{(j,i)}
% = \frac{1}{N-1} \sum_{\substack{ j=1 \\ j\neq i}}^{N} v_i^{(\sigma^*_i(m))^T}v_i^{(\sigma^*_{i}(m))} = v_i^{(\sigma^*_i(m))^T}v_i^{(\sigma^*_{i}(m))}
\end{equation}
followed by computing the best rank-1 approximation of each $(i,i)$ block of $\tilde H_{\sigma,m}$ using SVD.

%%% summary %%%
To conclude, we estimated three $3N\times 3N$ block matrices $\tilde{H}_{\sigma,m}$, $m=1,2,3$, whose $(i,j)$ $3\times3$ block is an estimate to the rank-1 matrix $v_i^{(\sigma_i^*(\tau_{12}(m)))^T}v_j^{(\sigma_{j}^*(\tau_{12}(m)))}$ which is estimated from $ (\tilde R_{ij}, \tilde{R}_{ji})$. 
We then factorize each matrix using SVD and obtain the estimates $\{ Og_i^*R_i \}_{i=1}^N$ for the orientations of the projection-images $\{P_{R_i}\}_{i=1}^N$, where $O$ is a rotation and $g_i^* \in G$.
% \rvc{Note that although the recovered estimation $\{ Og_i^*R_i \}_{i=1}^N$ is unknown, it is unique, and depends on the results of Algorithm 1 (which are not unique) and by the order of traversal for constructing the matrix.}{What do you mean? You have the $O$ and $g_{i}$ degrees of freedom. You are explaining why it is not unique.}

The construction of $\tilde H_{\sigma,m}$, $m=1,2,3$, and the estimation of the orientations of all projection-images
$\{P_{R_i}\}_{i=1}^N $ 
from the set 
$\{ (\tilde R_{ij}, \tilde{R}_{ji}) \}_{i<j \in [N]}$
is summarized in Algorithm~\ref{alg:rotest}.
%  {When the readers get here they are unlikely to remember what was the propose of $H_{\sigma,m}$. Maybe briefly repeat how to compute from it the rotations?}

The computational complexity of Algorithm~\ref{alg:relative} is quadratic in both the number
of images as well as in the size of $SO_G(3)$ (constructed in  Appendix~\ref{Constructing $SO_G(3)$}).
The computational complexity of Algorithm~\ref{alg:rotest} is quadratic in the
number of images.

%%% Algorithm 2 %%%

\begin{algorithm}
  \caption{Estimating $R_i$, $i = 1,\ldots,N$, for molecules with $\mathbb T$ or $\mathbb O$ symmetry.}\label{alg:rotest} 
  \begin{algorithmic}[1]
    \Input{$\{(\tilde{R}_{ij}, \tilde{R}_{ji})\}_{i<j \in [N]}$ estimated by Algorithm~\ref{alg:relative}.}
    \Initialize{Matrices  $\tilde H_{\sigma,m}$, $m=1,2,3$, of size $3N \times 3N$, with all entries set to zero.}
    \State $\{(\tilde R_{ij}, \tilde{R}_{ji})\}_{i<j \in [N]} \gets \mathit{handedness\ }\mathit{ synchronization(} \{(\tilde R_{ij}, \tilde{R}_{ji})\}_{i<j \in [N]}$)
    \For{$m = 1,2,3$}
    \State
       $\tilde H_{\sigma,m}^{(1,2)} = \tilde R_{12}^T e_{mm} \tilde R_{21}$. 
    \Comment{$\tilde H_{\sigma,m}^{(i,j)}$ denotes the $(i,j)$ $3 \times 3$ block of  $\tilde H_{\sigma,m}$. See~\eqref{block 12 est}}.
    % \EndFor
    % \For{$m = 1,2,3$}
    \For{$ i = 3,\ldots,N$}
    \State
    $
\tilde H_{\sigma,m}^{(1,i)}= 
\underset{\substack M \in \{\tilde R_{1i}^T e_{rr} \tilde R_{i1}\}_{r=1,2,3}}{\operatorname{\arg max}} \quad  
\lVert 
\tilde H_{\sigma,m}^{(1,2)^T}
M
\rVert_F.
$
		\Comment See~\eqref{block 1i est}.
    \EndFor
    % \EndFor
    % \For{$m = 1,2,3$}
    \For{$i < j = 2,\ldots,N$}
    \State 
    $ \tilde H_{\sigma,m}^{(i,j)} =
 \underset{\substack{M \in \{\pm \tilde R_{ij}^T e_{kl} \tilde R_{ji} \}_{k,l=1,2,3}}}{\operatorname{\arg min}} \quad  
\lVert
M - \tilde H_{\sigma,m}^{(1,i)^T} \tilde H_m^{(1,j)}
\rVert_F
    $.
    \Comment See~\eqref{block ij est}.
    \EndFor
    % \EndFor
    % \For{$m = 1,2,3$}
      \State $\tilde H_{\sigma,m} \gets \tilde H_{\sigma,m} + \tilde H_{\sigma,m}^T$
    % \EndFor 
    % \For{$m = 1,2,3$}
    \For{$i = 1,\ldots,N$}
      \State $\tilde H_{\sigma,m}^{(i,i)} = \frac{1}{N-1} 
      \sum_{j=1, j \neq i}^{N}
      \tilde H_{\sigma,m}^{(j,i)^T}\tilde H_{\sigma,m}^{(j,i)}$
      \Comment See~\eqref{block ii}.
    \EndFor
    % \EndFor
    % \For{$m = 1,2,3$}
      \State $V_m = \underset{\vert \vert v \vert \vert = 1}{\operatorname{argmax \ }}  v^T\tilde H_{\sigma,m}v$ \Comment{$V_m$ is the eigenvector of the leading eigenvalue of $\tilde H_{\sigma,m}$. } 
    \EndFor 
    \For{$i = 1,\ldots,N$}
    \For{$m = 1,2,3$}
    	\State $v_i^{(m)} = \frac{ V_m(3i-2:3i)}{\vert \vert V_m(3i-2:3i) \vert \vert}$
    	\Comment{$v_i^{(m)}$ is the $m$th row of the orthogonal matrix $\tilde R_i$.}
    \EndFor
    \State $\tilde R_i \gets \begin{pmatrix}
						-\  v_i^{(1)}\ -\\
						-\  v_i^{(2)}\ -\\
						-\  v_i^{(3)}\ -
					  \end{pmatrix}$
    % \EndFor
    % \For{$i = 1,\ldots,N$}
    	\If{ $\det \tilde  R_i < 0$} 
      	\State $\tilde R_i \gets - \tilde R_i$
      \EndIf 
    \EndFor
    \Output{$\tilde R_i$, $i = 1,\ldots,N$.}
  \end{algorithmic}
\end{algorithm}

% ===============================================================
% Chpter 7: Experimental results
% ===============================================================

\section{Experimental results}\label{sec:results}

We implemented the proposed algorithm in MATLAB, and tested it on both simulated and experimental data. We start with testing the algorithm on simulated data in Section~\ref{sec:simulated data}, to assess its robustness to noise. Then, in Section~\ref{sec:experimental data}, we test the algorithm on experimental cryo-electron microscopy data. All tests were executed on a dual Intel Xeon E5-2683 CPU (32 cores in total), with 768GB of RAM running Linux, and one nVidia GTX TITAN XP GPU (used for Algorithm~\ref{alg:relative}). The implementation of the algorithms is available as part of the ASPIRE software package~\cite{aspire}. To assess the actual memory consumption of the algorithm, we monitored it through the operating system during its execution. The maximal amount of memory used by the algorithm is on the order of storing the reconstructed volume.

\subsection{Simulated data}\label{sec:simulated data}

\begin{figure}
	\begin{center}
	\includegraphics[width=0.2\textwidth]{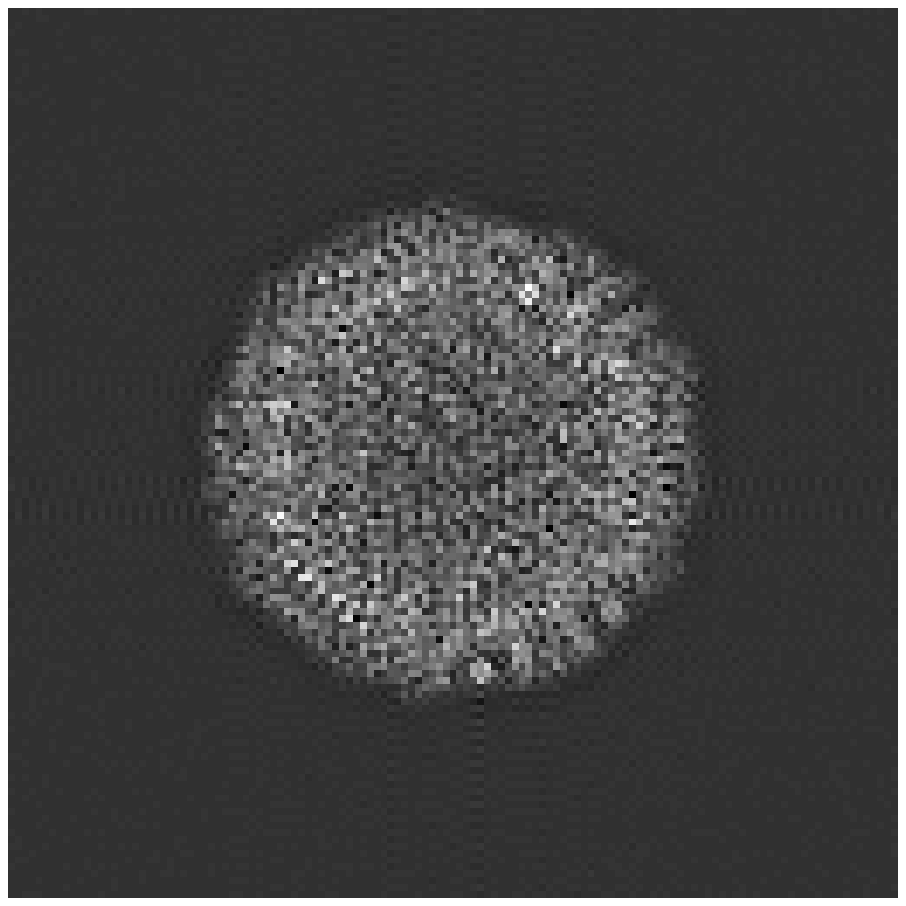}
	\includegraphics[width=0.2\textwidth]{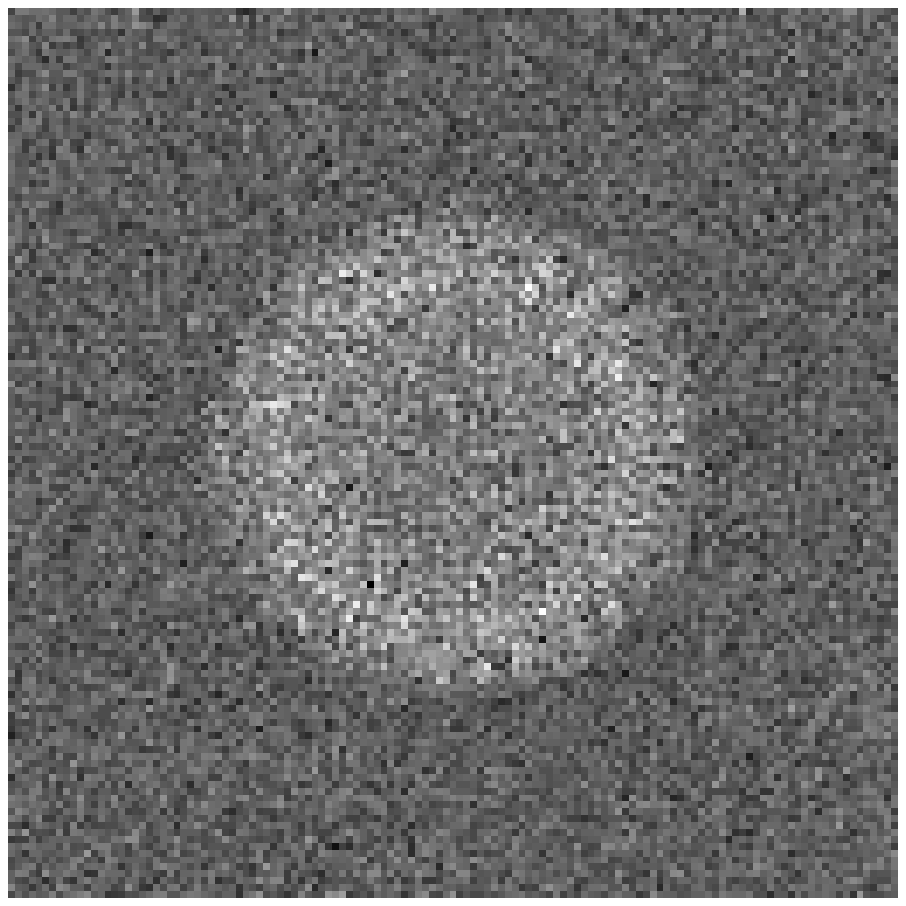}
	\includegraphics[width=0.2\textwidth]{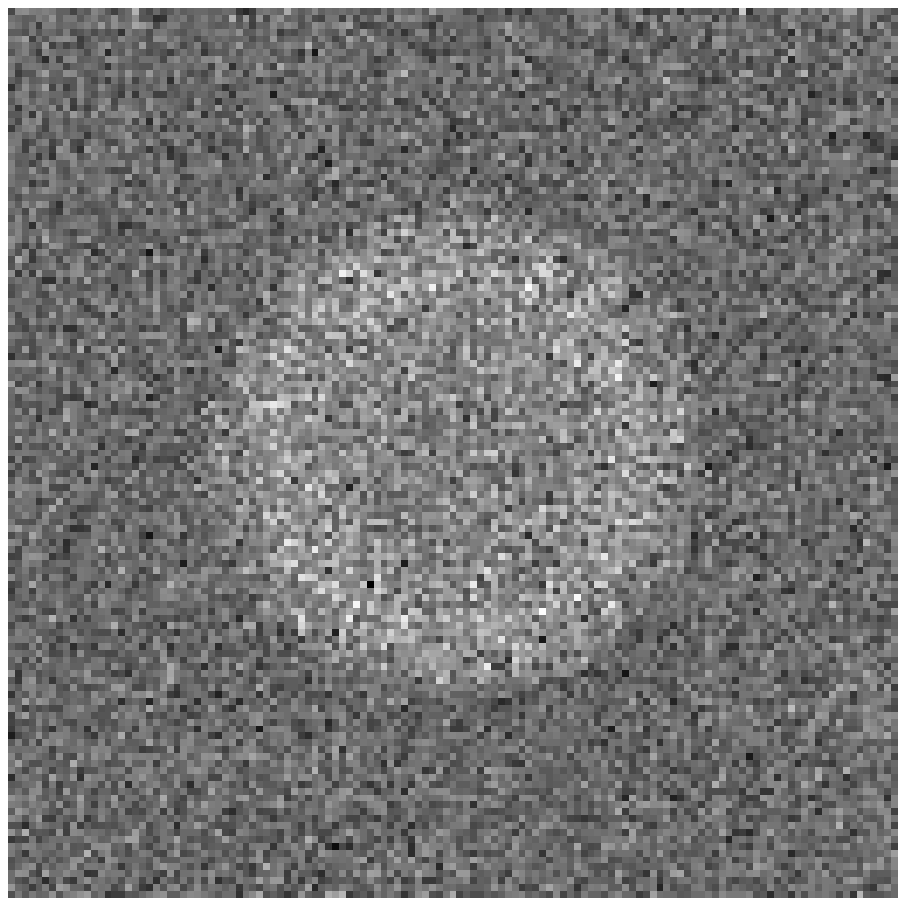}
	\includegraphics[width=0.2\textwidth]{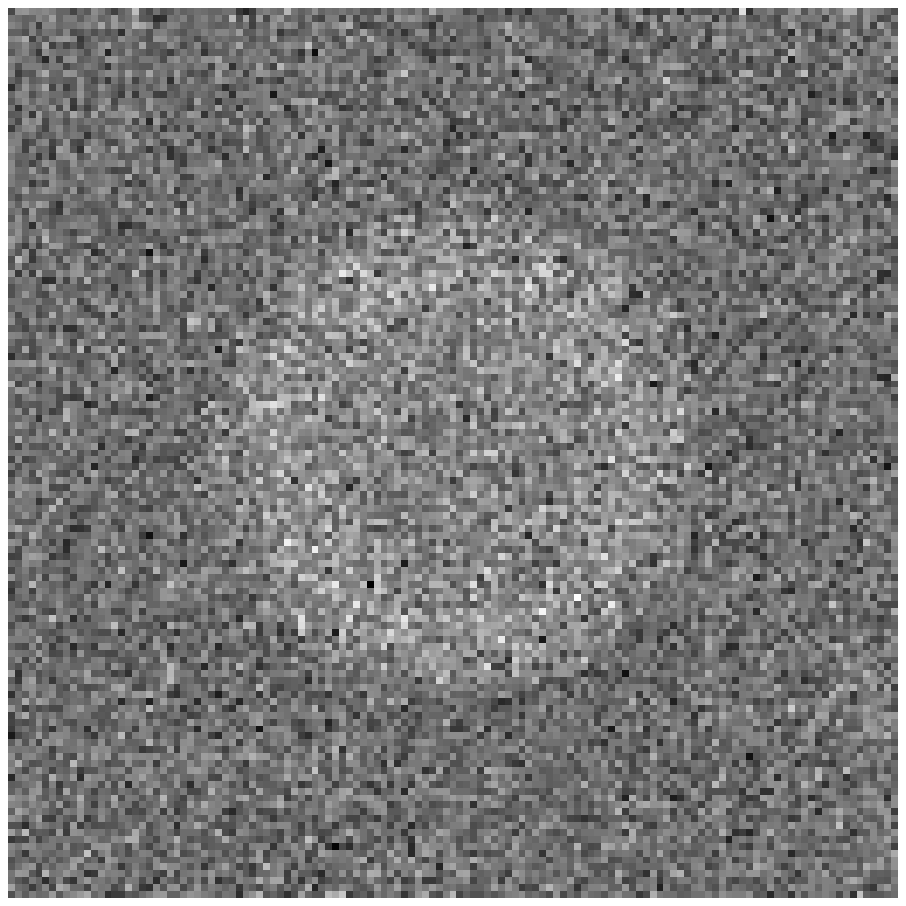}\\
	\includegraphics[width=0.2\textwidth]{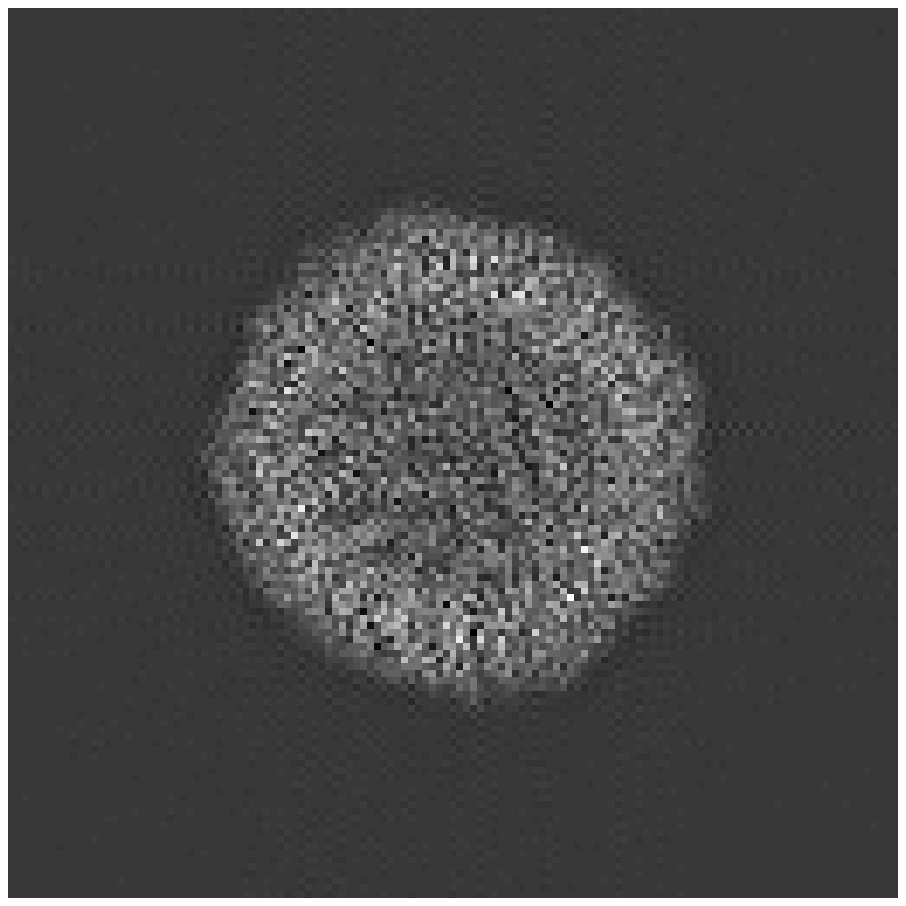}
	\includegraphics[width=0.2\textwidth]{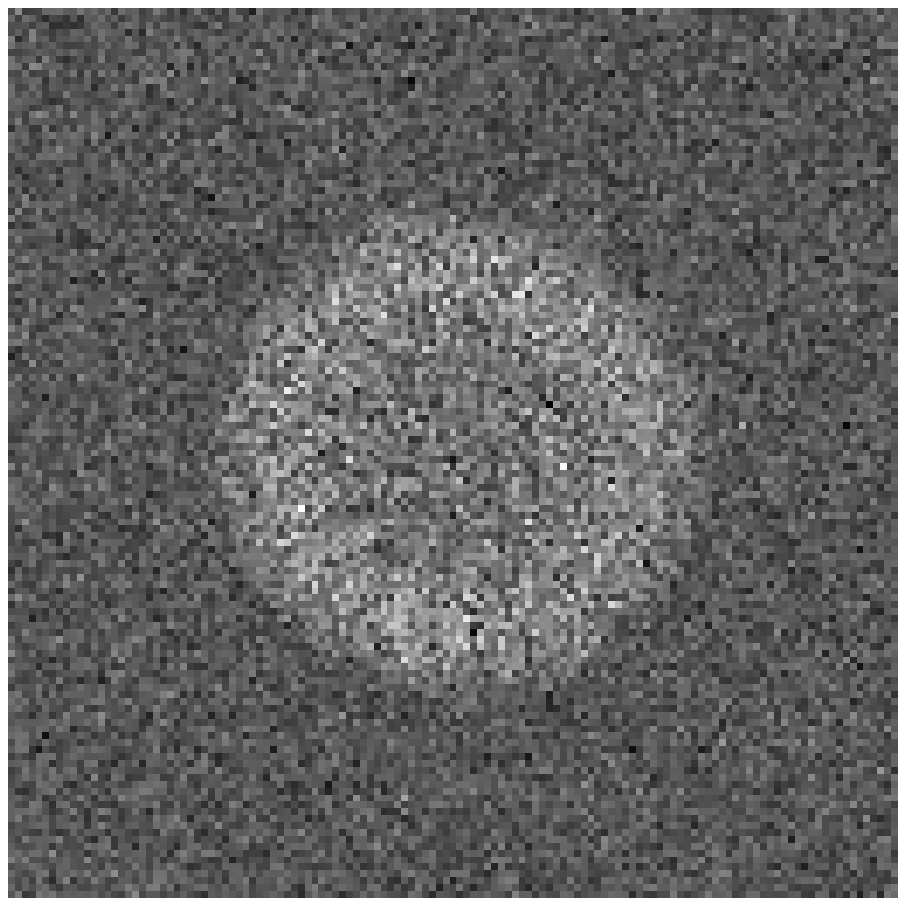}
	\includegraphics[width=0.2\textwidth]{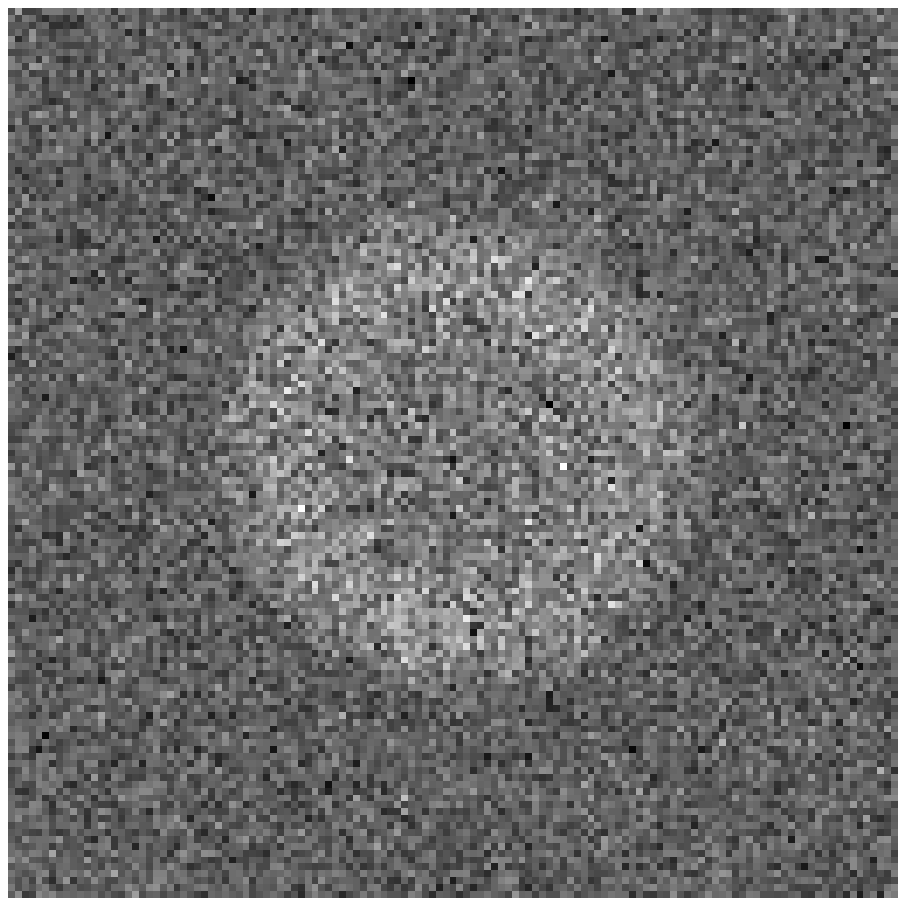}
	\includegraphics[width=0.2\textwidth]{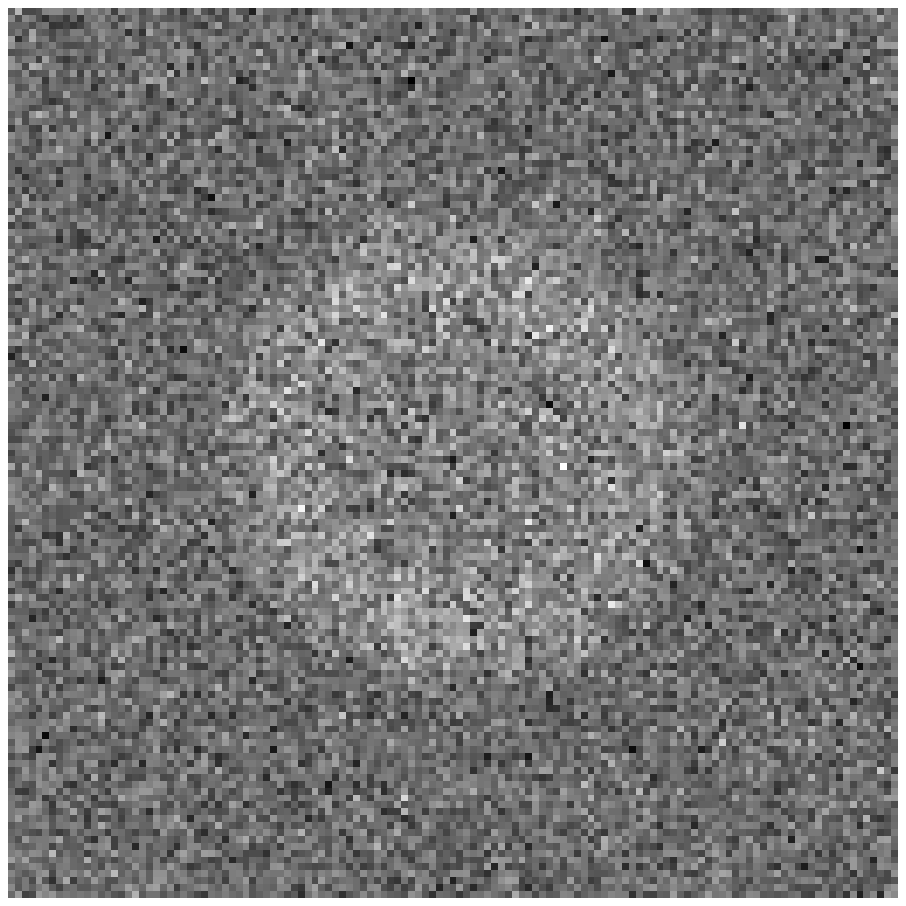}\\
	\includegraphics[width=0.2\textwidth]{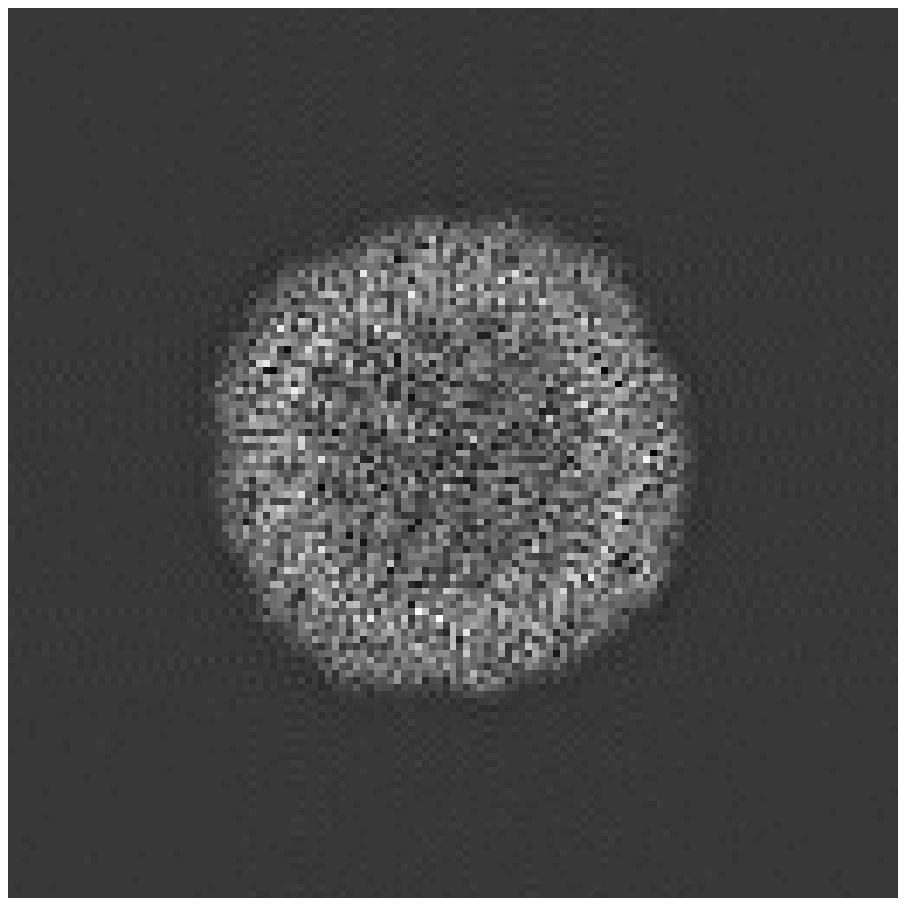}
	\includegraphics[width=0.2\textwidth]{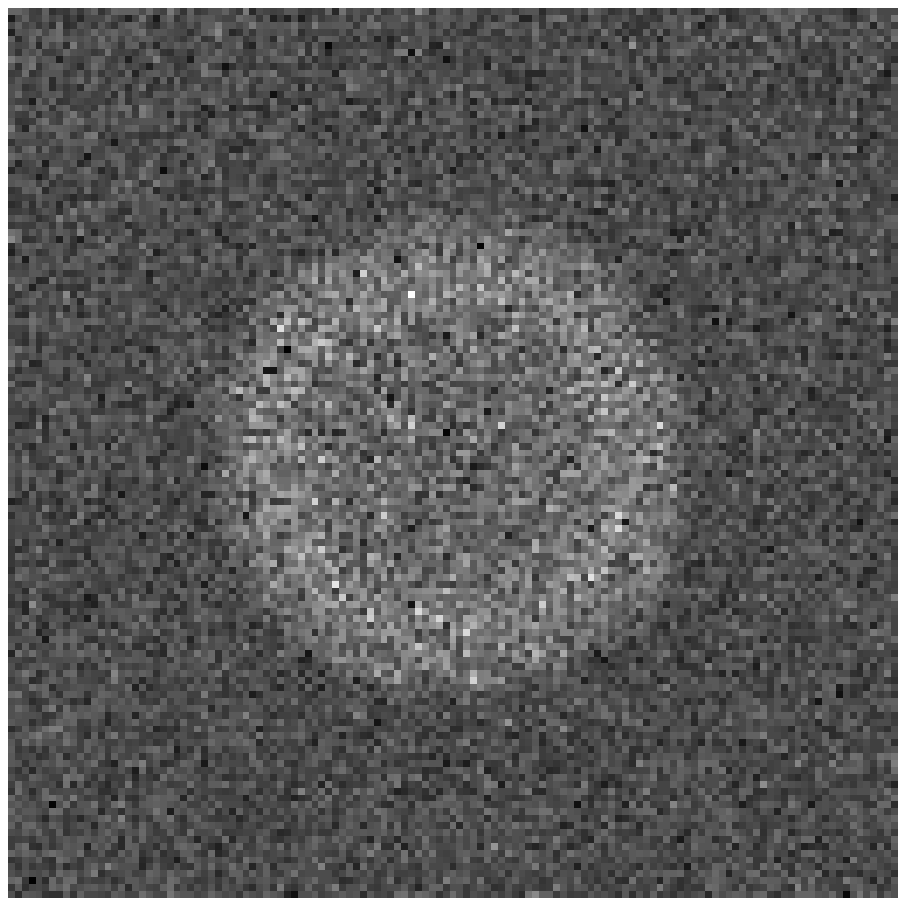}
	\includegraphics[width=0.2\textwidth]{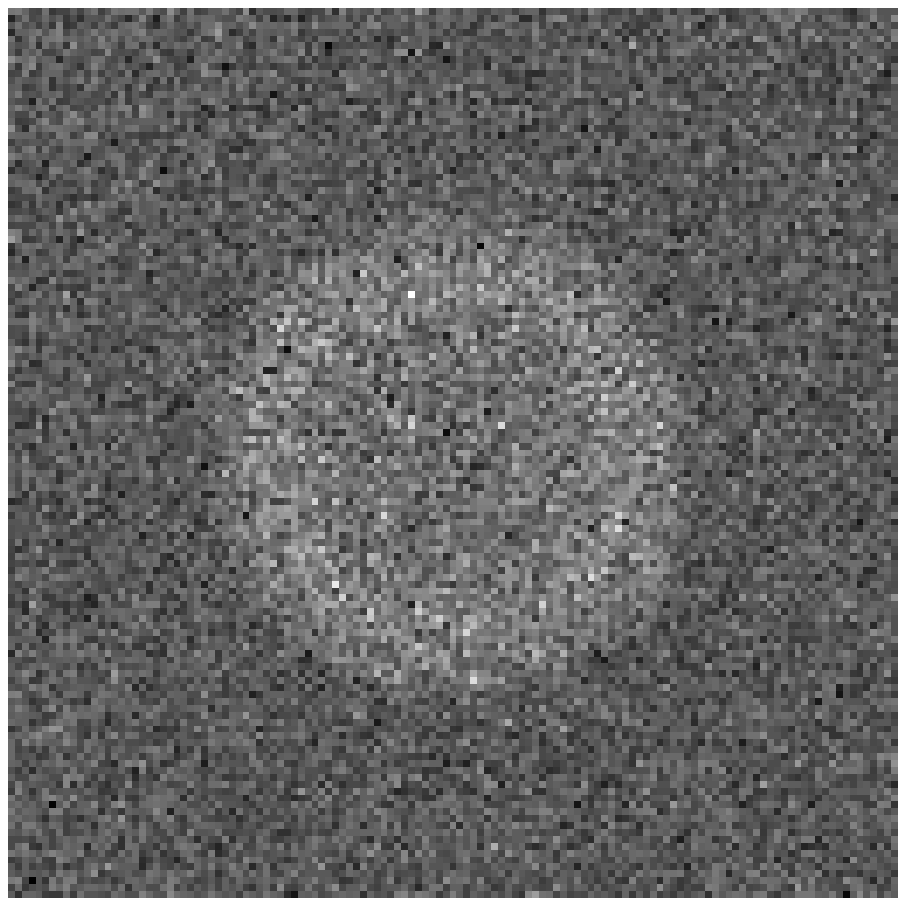}
	\includegraphics[width=0.2\textwidth]{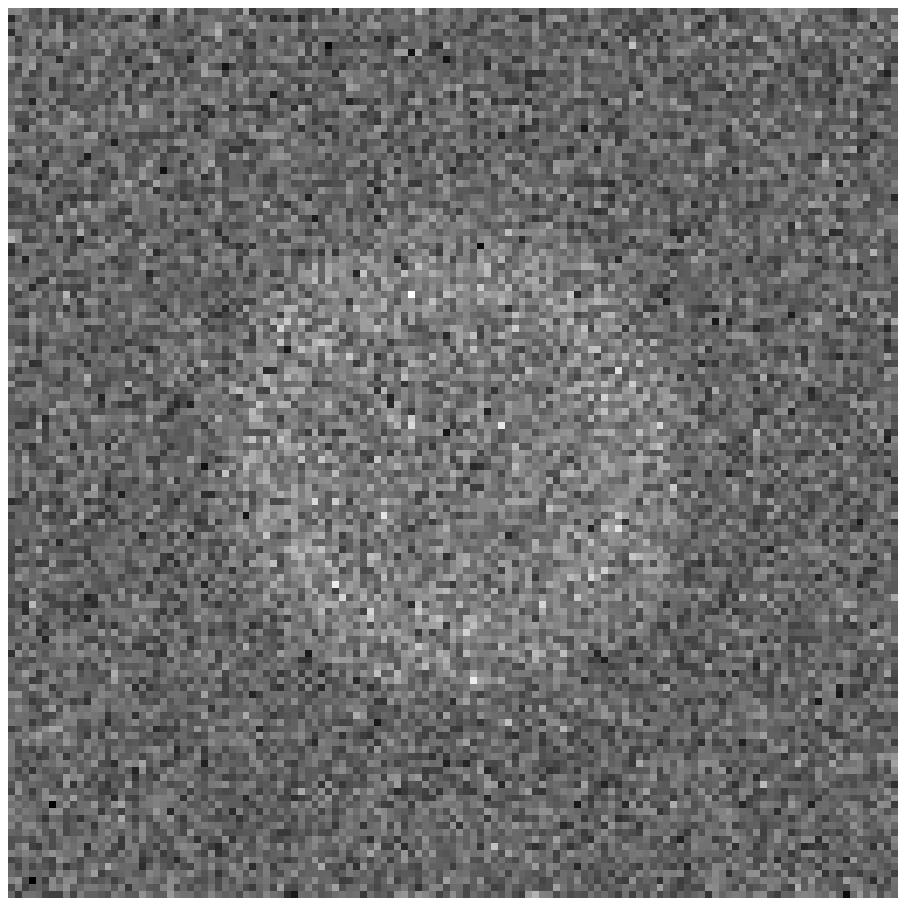}\\
	\includegraphics[width=0.2\textwidth]{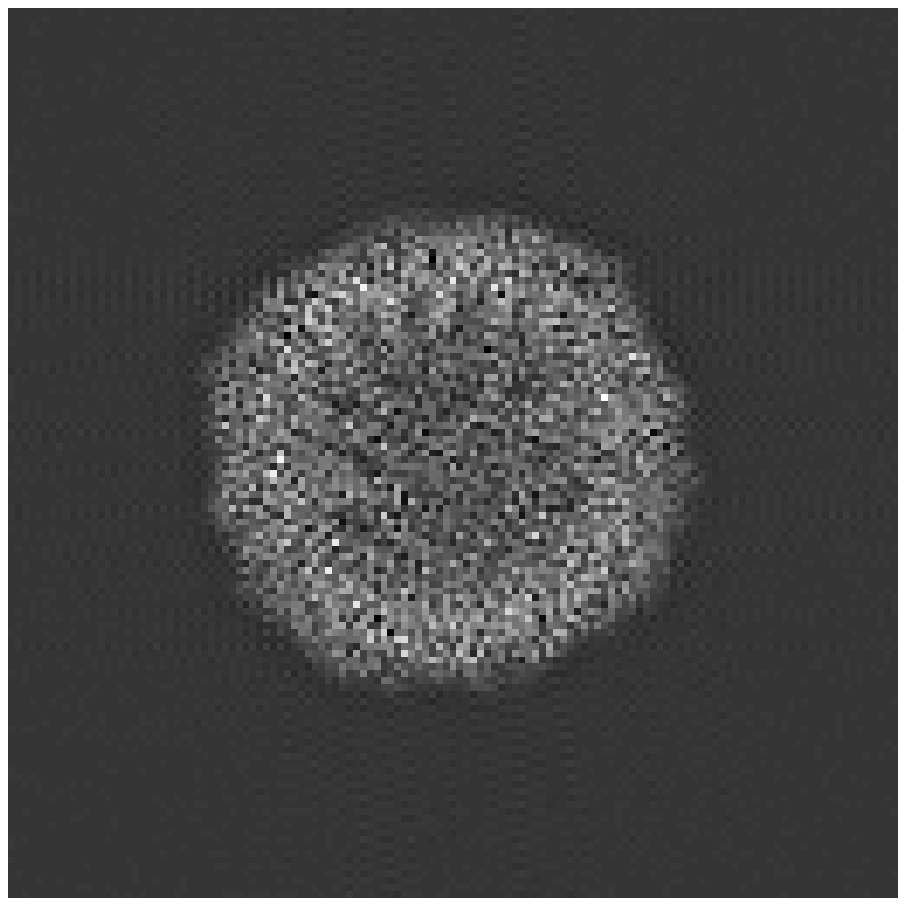}
	\includegraphics[width=0.2\textwidth]{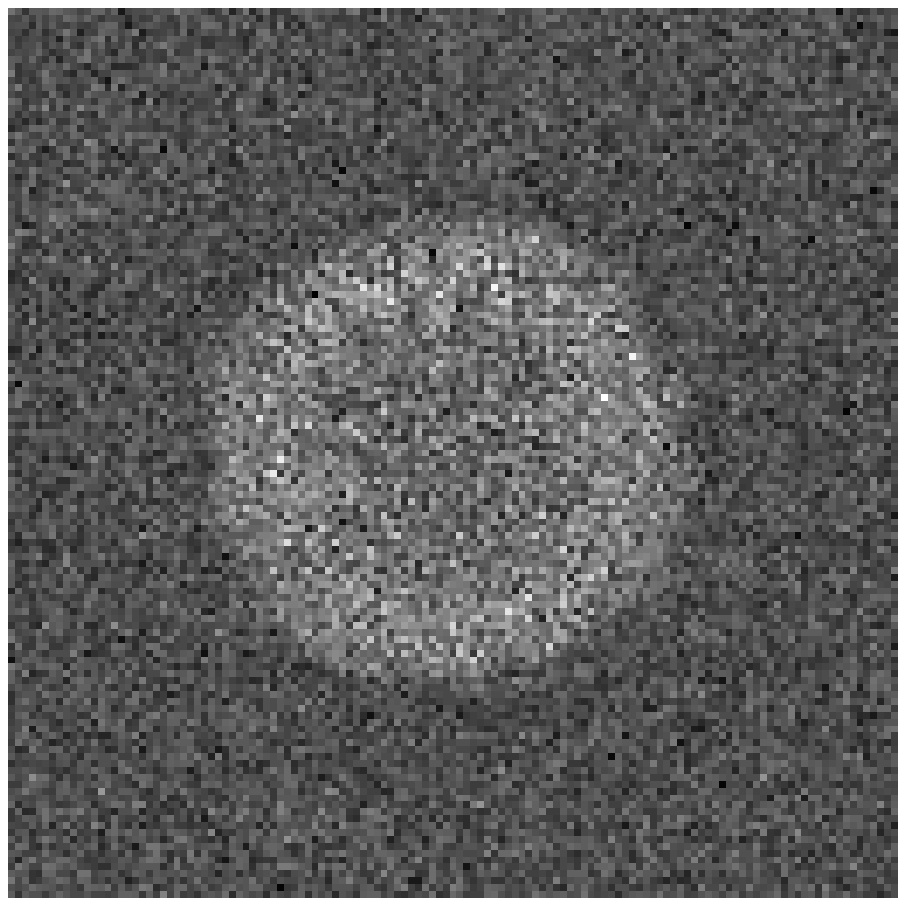}
	\includegraphics[width=0.2\textwidth]{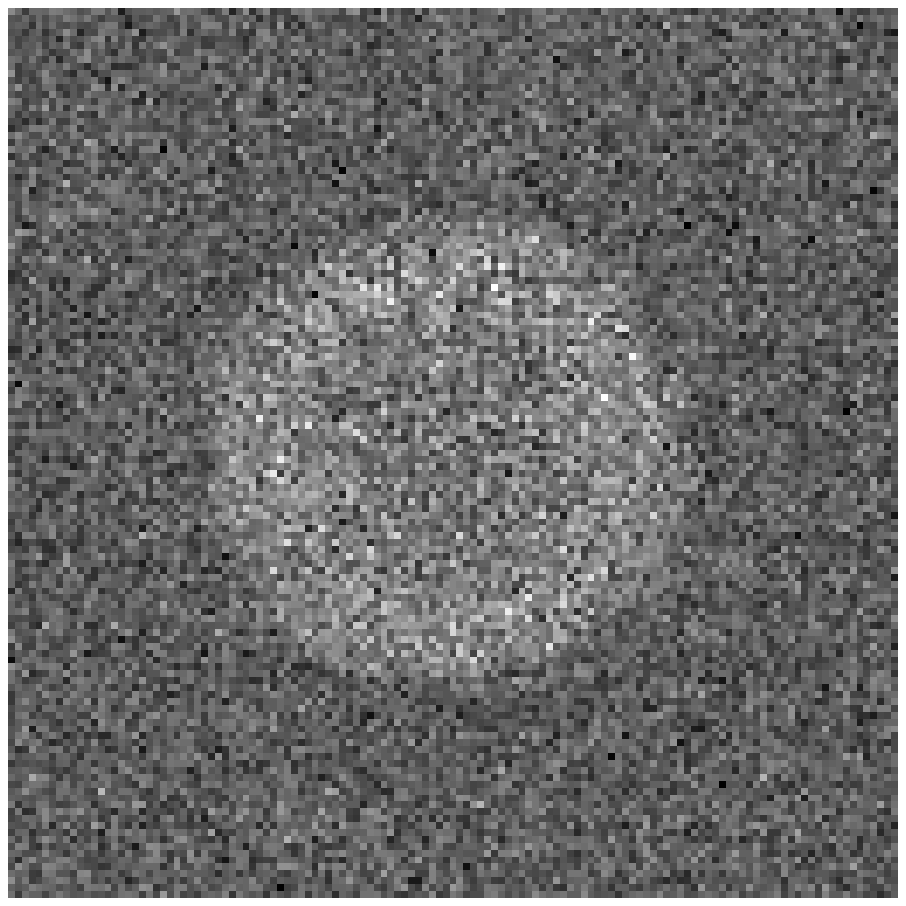}
	\includegraphics[width=0.2\textwidth]{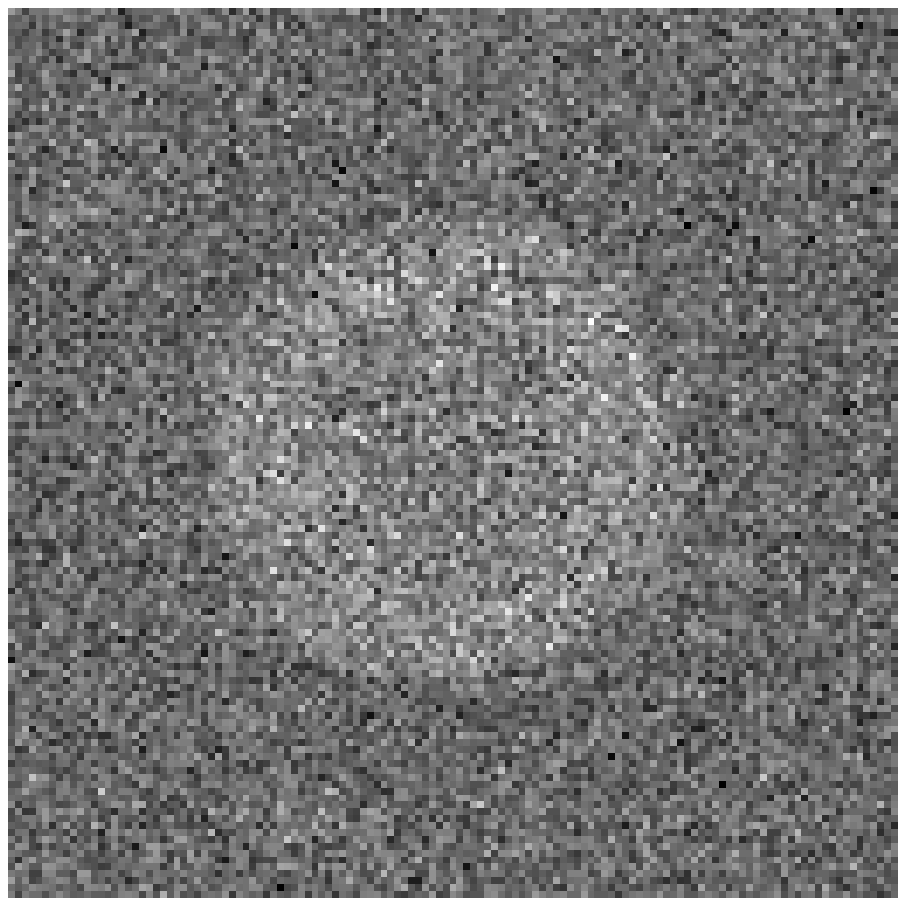}
	\end{center}
	\caption{Examples of simulated projection-images of EMD-4905~\protect\cite{EMD4905} ($\mathbb{O}$ symmetry) with signal to noise ratio of (from left to right) 1000, 1, 1/2, 1/4.}
	\label{fig:simulated 4905 projections}
\end{figure}

To test the performance of our algorithm in the presence of noise, we applied it to noisy simulated projection-images as follows. For $\mathbb{O}$ symmetry, we downloaded from EMDB the map EMD-4905~\cite{EMD4905}, generated from it clean projection-images of size $129\times129$ pixels, and added to the clean images Gaussian noise with zero mean and variance that results in signal to noise ratio (SNR) of the images the is equal to 1000 (considered as clean images for reference), $1$, $1/2$, and $1/4$ (the signal to noise ratio is defined as the ratio between the energy of the signal in the image and the energy of the noise). Figure~\ref{fig:simulated 4905 projections} shows several examples of projection-images of EMD-4905 at these noise levels.

\begin{figure}
	\begin{center}
		\subfloat[$N=25$]{
			\includegraphics[width=0.4\textwidth]{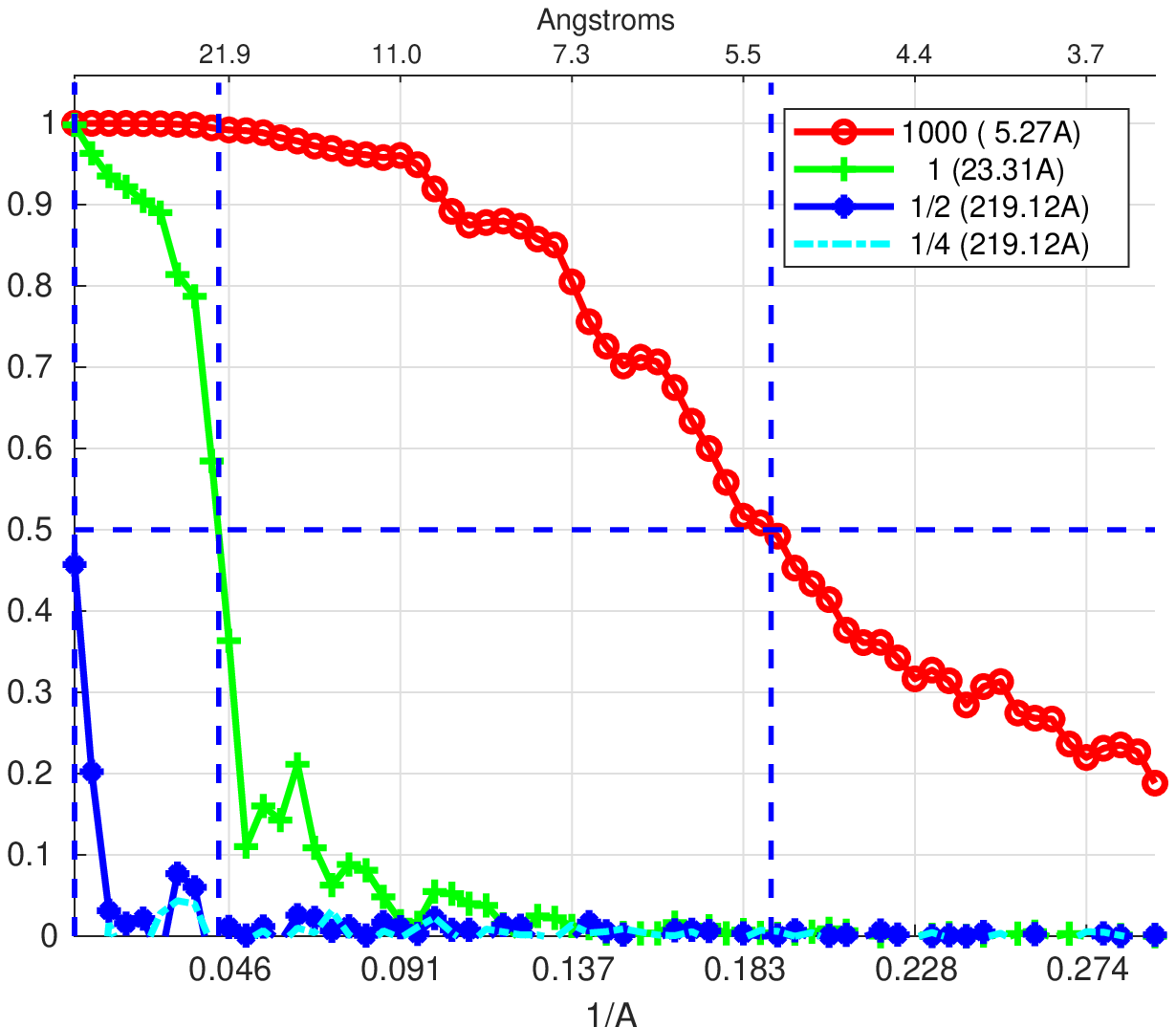}
		}
		\subfloat[$N=50$]{
	        \includegraphics[width=0.4\textwidth]{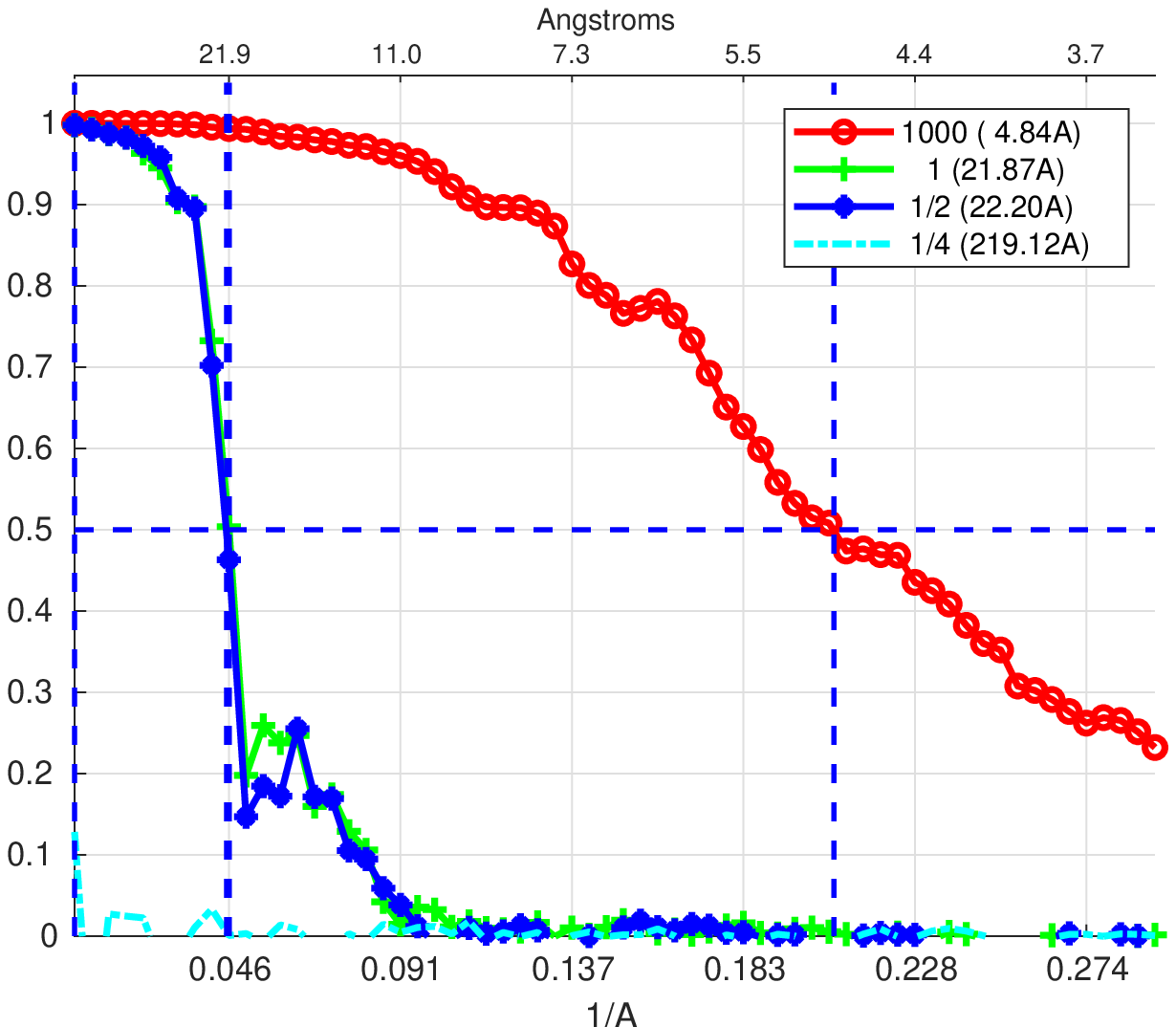}
        }\\
		\subfloat[$N=100$]{
			\includegraphics[width=0.4\textwidth]{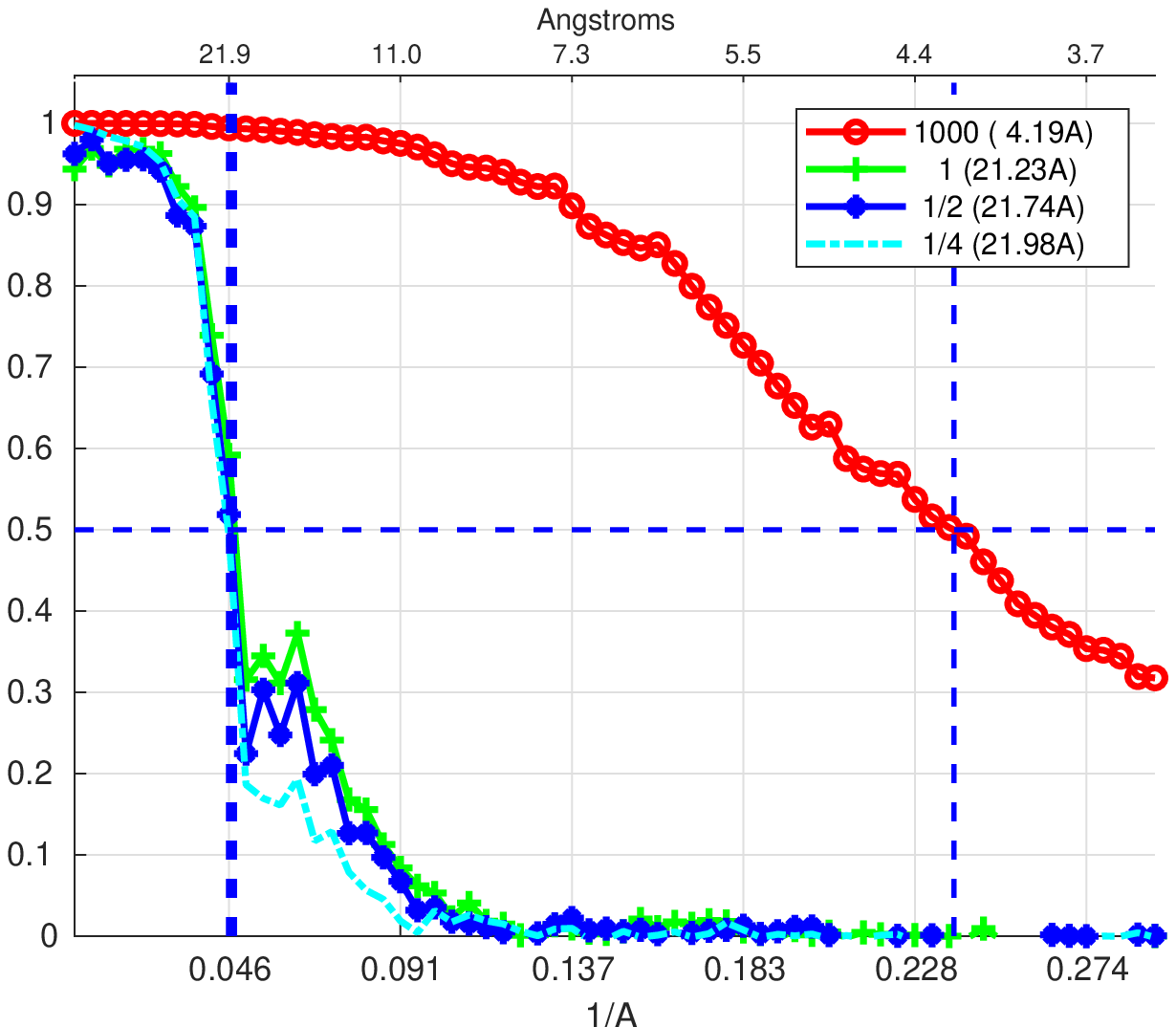}
		}
		\subfloat[$N=200$]{
			\includegraphics[width=0.4\textwidth]{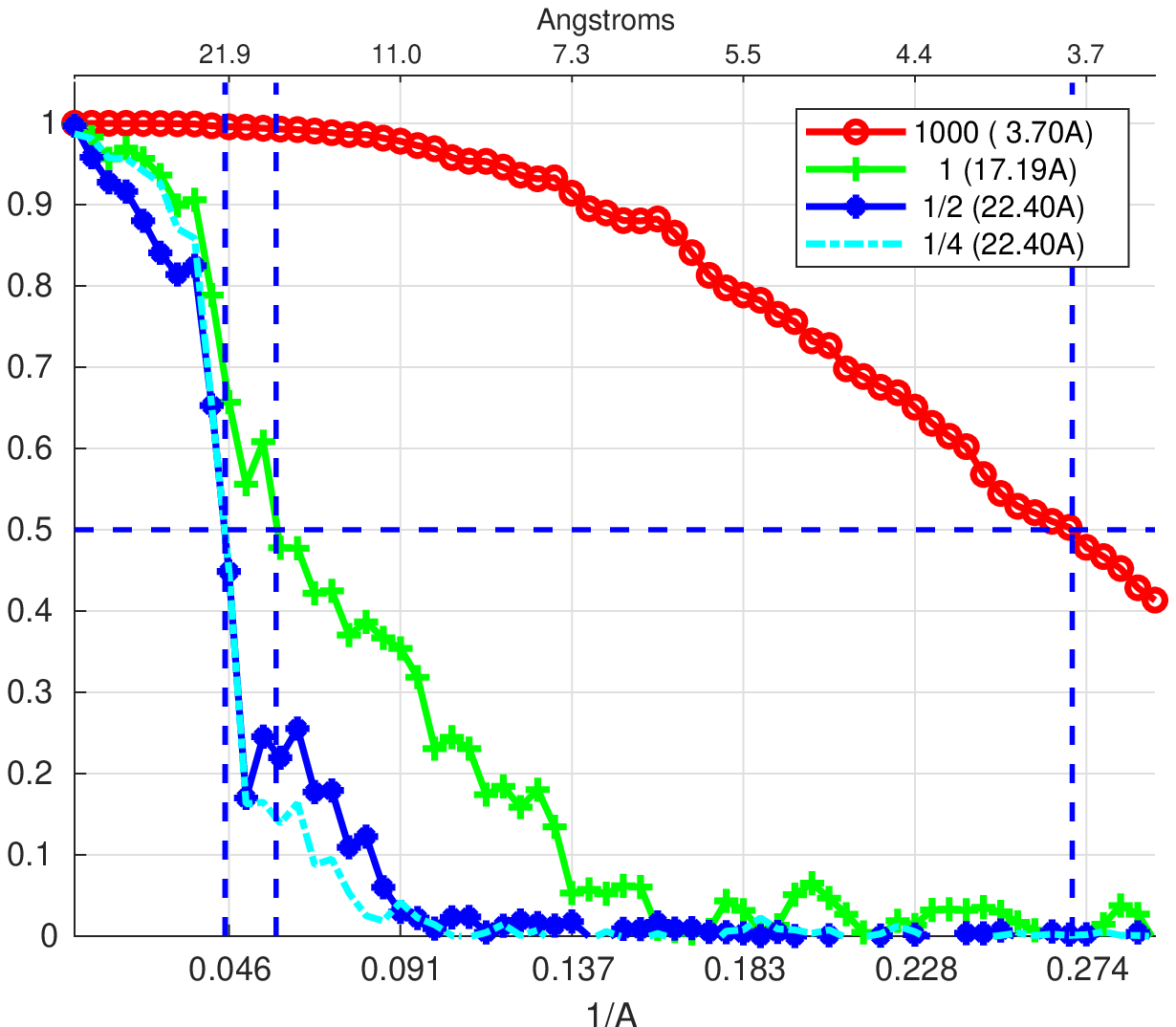}
		}	
	\end{center}
	\caption{Fourier shell correlation curves for volumes reconstructed from simulated projection-images of EMD-4905 ($\mathbb{O}$ symmetry). Each volume is reconstructed from a set of images whose size is specified in the caption of the panels.}
	\label{fig:ON}
\end{figure}

\begin{figure}
	\begin{center}
		\subfloat[$SNR=1000$]{
			\includegraphics[width=0.4\textwidth]{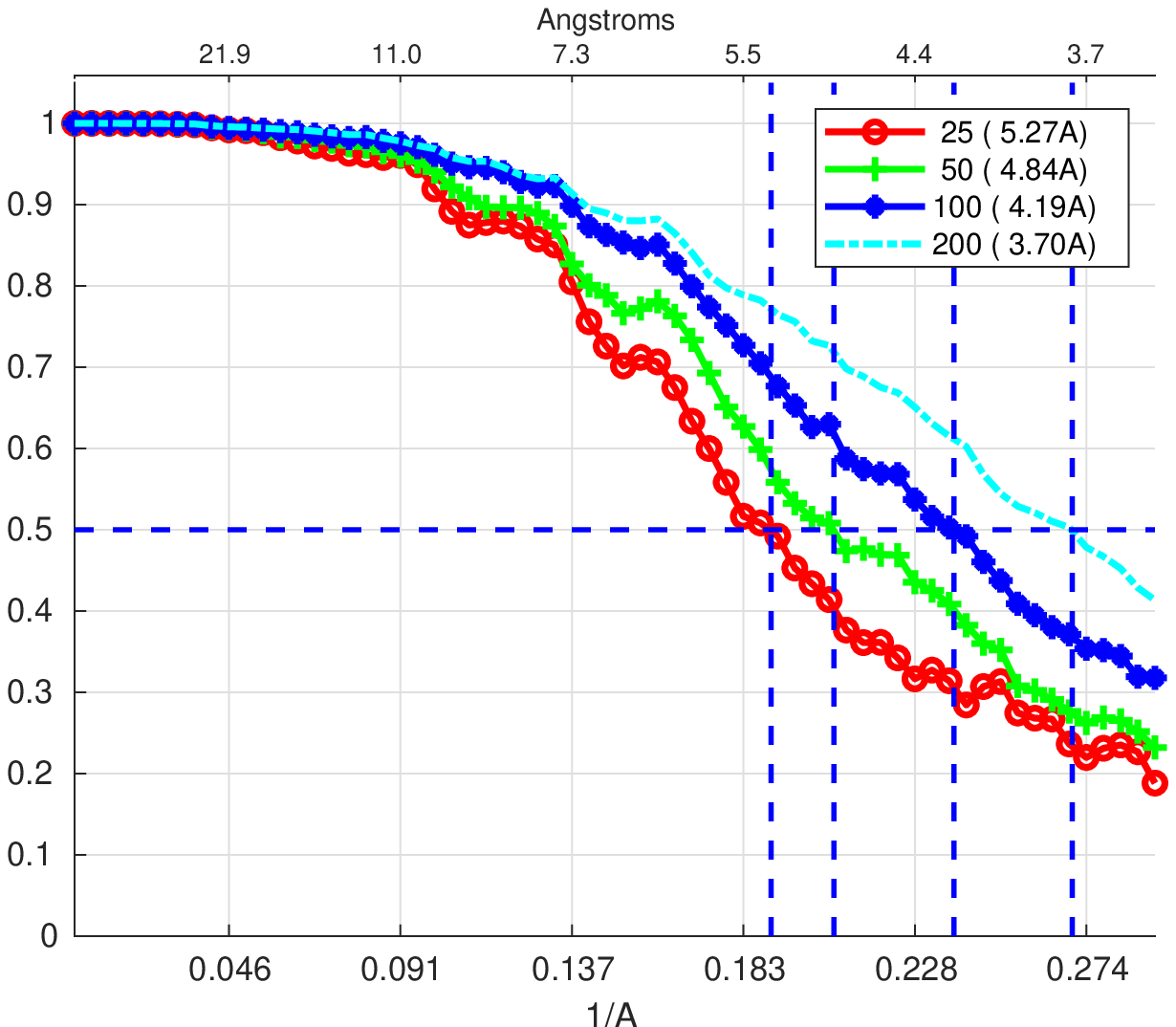}
		}
		\subfloat[$SNR=1$]{
			\includegraphics[width=0.4\textwidth]{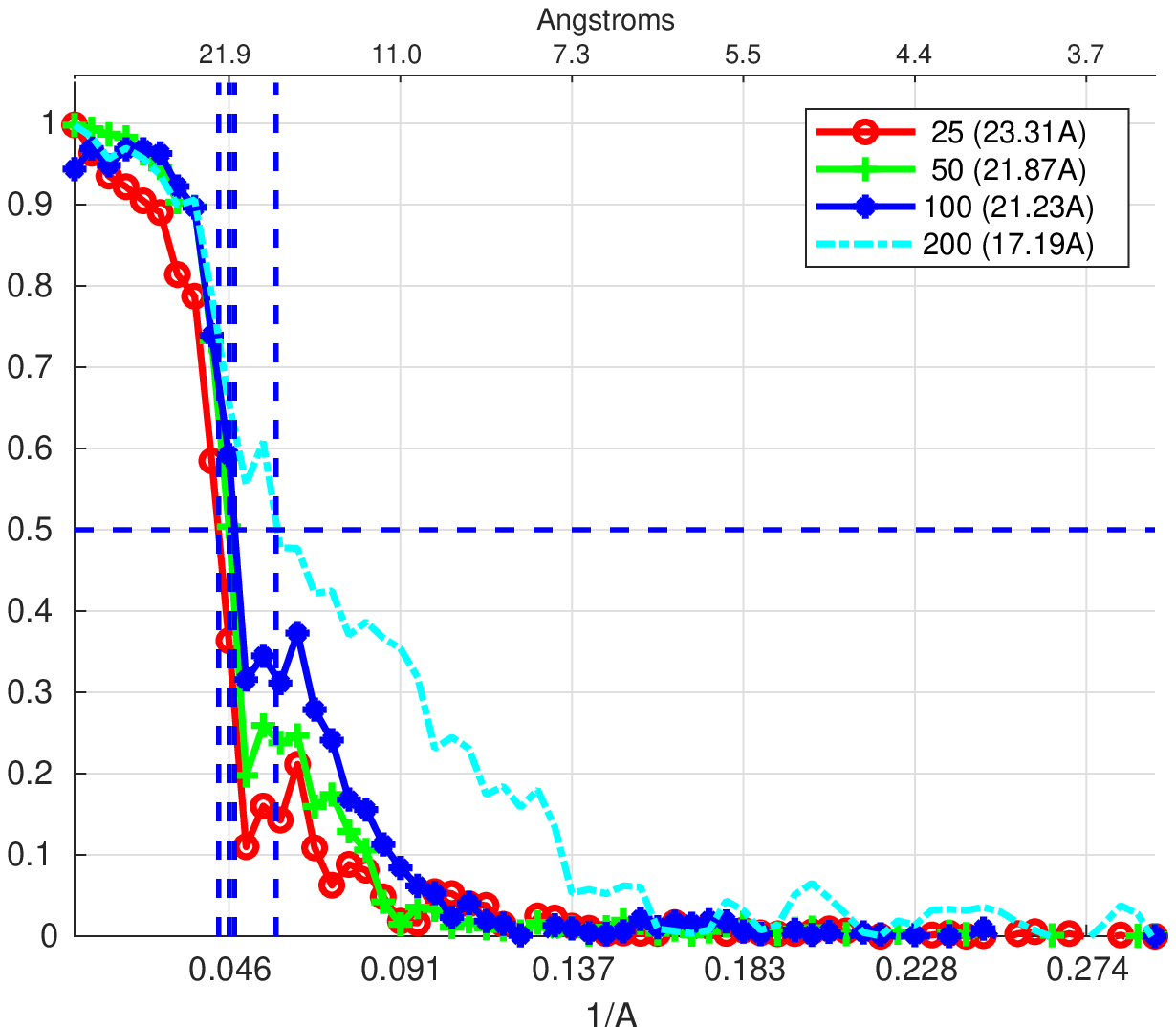}
		}\\
		\subfloat[$SNR=1/2$]{
			\includegraphics[width=0.4\textwidth]{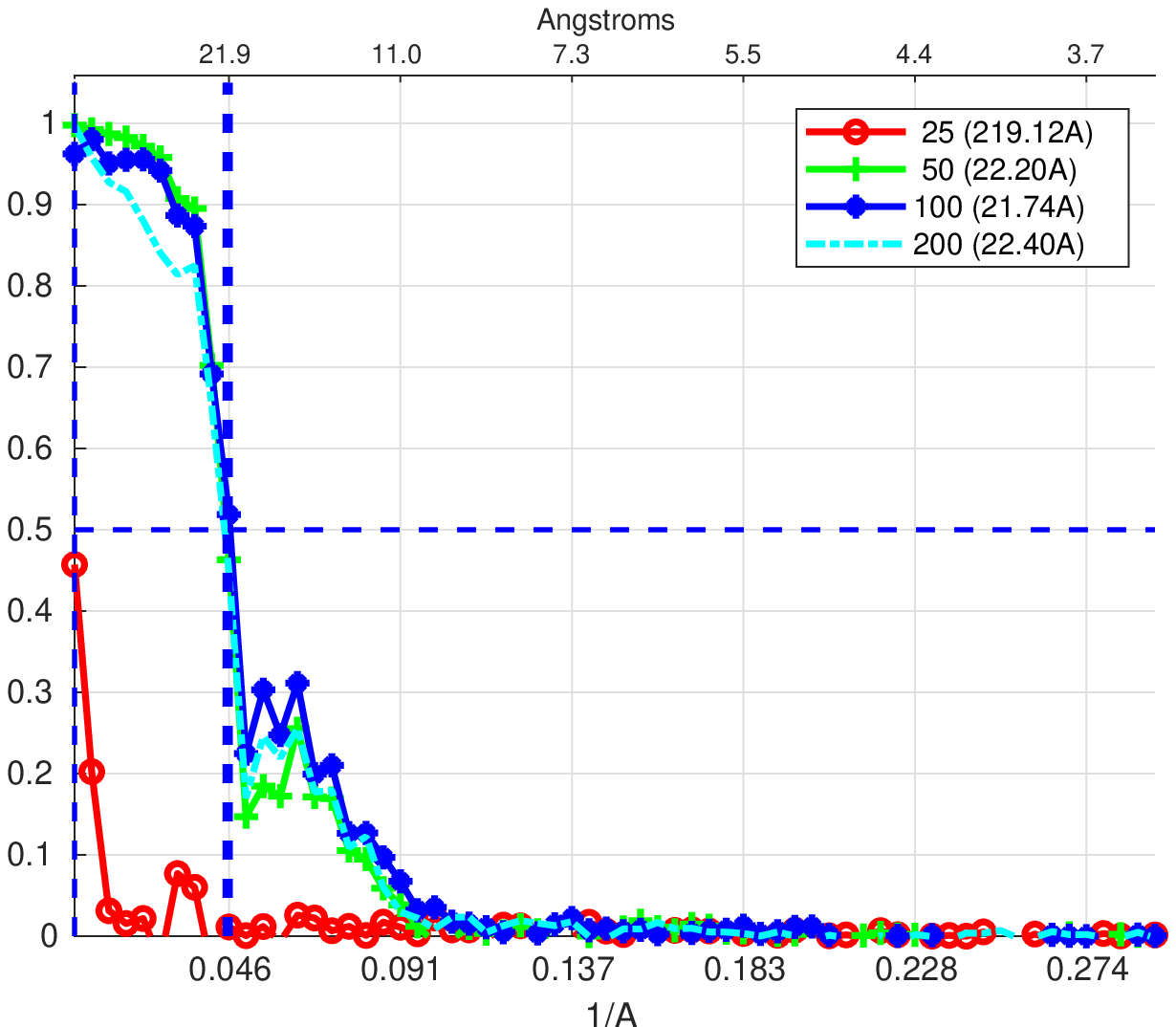}
		}
		\subfloat[$SNR=1/4$]{
			\includegraphics[width=0.4\textwidth]{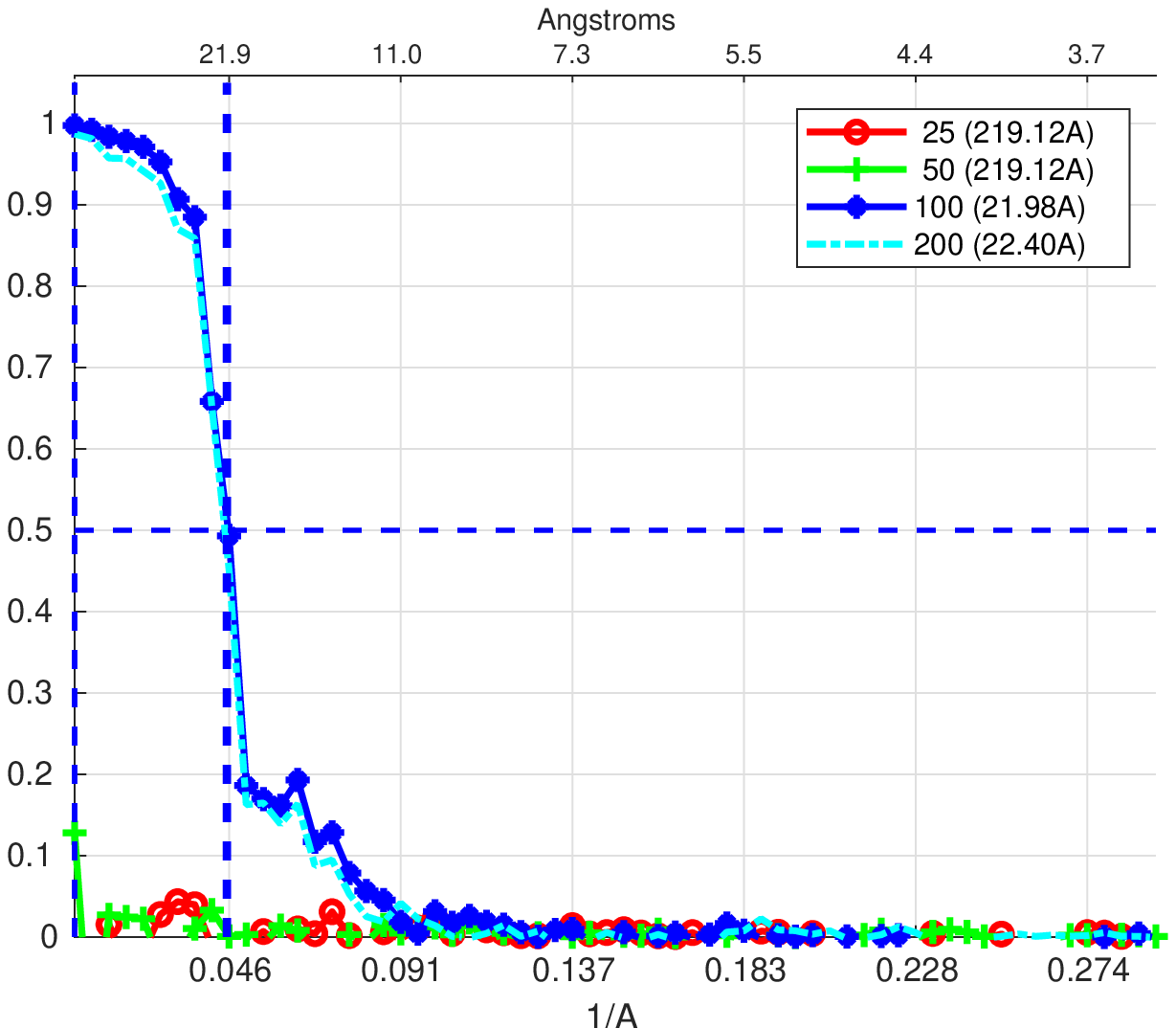}
		}	
	\end{center}
	\caption{Fourier shell correlation curves for volumes reconstructed from simulated projection-images of EMD-4905 ($\mathbb{O}$ symmetry).  Each panel corresponds to a given SNR, and a number of projections as appears in the legend.}
	\label{fig:OSNR}
\end{figure}

We then applied our algorithm on sets of $N=25, 50, 100, 200$ images at these noise levels. For each group of $N$ images, we plot in Fig.~\ref{fig:ON} the Fourier shell correlation curve (FSC)~\cite{vanHeel_Schatz} of the volume reconstructed by our algorithm relative to the ground truth volume. In a nutshell, the FSC measures the size of the smallest feature in the reconstructed volume that can be resolved. Figure~\ref{fig:ON} shows that increasing the number of input images improves the performance of the algorithm in terms of the achieved resolution. Moreover, we see that the algorithm fails for some of the noise levels (achieved resolution worse than 30~\AA), but as we increase the number of images, the algorithm successfully reconstructs a three-dimensional model of the molecule, with resolution of about 20~\AA. To further demonstrate this point, we show in Figure~\ref{fig:OSNR} the Fourier shell correlation curves for a fixed SNR and a variable number of images. As before, increasing the number of images improves the achieved resolution. The timing (in seconds) of the algorithm is summarized in Table~\ref{tbl:timing O}. These timings were computed by averaging for each $N$ the timing results for all SNRs (as the running time is independent of the noise level).

\begin{table}
	\begin{center}
		\begin{tabular}{lrrrr}
			\hline
			N & 25  & 50 & 100 & 200 \\
			Time (sec) & 282 &  852 & 2,842 & 11,177\\
			\hline
		\end{tabular}
	\end{center}
	\caption{Timing (in seconds) for $\mathbb{O}$ symmetry.}
	\label{tbl:timing O}
\end{table}

\begin{figure}
	\begin{center}
		\includegraphics[width=0.2\textwidth]{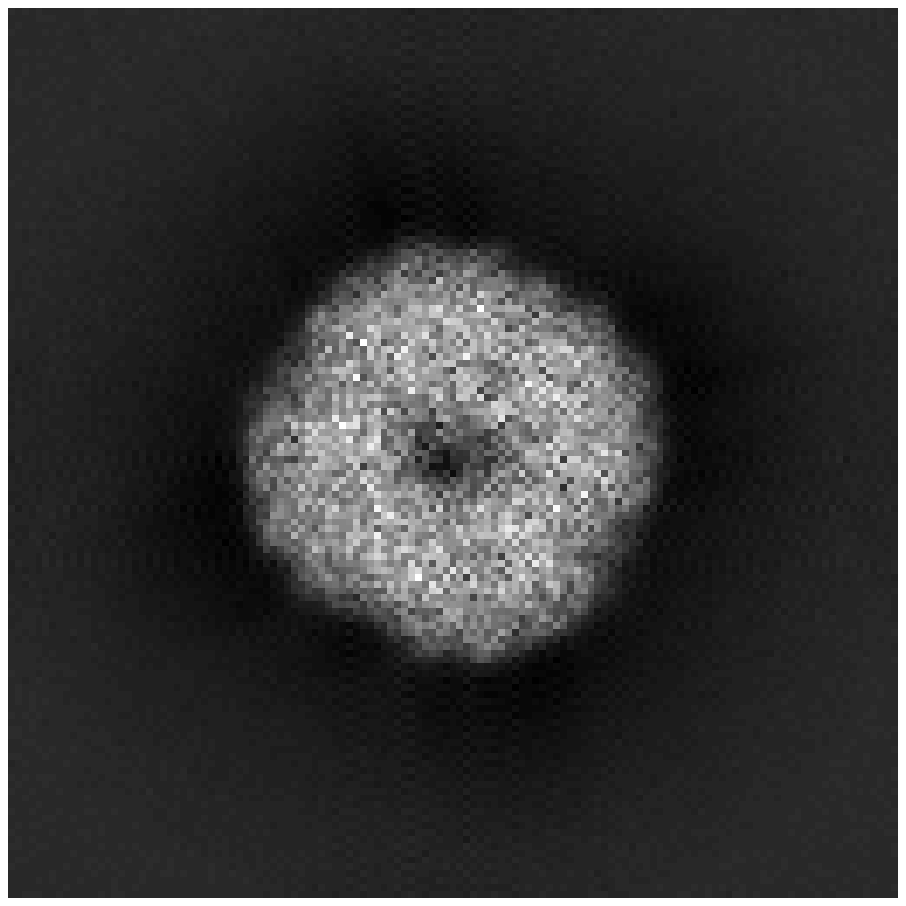}
		\includegraphics[width=0.2\textwidth]{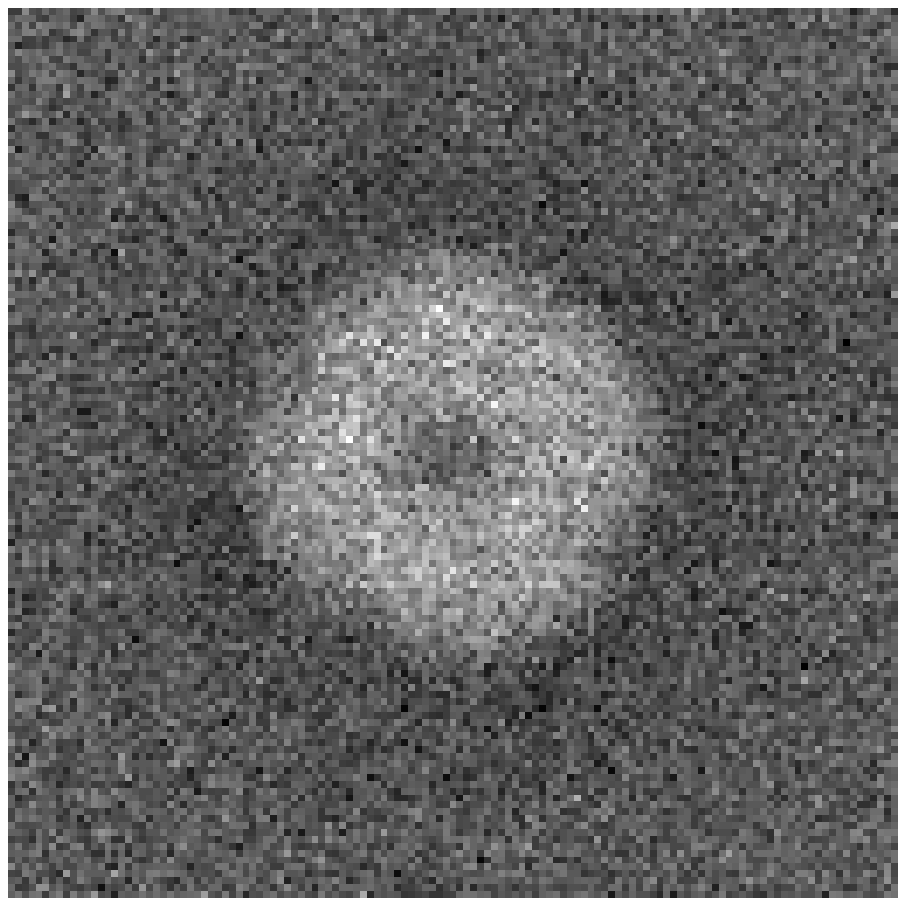}
		\includegraphics[width=0.2\textwidth]{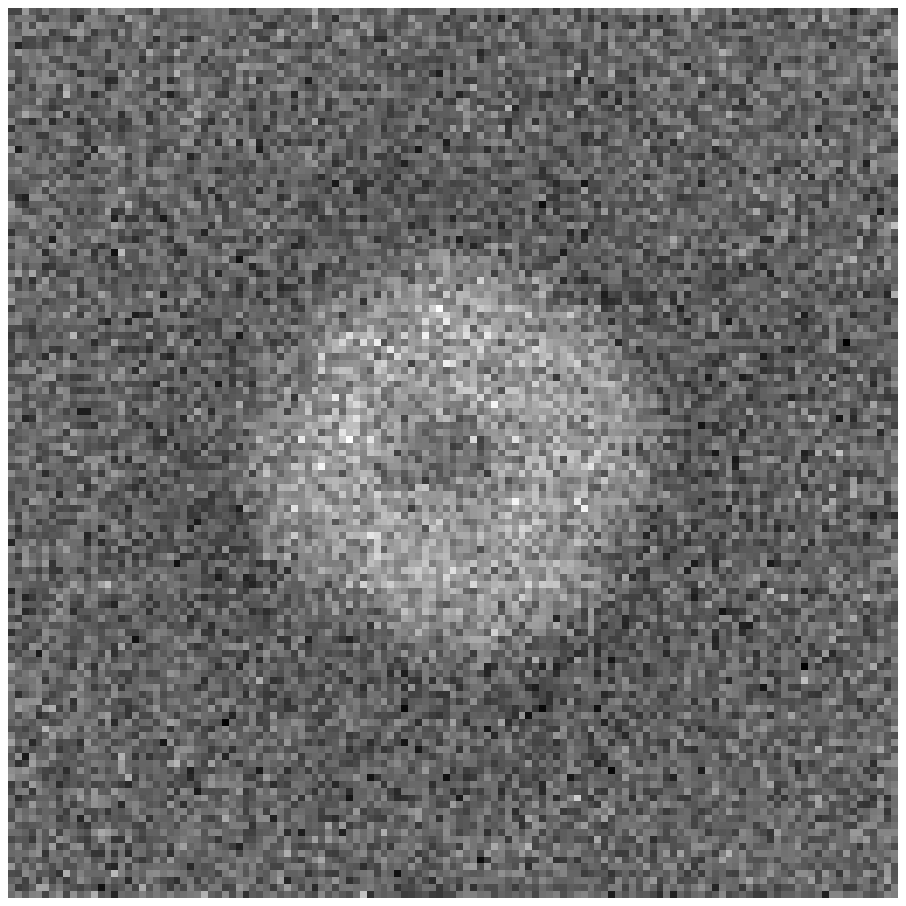}
		\includegraphics[width=0.2\textwidth]{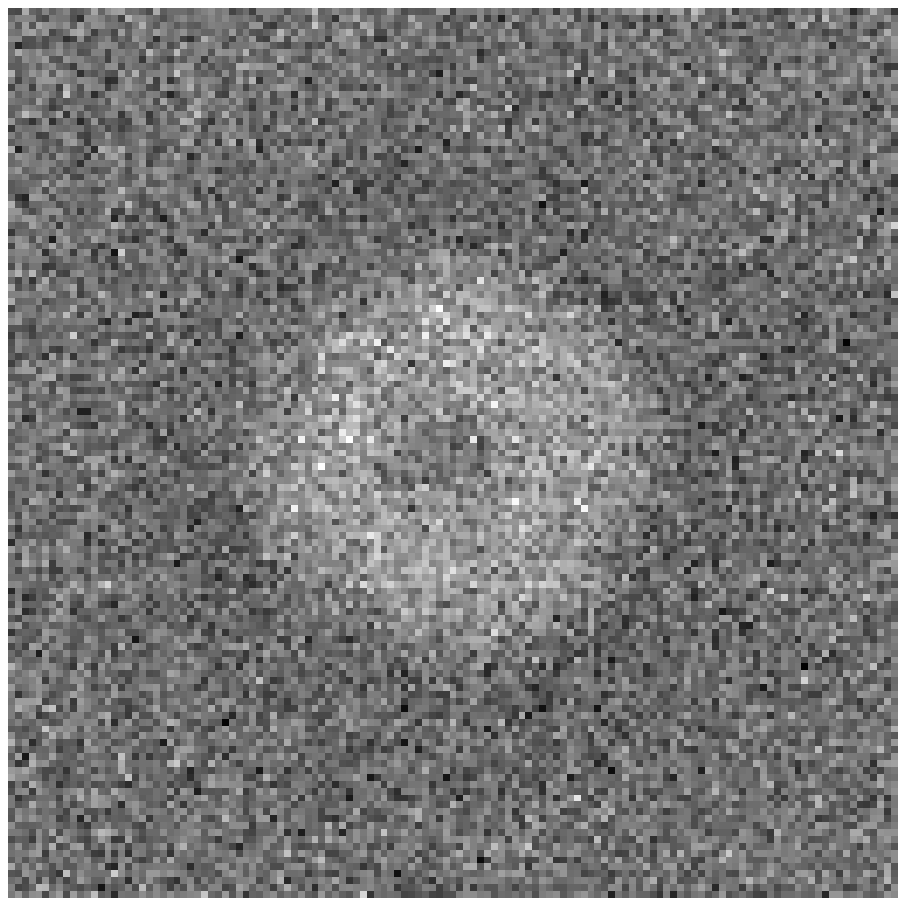}\\
		\includegraphics[width=0.2\textwidth]{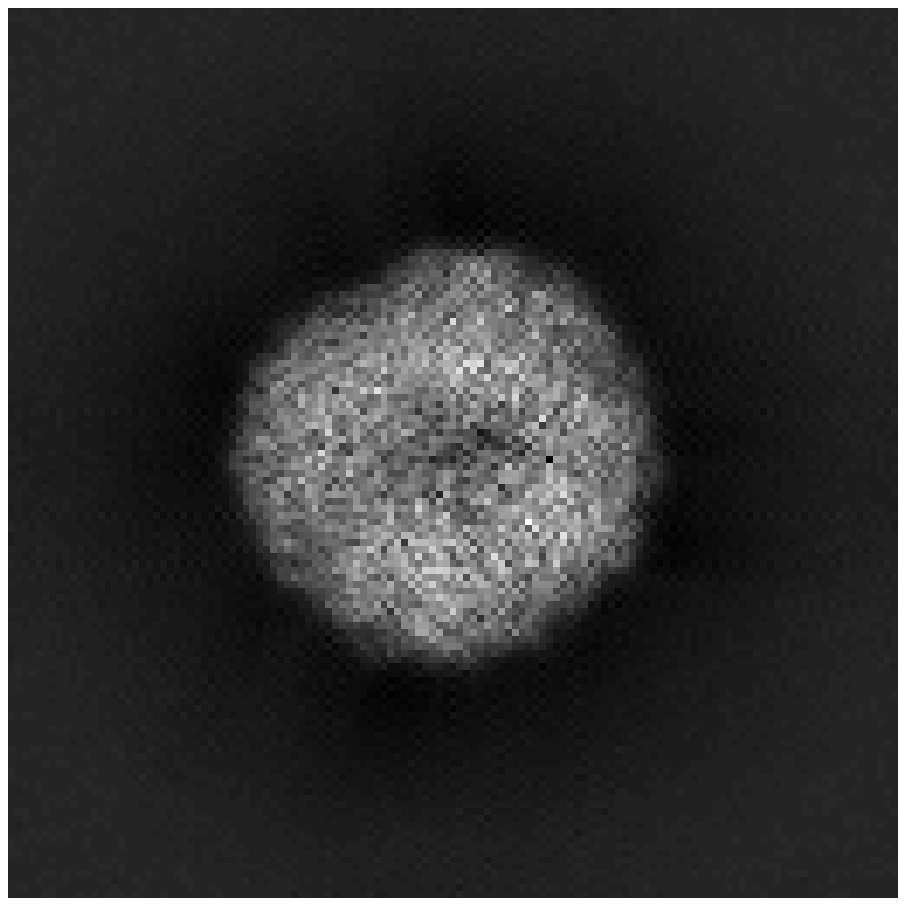}
		\includegraphics[width=0.2\textwidth]{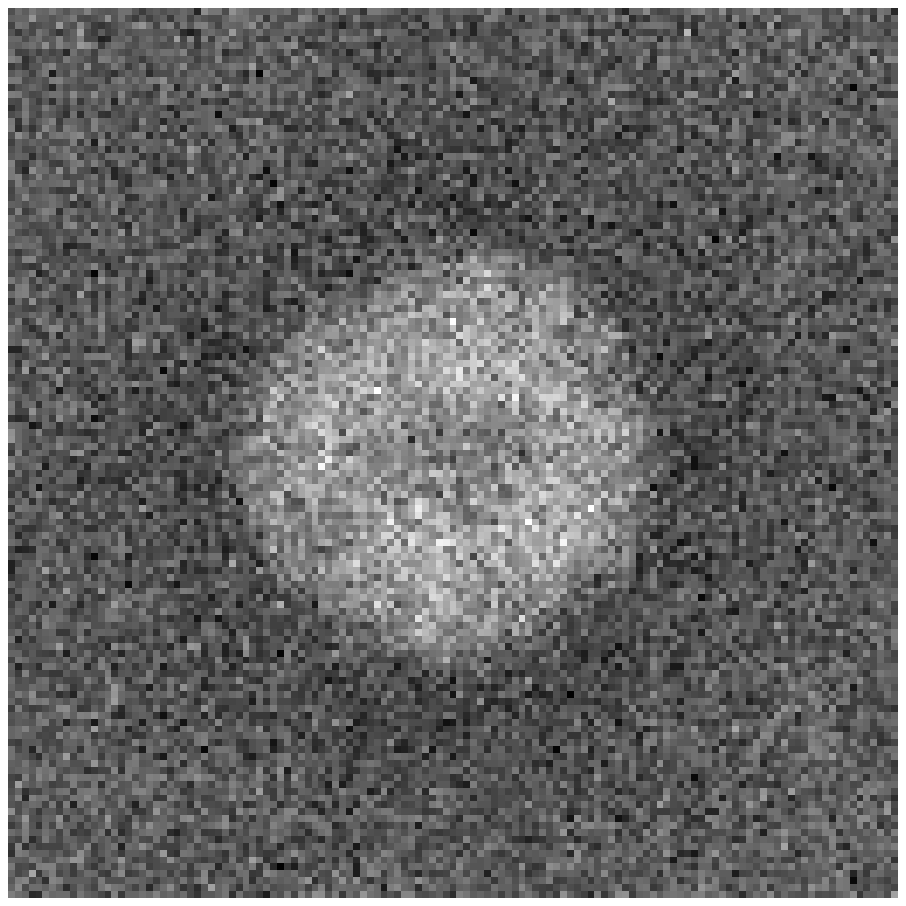}
		\includegraphics[width=0.2\textwidth]{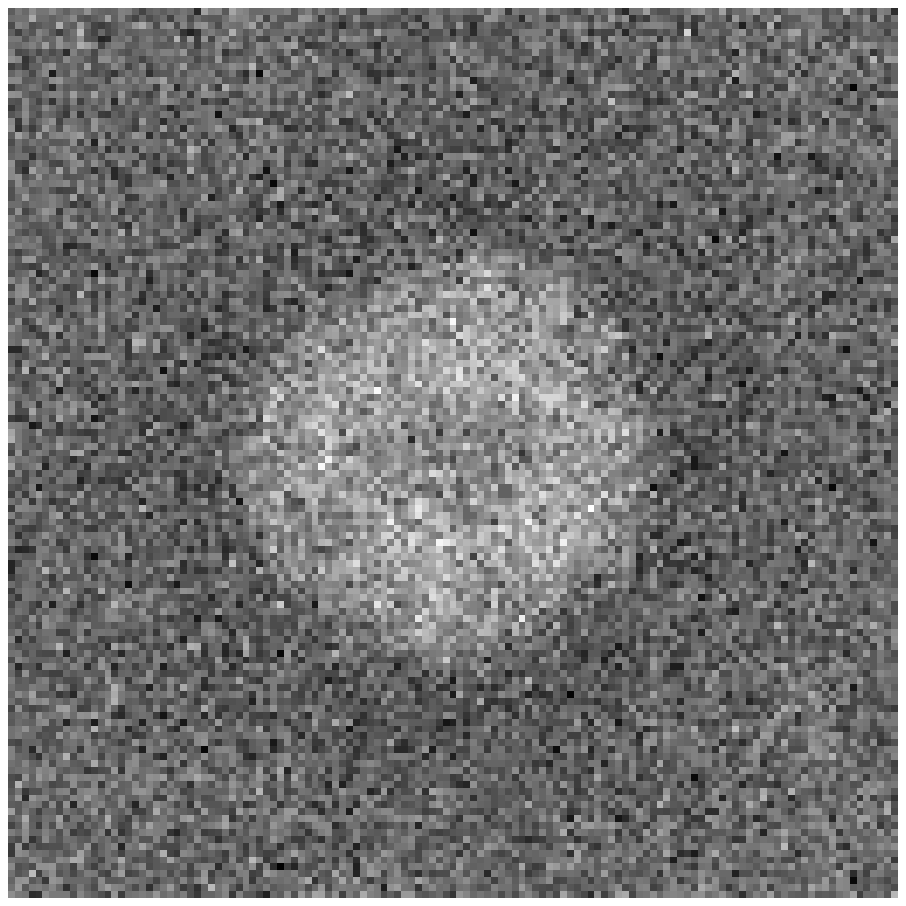}
		\includegraphics[width=0.2\textwidth]{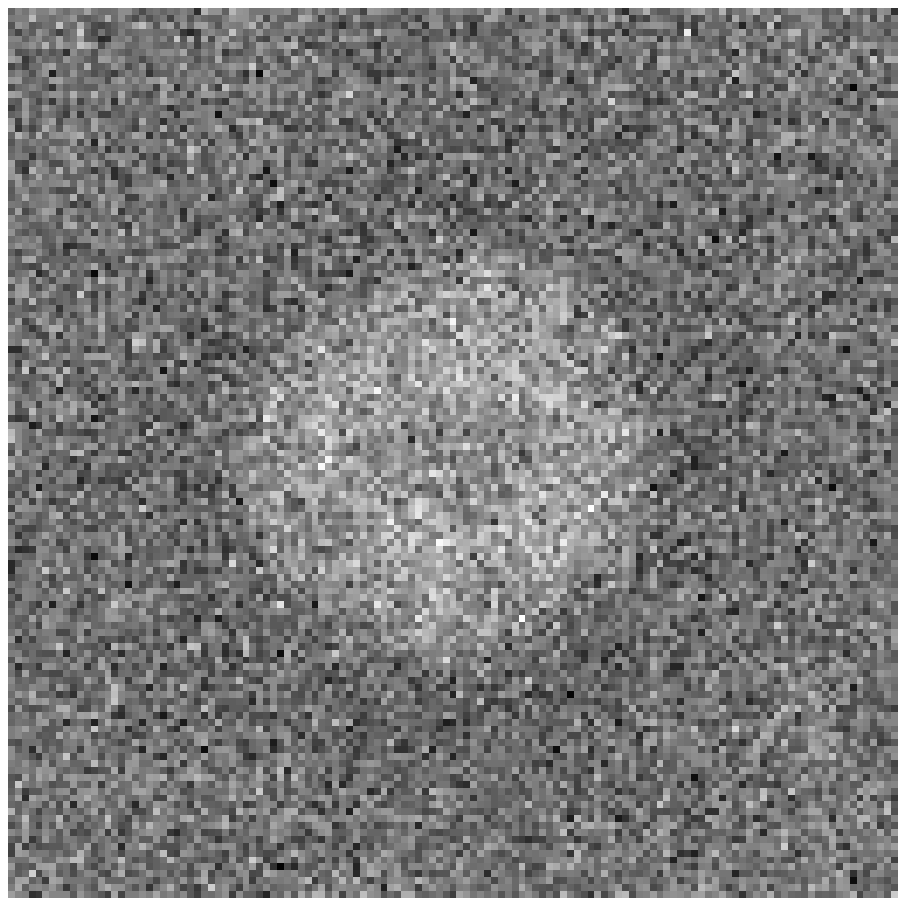}\\
		\includegraphics[width=0.2\textwidth]{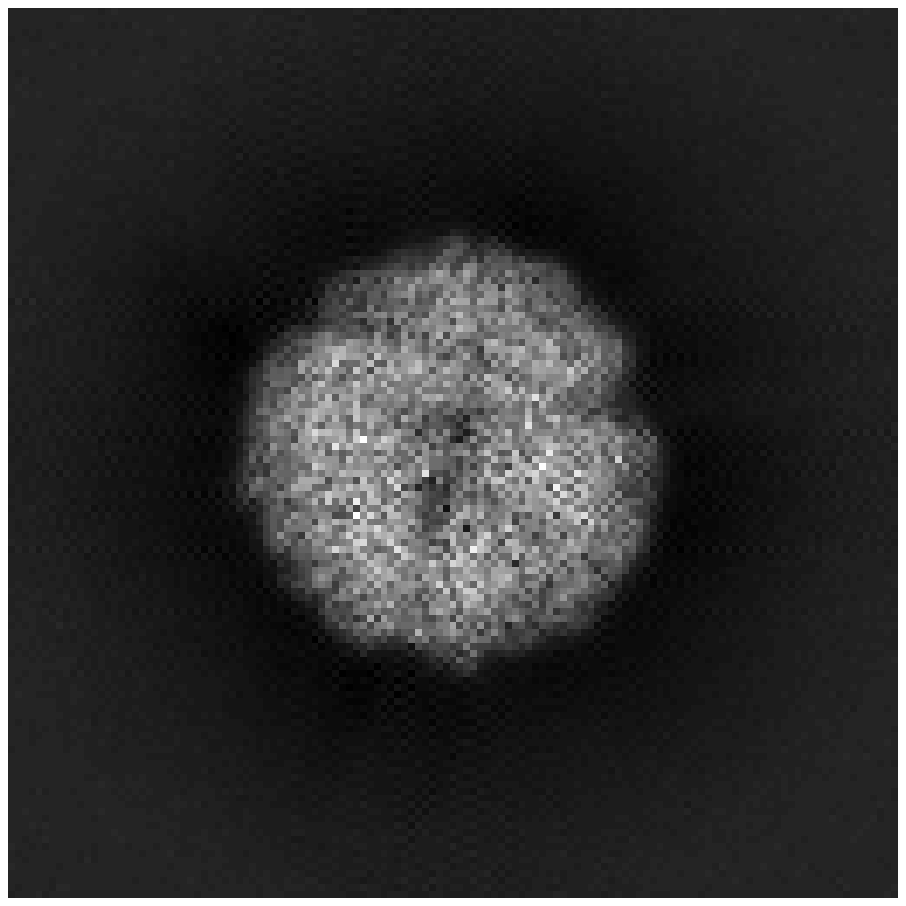}
		\includegraphics[width=0.2\textwidth]{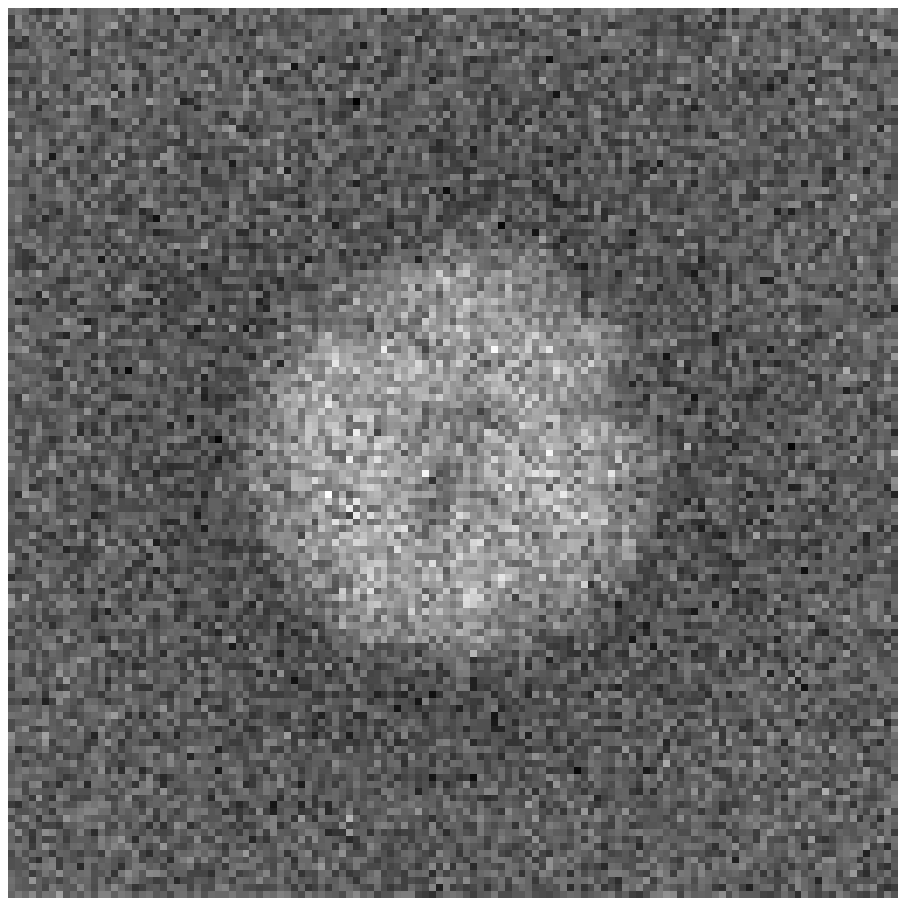}
		\includegraphics[width=0.2\textwidth]{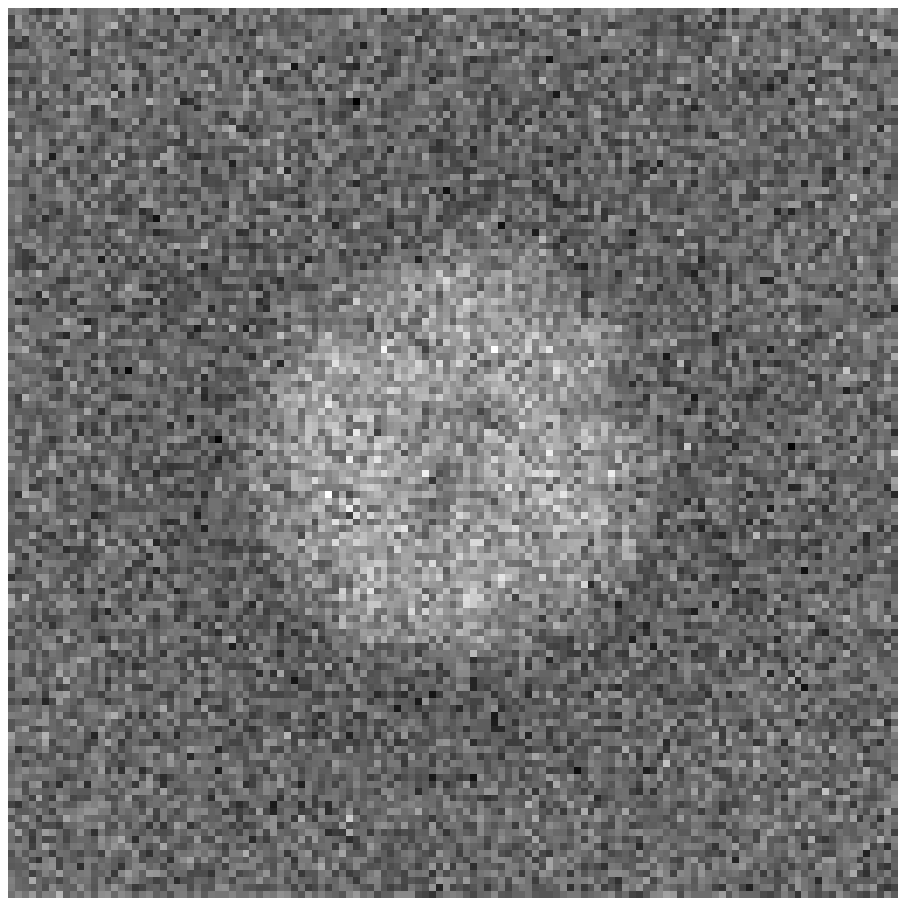}
		\includegraphics[width=0.2\textwidth]{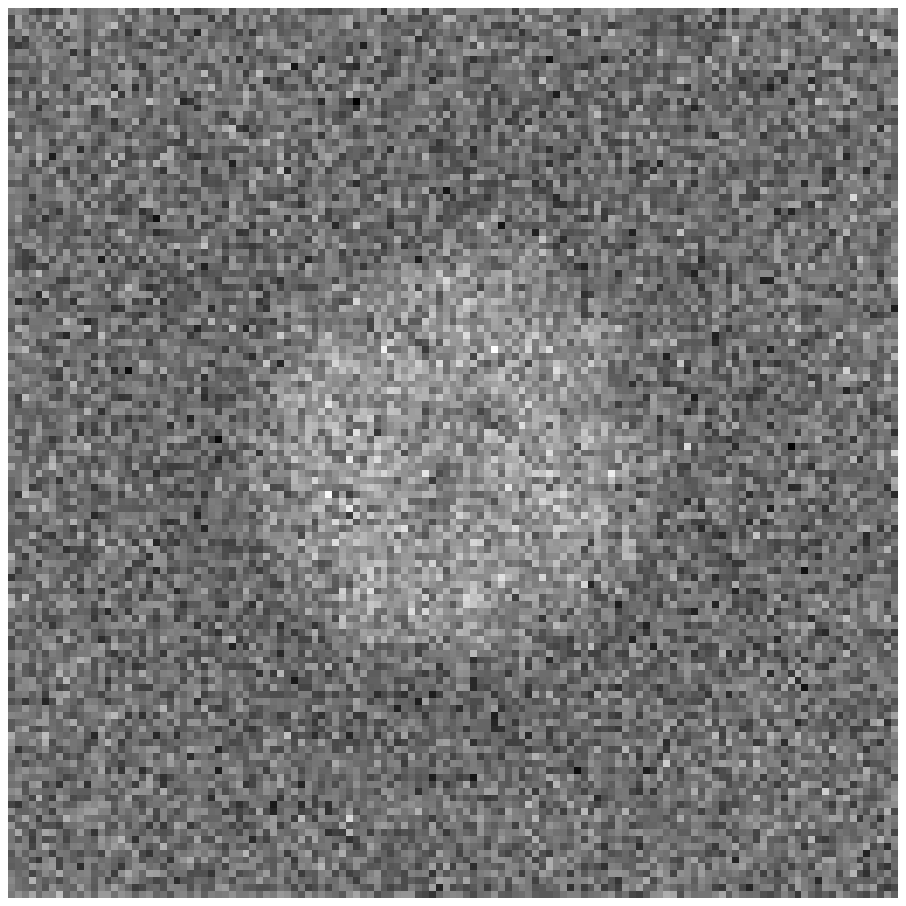}\\
		\includegraphics[width=0.2\textwidth]{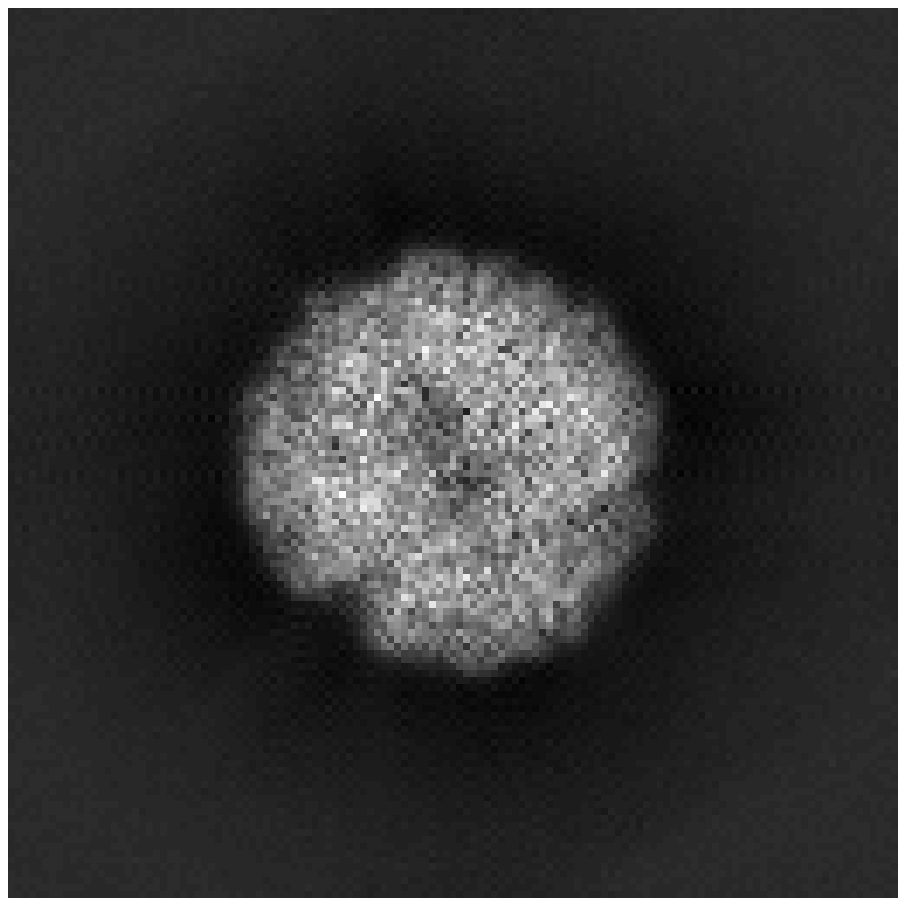}
		\includegraphics[width=0.2\textwidth]{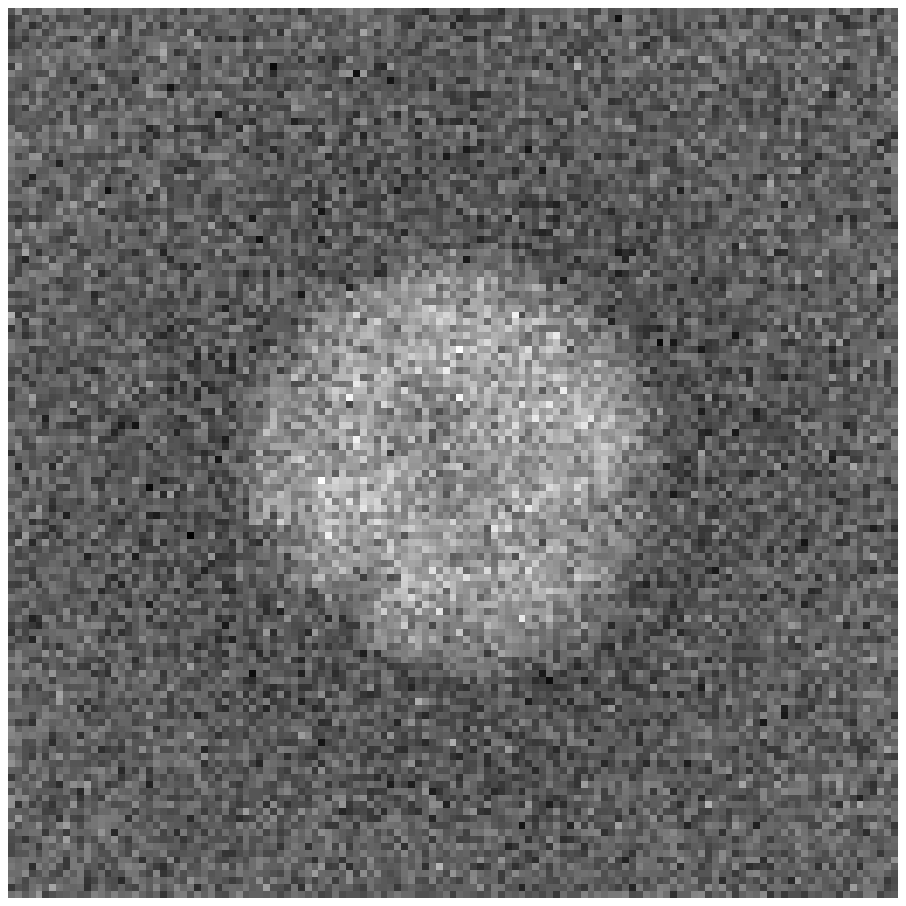}
		\includegraphics[width=0.2\textwidth]{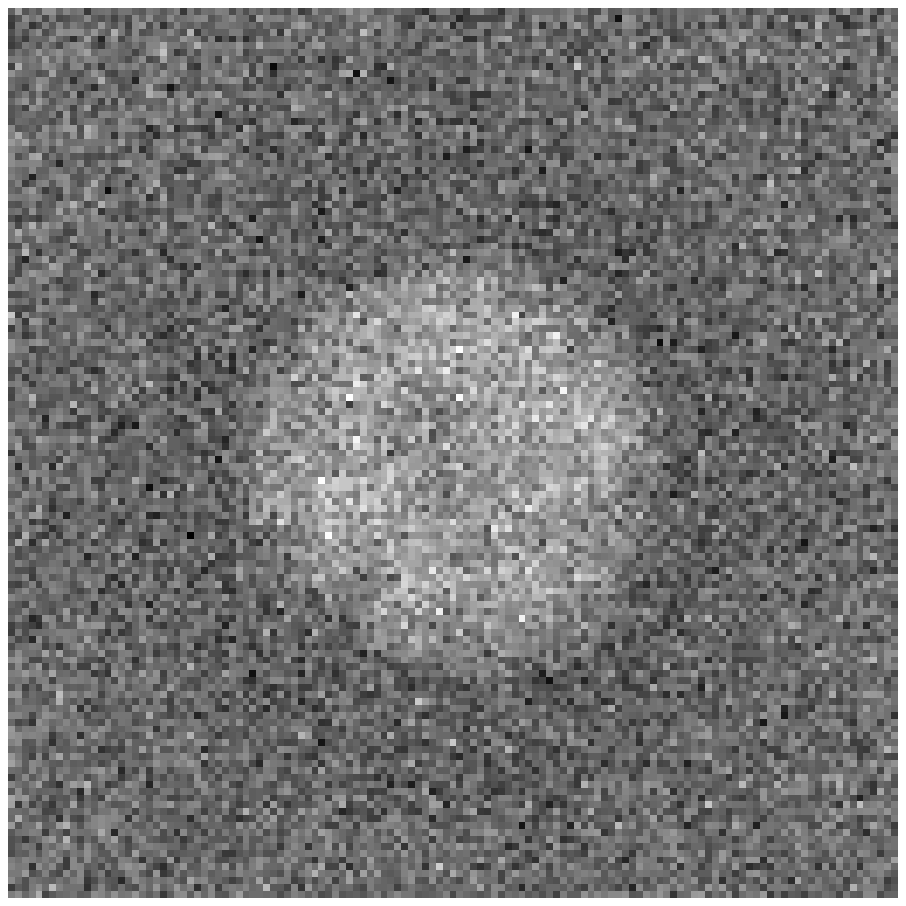}
		\includegraphics[width=0.2\textwidth]{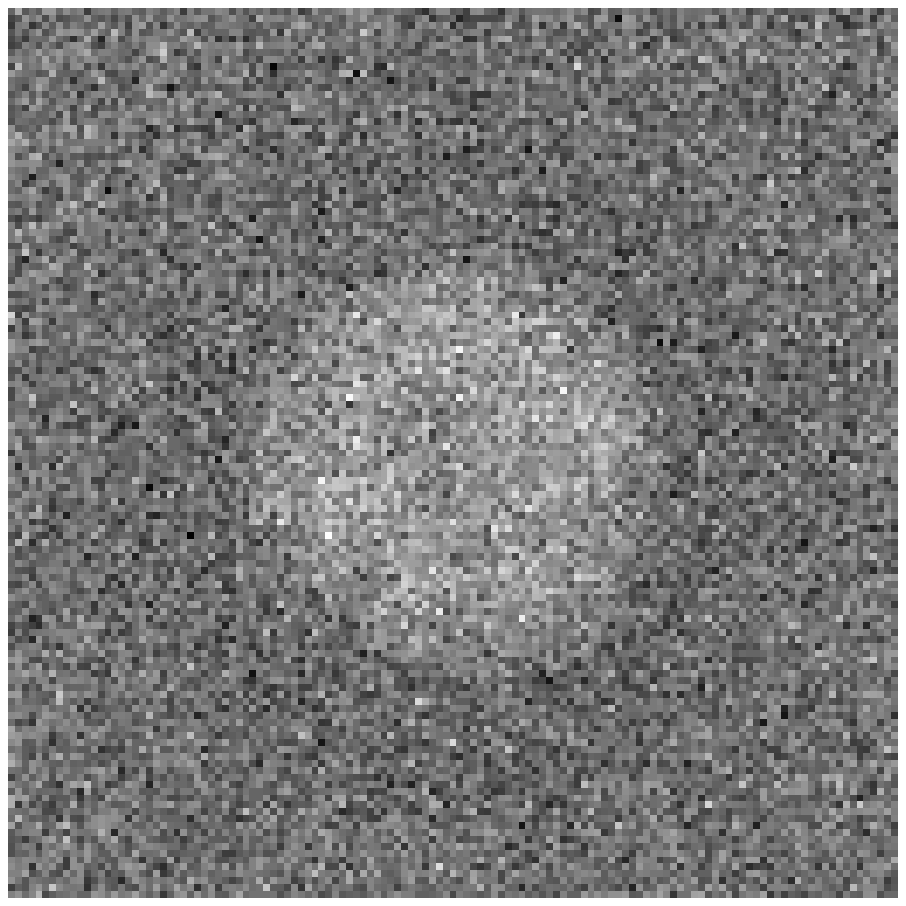}
	\end{center}
	\caption{Examples of simulated projection-images of EMD-10835~\protect\cite{EMD10835} ($\mathbb{T}$ symmetry) with signal to noise ratio of (from left to right) 1000, 1, 1/2, 1/4.}
	\label{fig:simulated 10835 projections}
\end{figure}

\begin{figure}
	\begin{center}
		\subfloat[$N=25$]{
			\includegraphics[width=0.4\textwidth]{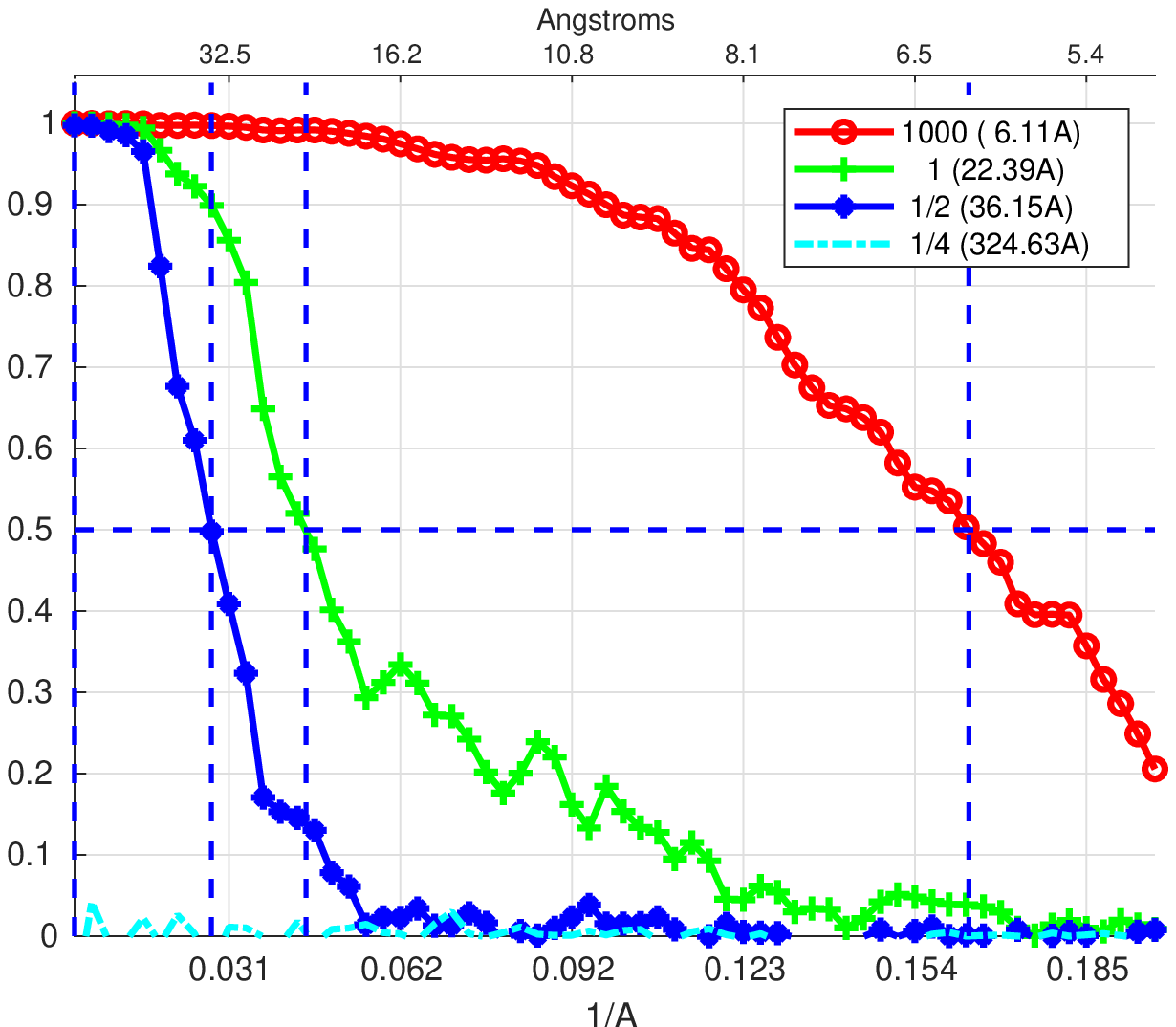}
		}
		\subfloat[$N=50$]{
			\includegraphics[width=0.4\textwidth]{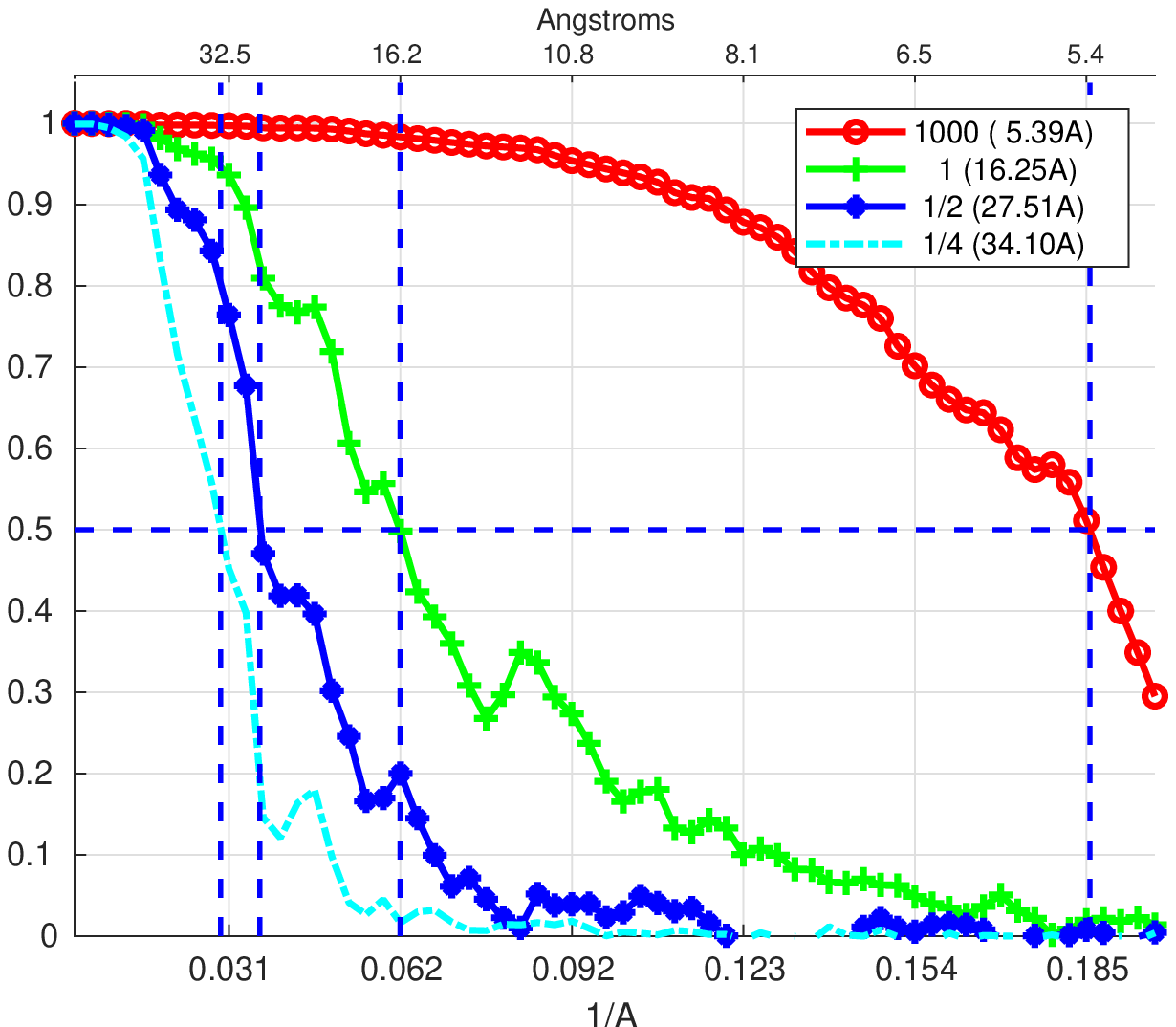}
		}\\
		\subfloat[$N=100$]{
			\includegraphics[width=0.4\textwidth]{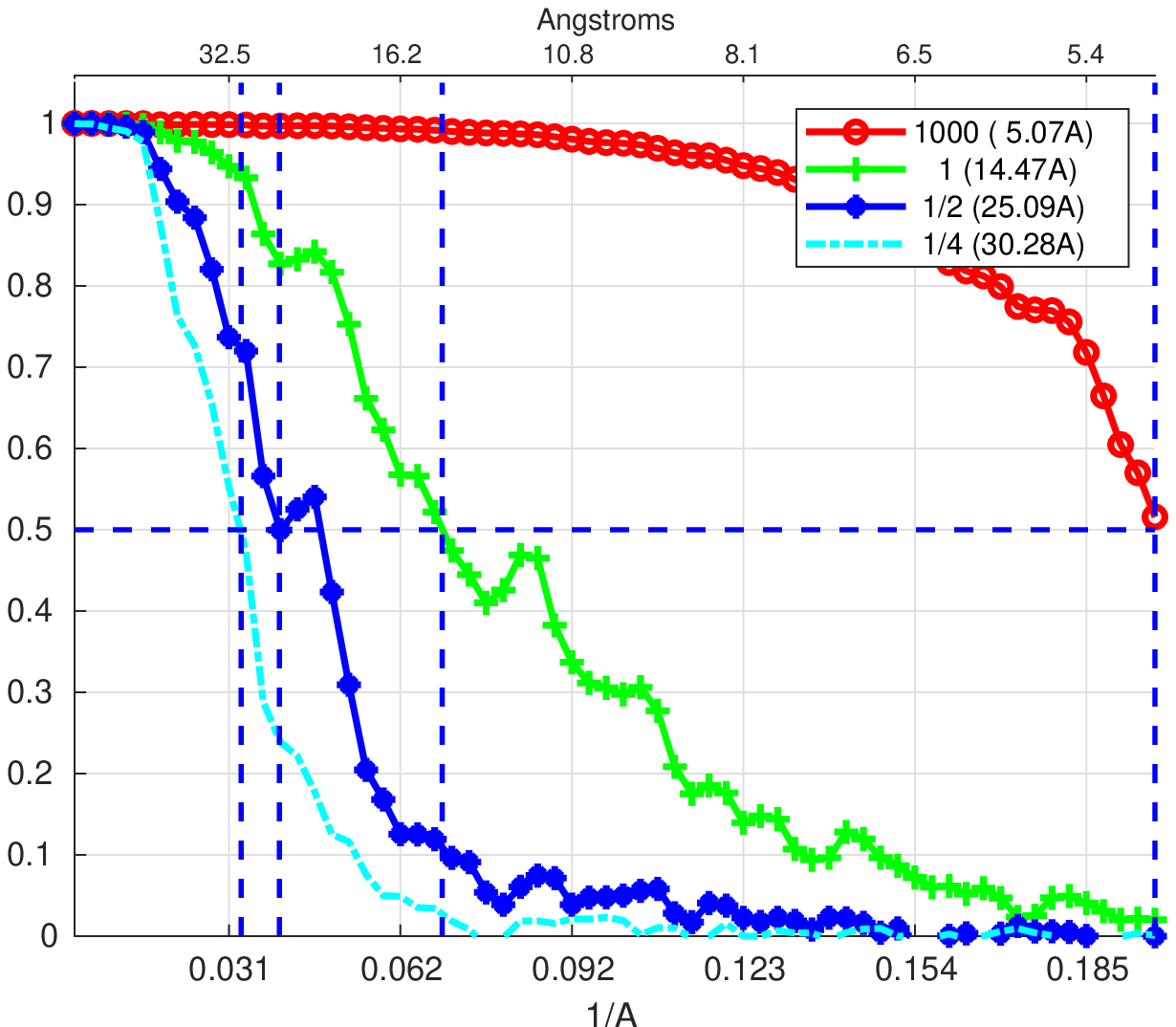}
		}
		\subfloat[$N=200$]{
			\includegraphics[width=0.4\textwidth]{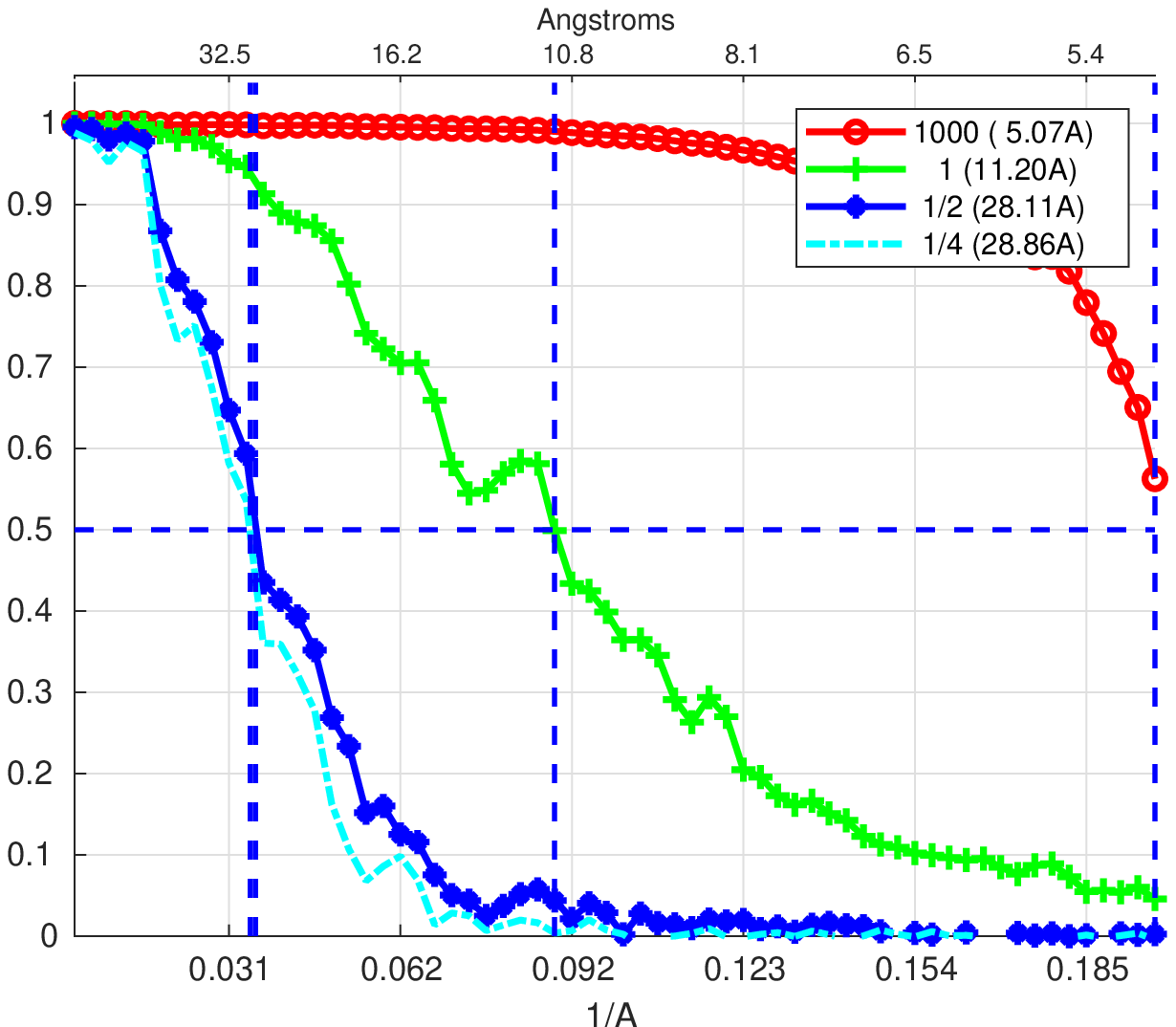}
		}	
	\end{center}
	\caption{Fourier shell correlation curves for volumes reconstructed from simulated projection-images of EMD-10835 ($\mathbb{T}$ symmetry).  See Fig.~\protect\ref{fig:ON} for details.}
	\label{fig:TN}
\end{figure}

\begin{figure}
	\begin{center}
		\subfloat[$SNR=1000$]{
			\includegraphics[width=0.4\textwidth]{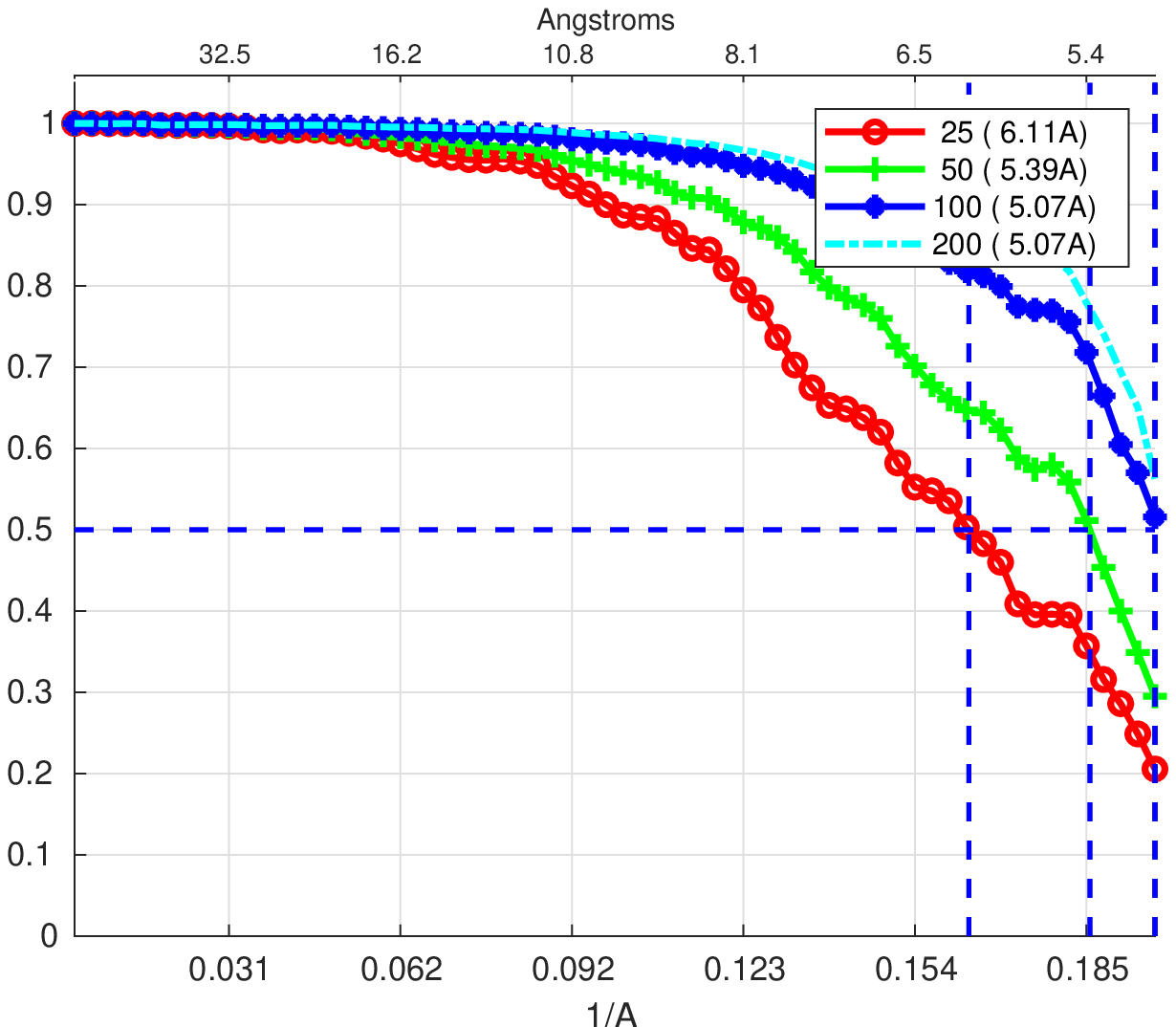}
		}
		\subfloat[$SNR=1$]{
			\includegraphics[width=0.4\textwidth]{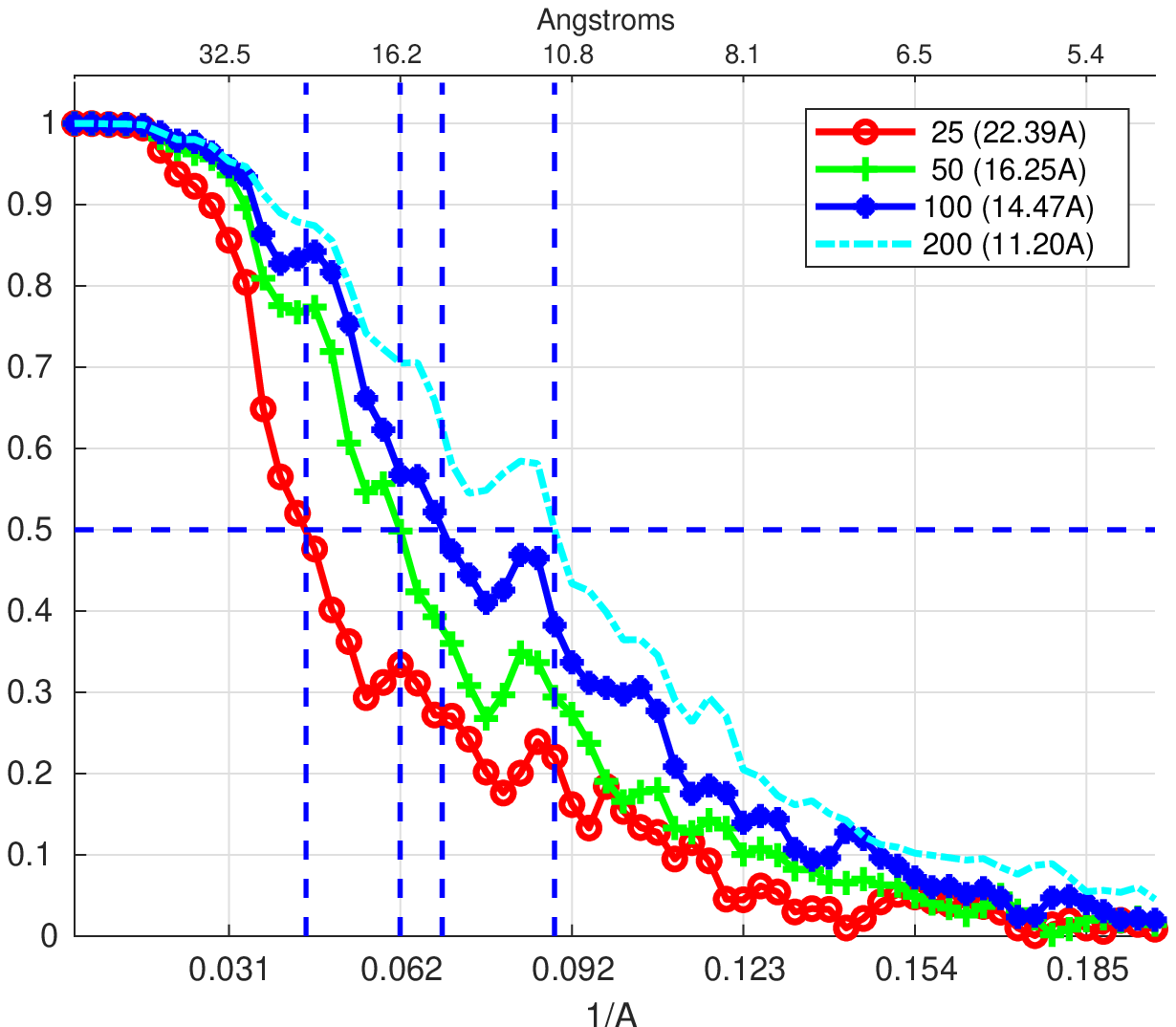}
		}\\
		\subfloat[$SNR=1/2$]{
			\includegraphics[width=0.4\textwidth]{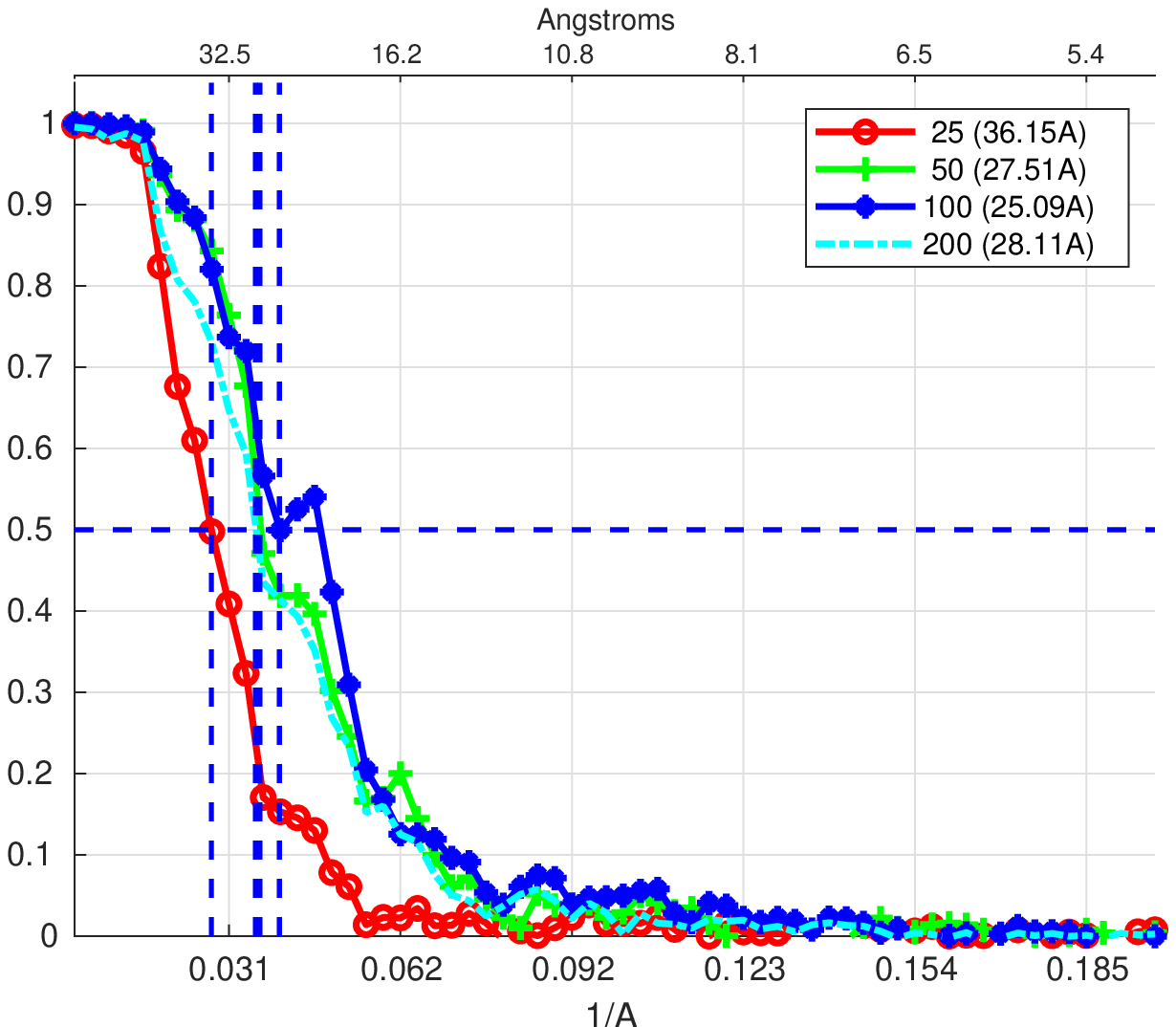}
		}
		\subfloat[$SNR=1/4$]{
			\includegraphics[width=0.4\textwidth]{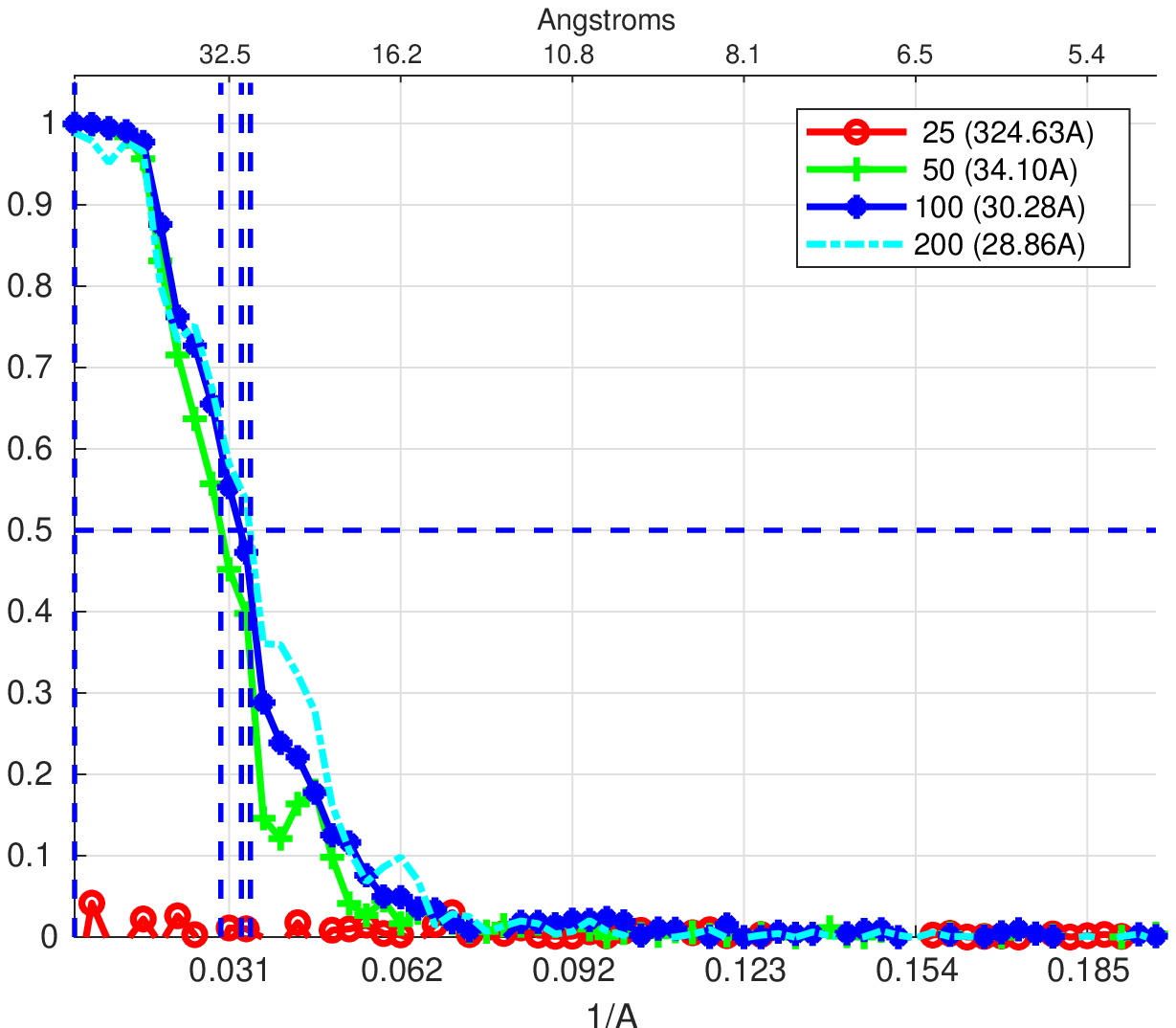}
		}	
	\end{center}
	\caption{Fourier shell correlation curves for volumes reconstructed from simulated projection-images of EMD-10835 ($\mathbb{T}$ symmetry). See Fig.\protect\ref{fig:OSNR} for details.}
	\label{fig:TSNR}
\end{figure}

The results of the same experiment for $\mathbb{T}$ symmetry with projections generated from EMD-10835~\cite{EMD10835} are shown in Figs.~\ref{fig:simulated 10835 projections},~\ref{fig:TN} and~\ref{fig:TSNR}. The timing for $\mathbb{T}$ symmetry is summarized in Table~\ref{tbl:timing T}.

\begin{table}
	\begin{center}
		\begin{tabular}{lrrrr}
			\hline
			N & 25  & 50 & 100 & 200 \\
			Time (sec) & 397 & 1,352 &  5,178 & 20,228 \\			
			\hline
		\end{tabular}
	\end{center}
	\caption{Timing (in seconds) for $\mathbb{T}$ symmetry.}
	\label{tbl:timing T}
\end{table}

\subsection{Experimental data}\label{sec:experimental data}

Next, we applied our algorithm to two experimental data sets -- EMPIAR-10272 and EMPIAR-10389 from the EMPIAR repository~\cite{iudin2016empiar}. The EMPIAR-10272 data set corresponds to EMD-4905~\cite{EMD4905} that has $\mathbb O$ symmetry, and the EMPIAR-10389 data set corresponds to EMD-10835~\cite{EMD10835} that has $\mathbb T$ symmetry. For comparison, we also generated an ab-initio models from these data sets using Relion~\cite{relion3}.

\begin{figure}
\begin{center}
\includegraphics[width=0.15\textwidth]{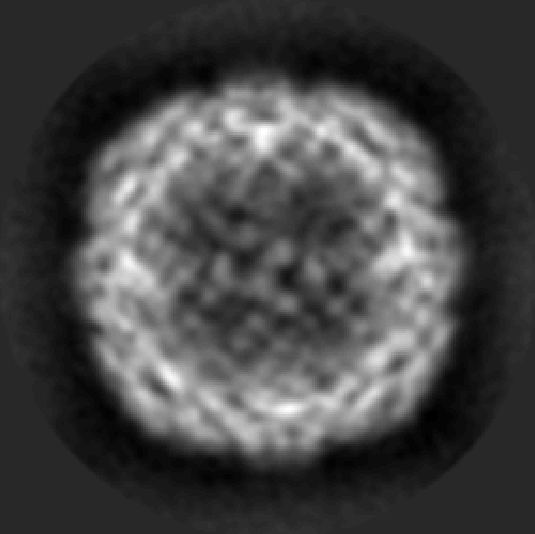}
\includegraphics[width=0.15\textwidth]{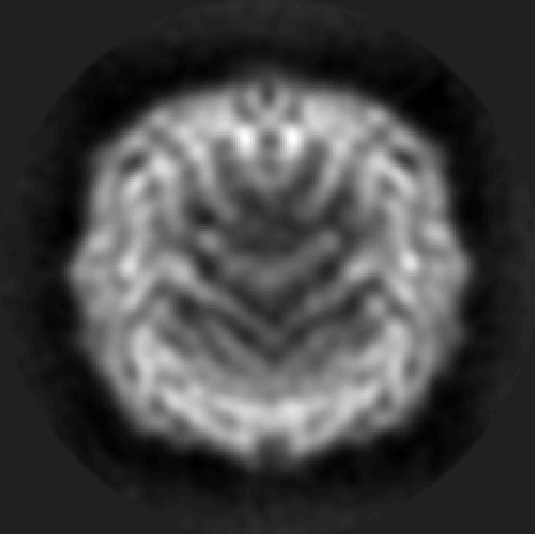}
\includegraphics[width=0.15\textwidth]{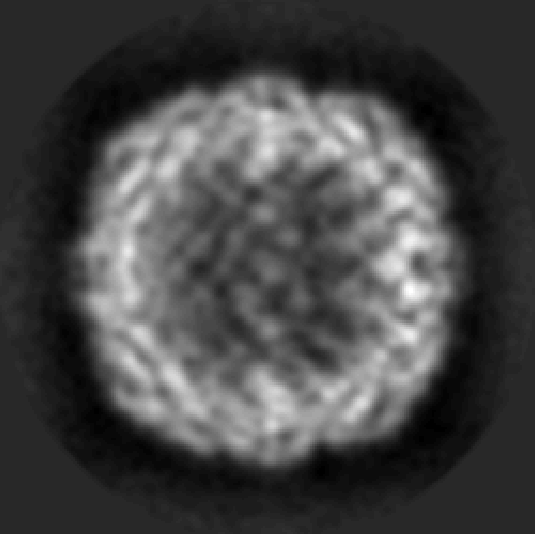}
\includegraphics[width=0.15\textwidth]{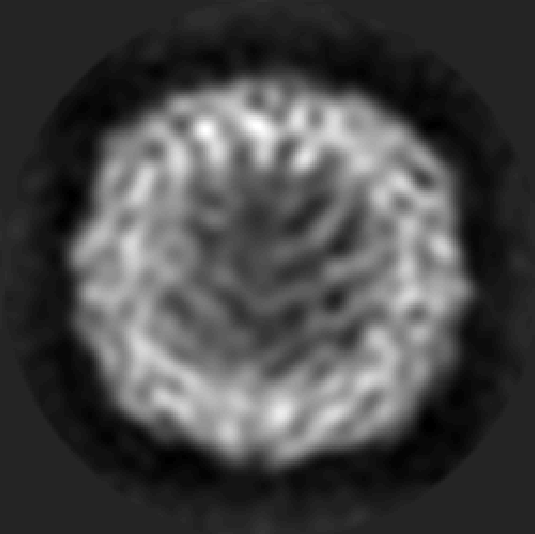}\\
\includegraphics[width=0.15\textwidth]{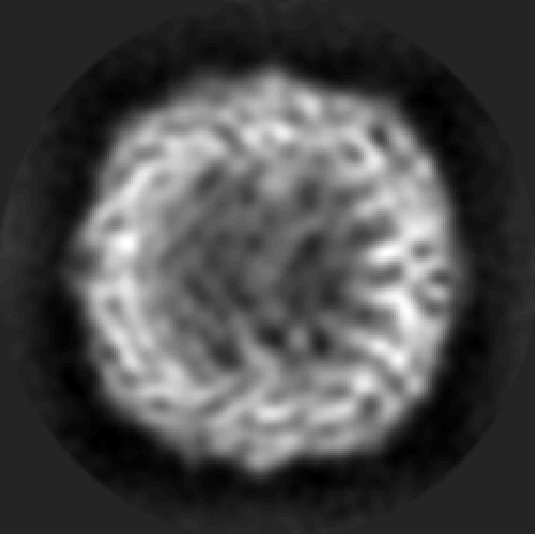}
\includegraphics[width=0.15\textwidth]{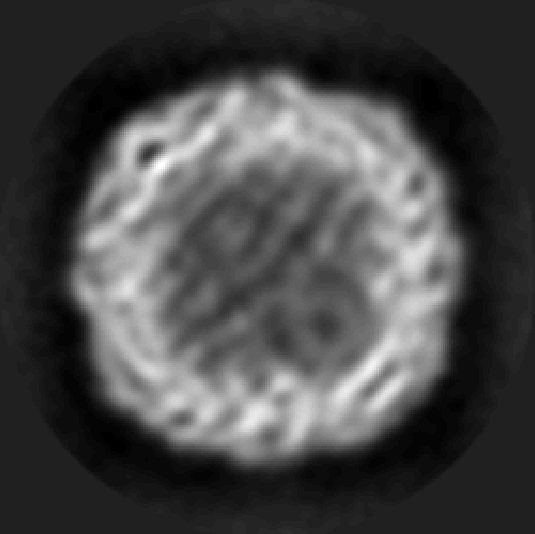}
\includegraphics[width=0.15\textwidth]{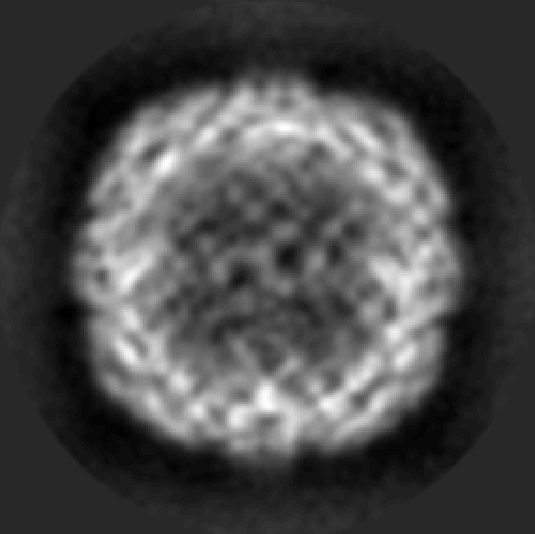}
\includegraphics[width=0.15\textwidth]{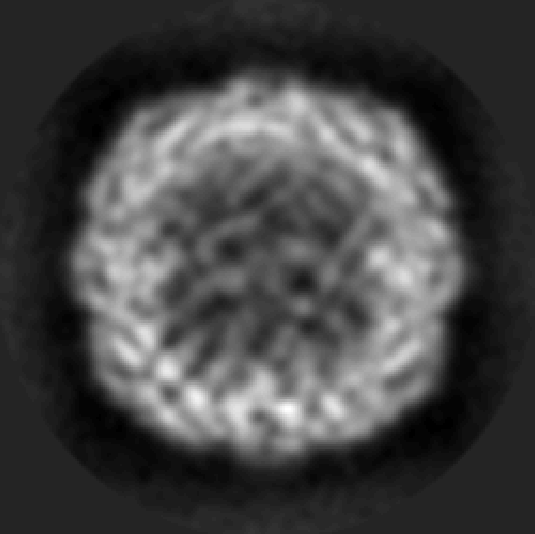}\\
\includegraphics[width=0.15\textwidth]{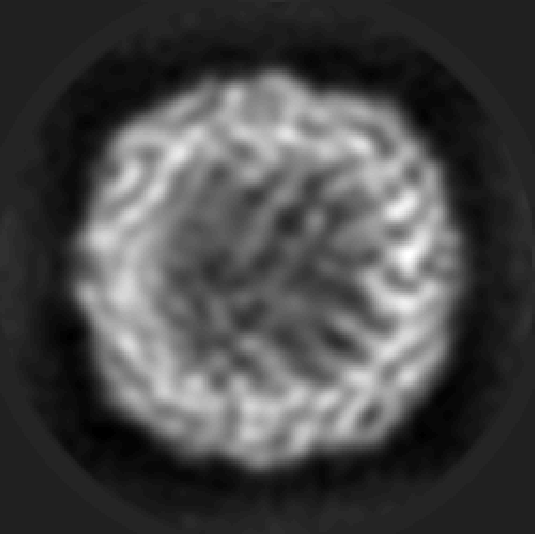}
\includegraphics[width=0.15\textwidth]{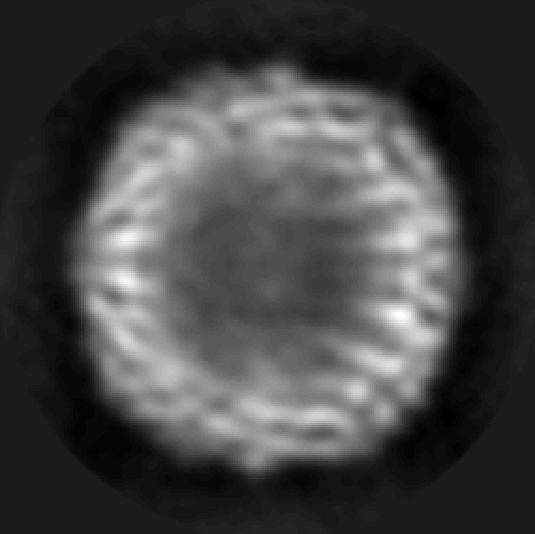}
\includegraphics[width=0.15\textwidth]{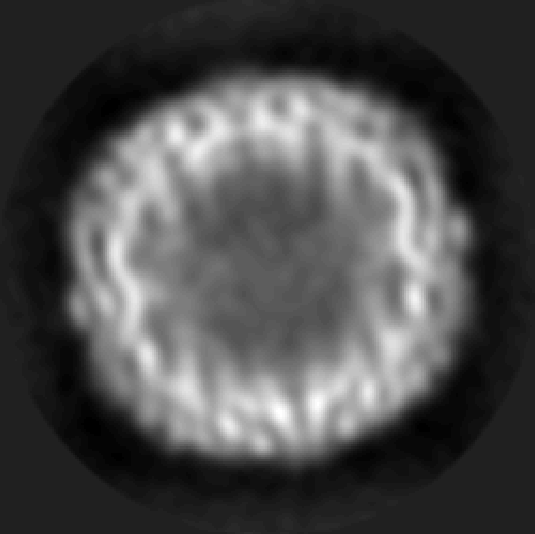}
\includegraphics[width=0.15\textwidth]{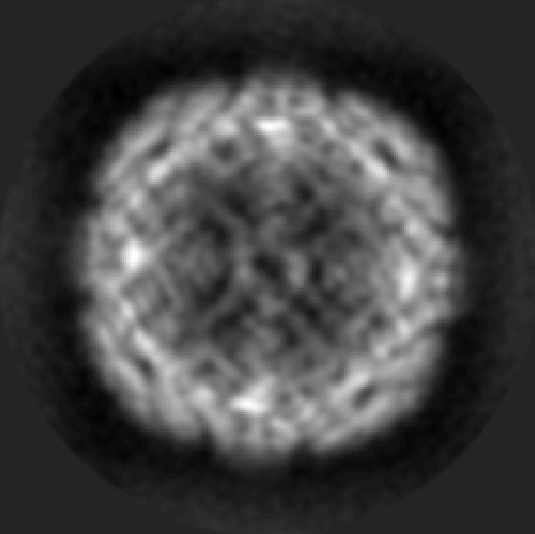}\\
\includegraphics[width=0.15\textwidth]{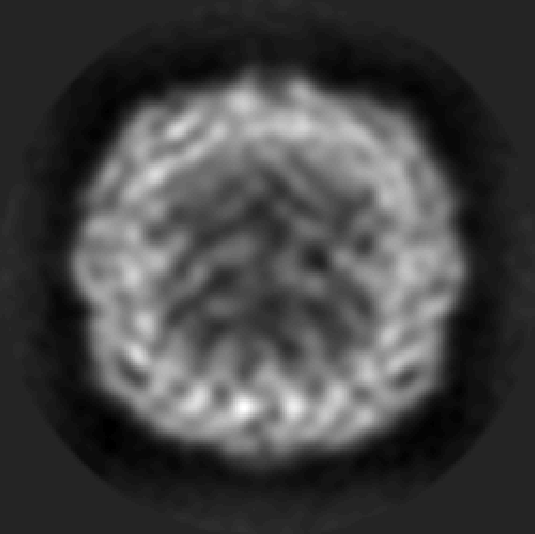}
\end{center}
\caption{Class averages used to generate an ab-initio model for EMPIAR-10272.}
\label{fig:classes102727}
\end{figure}

\begin{figure}
	\begin{center}
		\includegraphics[width=0.45\textwidth]{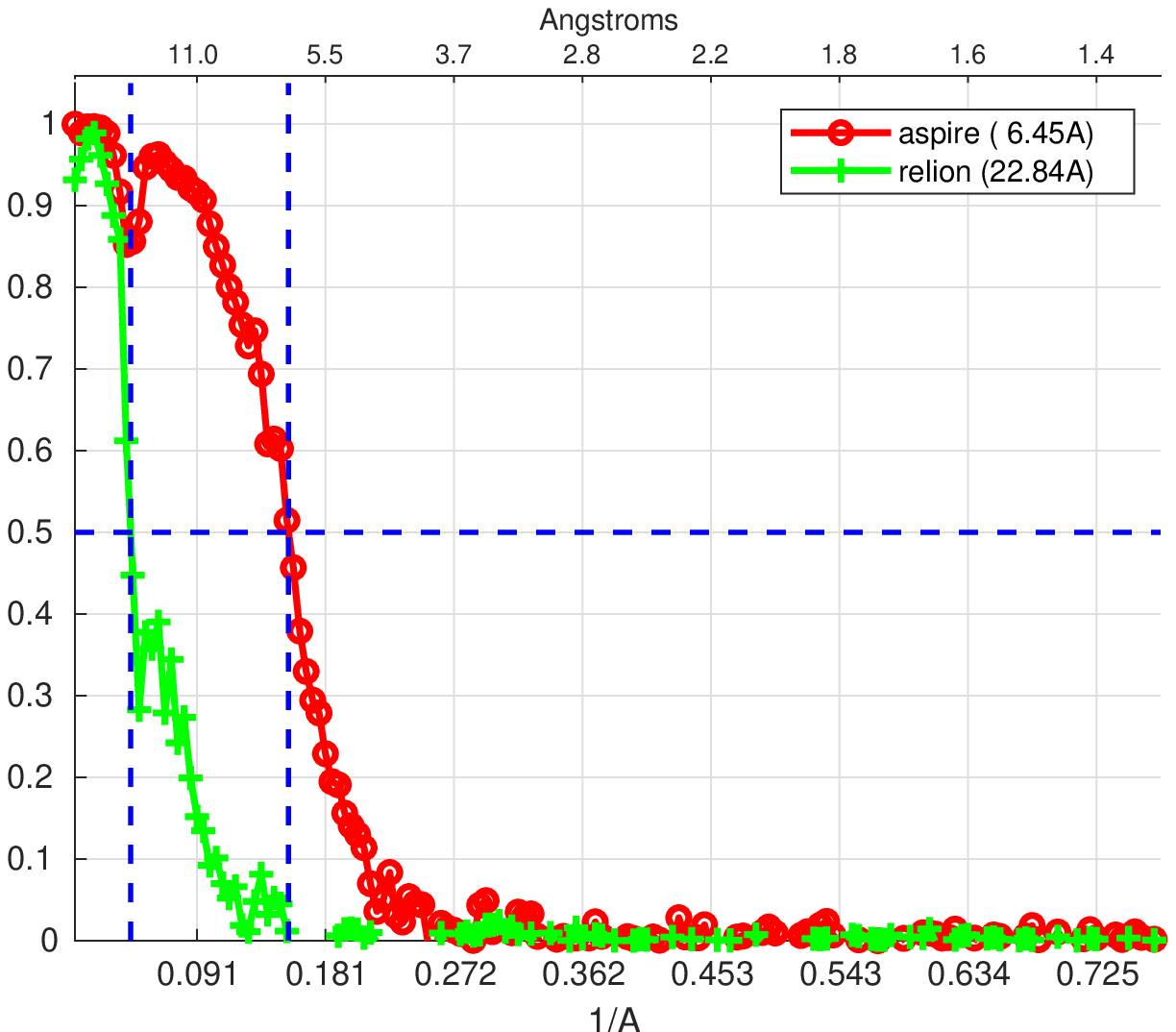}
	\end{center}
	\caption{Fourier shell correlation curves for ab-initio models for EMPIAR-10272.}
	\label{fig:abinitio10272_fsc}
\end{figure}

\begin{figure}
	\begin{center}
		\subfloat[Reference]{
			\includegraphics[width=0.3\textwidth]{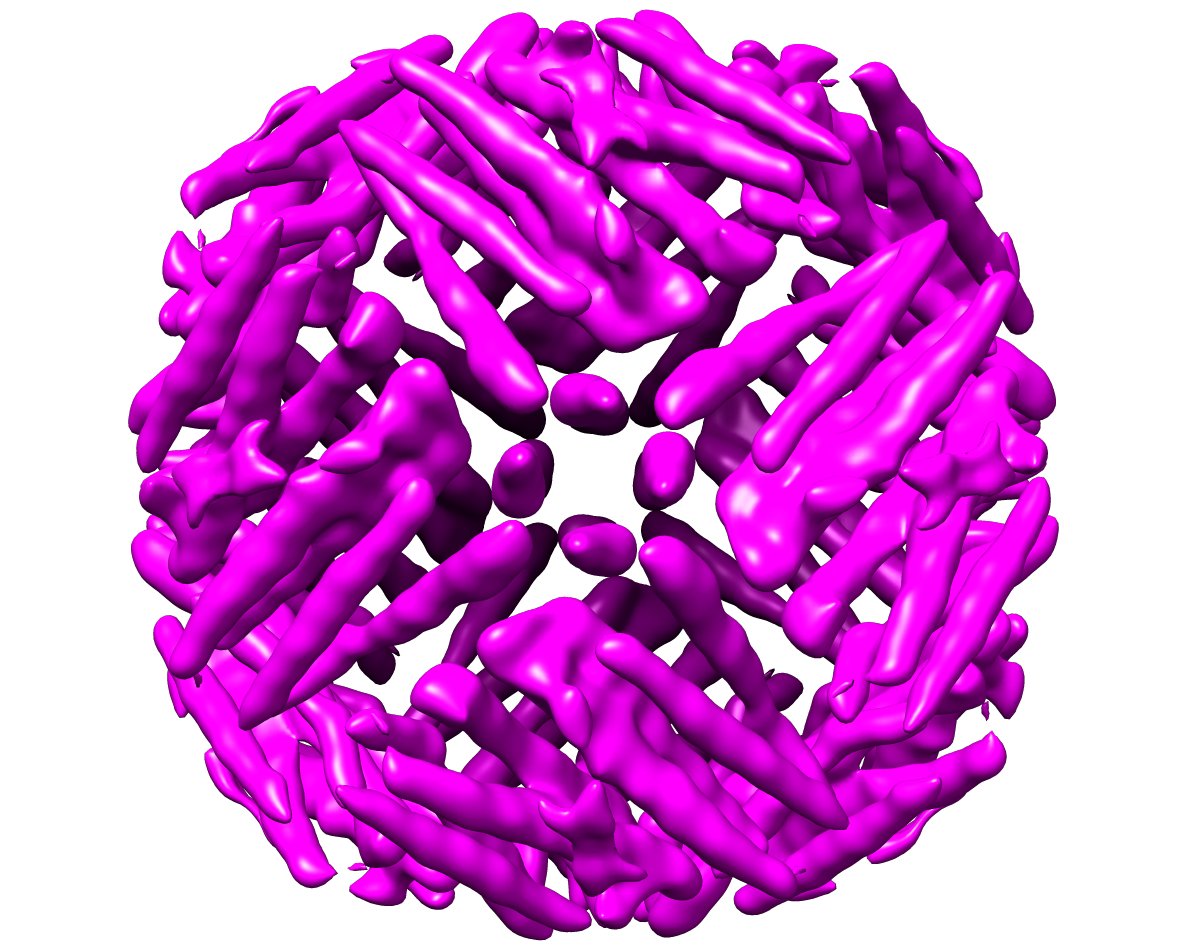} \label{fig:10272_ref}}
		\hfill
		\subfloat[ASPIRE]{
			\includegraphics[width=0.3\textwidth]{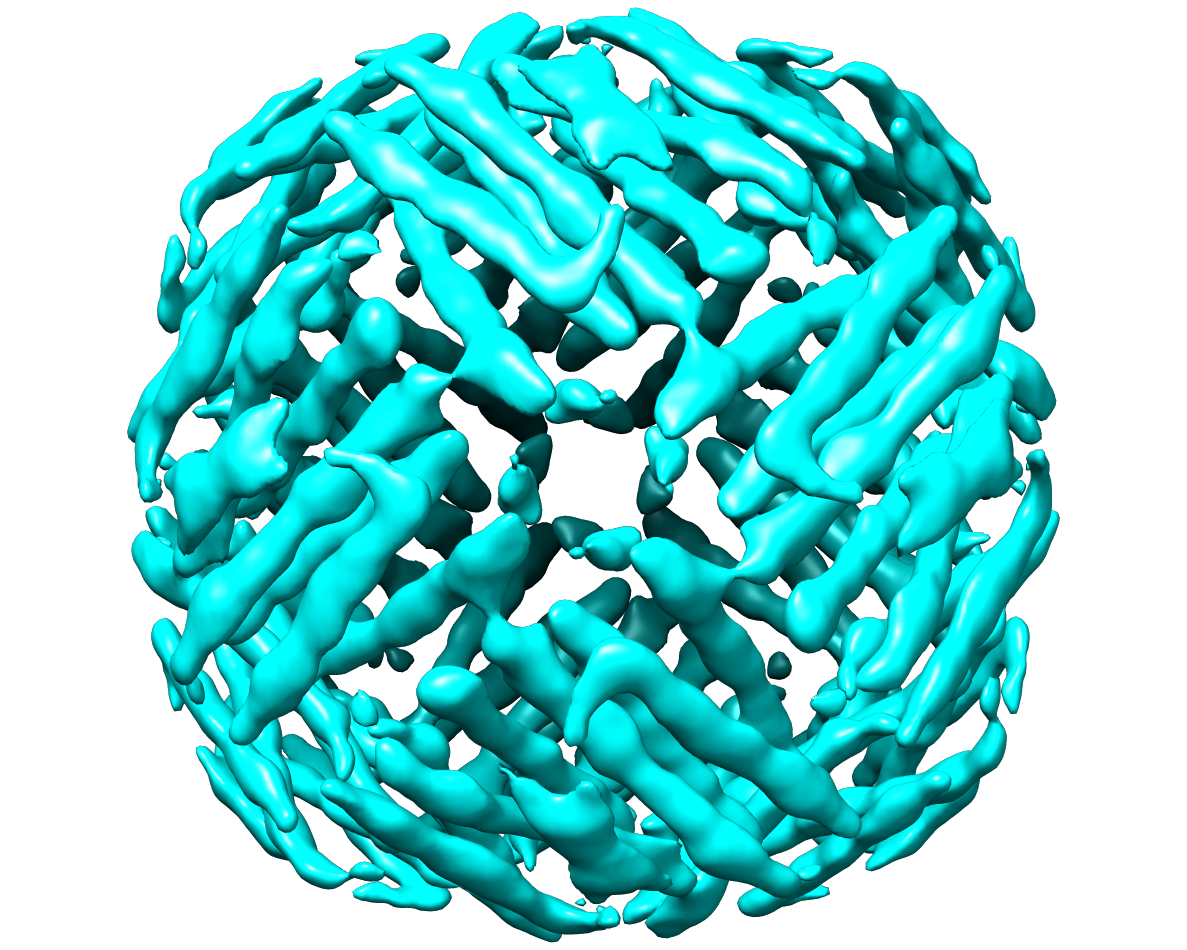} \label{fig:10272_aspire}}
		\hfill
		\subfloat[Relion]{
		\includegraphics[width=0.3\textwidth]{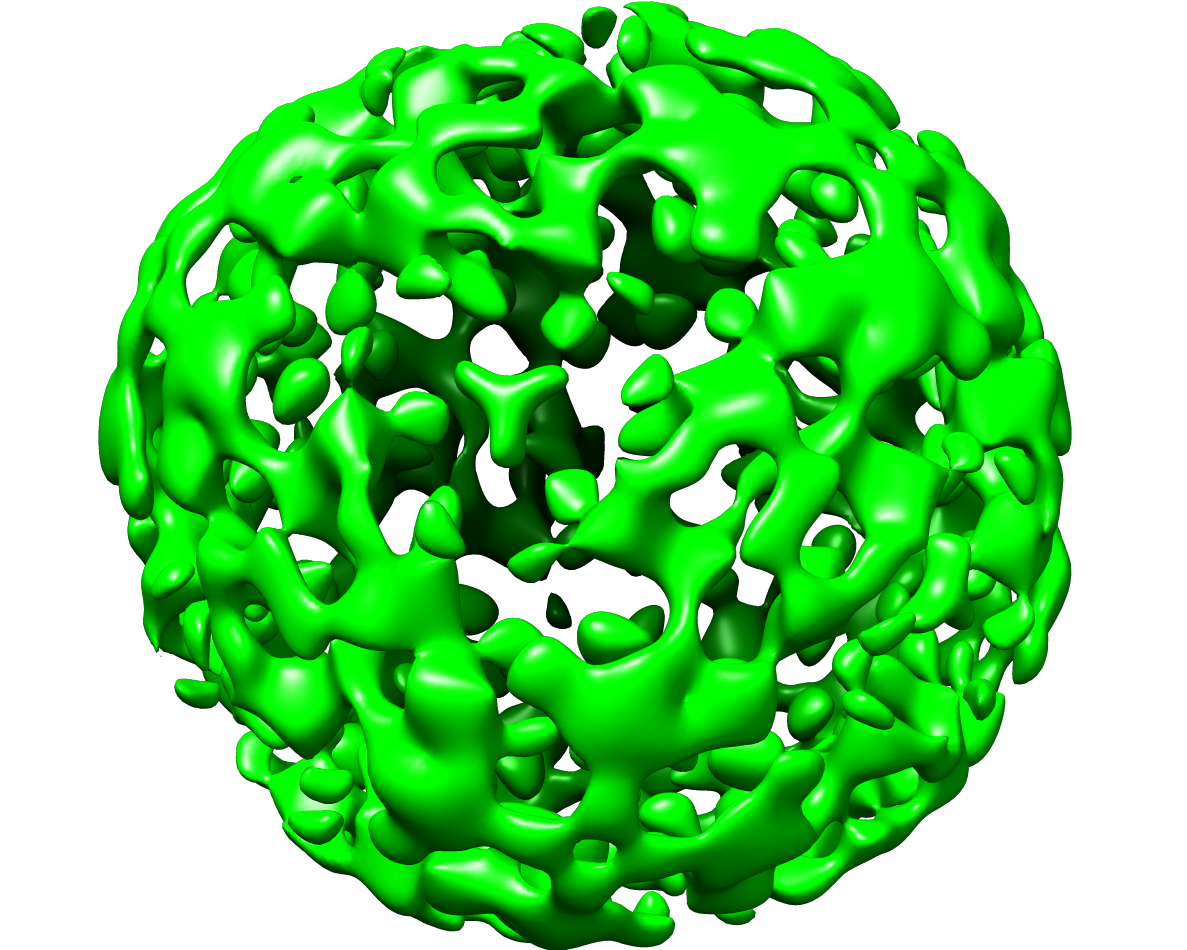} \label{fig:10272_relion}}
		\hfill
	\end{center}
\caption{Ab-initio models for EMPIAR-10272.}
\label{fig:abinitio10272}
\end{figure}

The EMPIAR-10272 data set consists of 480 micrographs, each comprised of 38 raw unaligned movie frames, with pixel size of 0.65~\AA/pixel. We first applied motion correction to the movie frames using MotionCor2~\cite{Zheng2017}, resulting in aligned micrographs, to which we applied CTF estimation~\cite{Frank2006} using CTFFind4~\cite{ROHOU2015216}. All subsequent processing steps were executed in Relion~\cite{relion3}. We  used Laplacian auto-picking followed by one round of 2D classification to generate templates for template-based picking. Auto-picking resulted in 80,806 particles, which were subjected to~15~rounds of 2D classification, until 24,540 particles in 13 classes were retained. These 13 classes (Fig.~\ref{fig:classes102727}) were the input to our algorithm, and resulted in an ab-initio model whose resolution is 6.45~\AA~(compared to the ground-truth density map EMD-4905~\cite{EMD4905}). For comparison, we generated a three-dimensional ab-initio model using Relion, using as an input the same particles that were used to generate the class averages for our algorithm. The resolution of the model estimated by Relion 22.84~\AA~(also compared to the ground-truth density map EMD-4905~\cite{EMD4905}). The Fourier shell correlation curves~\cite{vanHeel_Schatz} for the initial models generated by our algorithm and by Relion are shown in Fig.~\ref{fig:abinitio10272_fsc}. To assess visually the two models, we show in Fig.~\ref{fig:abinitio10272} a two-dimensional view of the ground-truth volume, the volume reconstructed by our algorithm (denoted ASPIRE in the figure), and the volume reconstructed by Relion. It can be observed that for this data set, the initial model generated by Relion is clearly inferior to the one generated by our algorithm.

%Figure~\ref{fig:abinitio10272_vol} shows a 3D rendering of the ab-initio model and Figure~\ref{fig:abinitio10272_fsc} shows its FSC curve, computed against the ground truth volume EMD-4905~\cite{EMD4905}. As a further verification of our ab-initio model, we used it for high resolution refinement as follows. We applied to the 24,540 particles one round of 3D~classification with  three classes using our ab-initio model as a reference, and retained the particles in two of the three classes, which together consist of 17,418 particles. We then ran 3D refinement, followed by CTF refinement, and by another round of 3D refinement, resulting in a final resolution of 2.33~\AA. 

% Filter volref_10272.mrc to 2.4
% Colors: REF (1,0,1) ASPIRE (0,1,1) RELION (0,1,0)

\begin{figure}
	\begin{center}
		\includegraphics[width=0.15\textwidth]{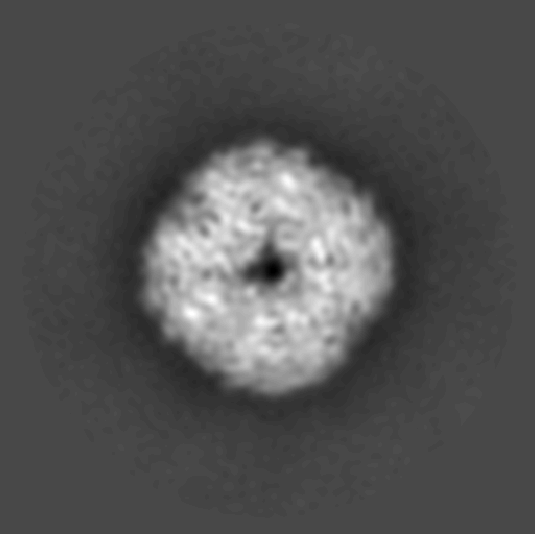}
		\includegraphics[width=0.15\textwidth]{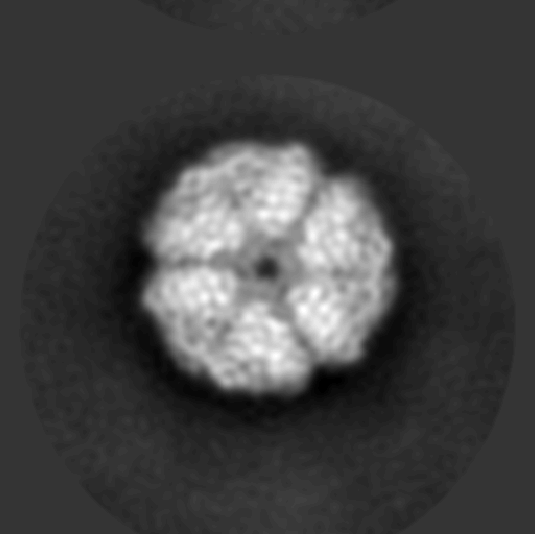}
		\includegraphics[width=0.15\textwidth]{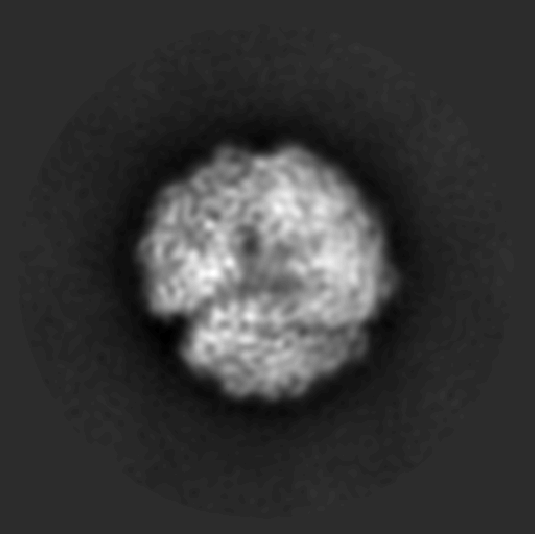}
		\includegraphics[width=0.15\textwidth]{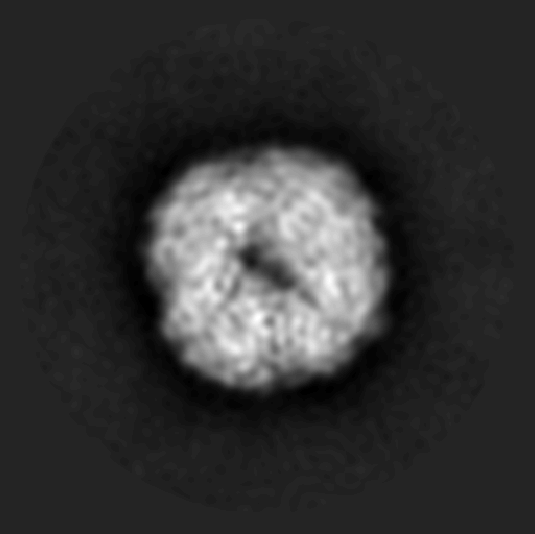}\\
		\includegraphics[width=0.15\textwidth]{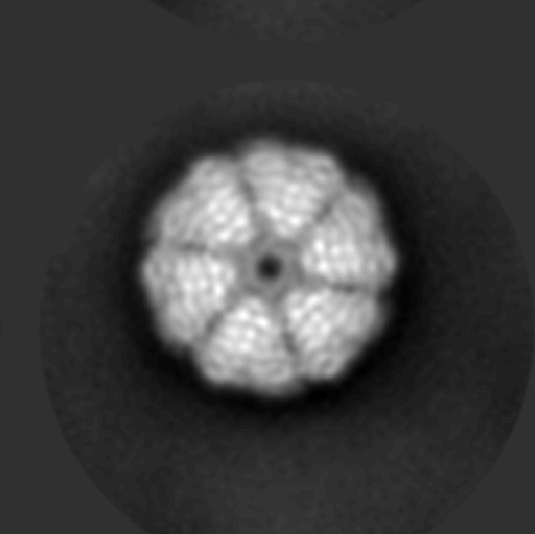}
		\includegraphics[width=0.15\textwidth]{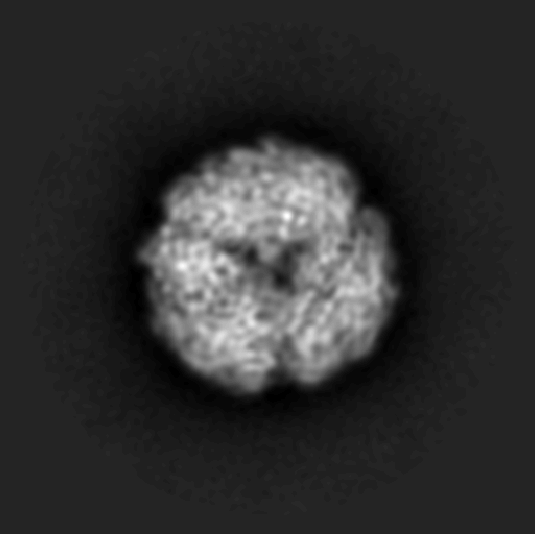}
		\includegraphics[width=0.15\textwidth]{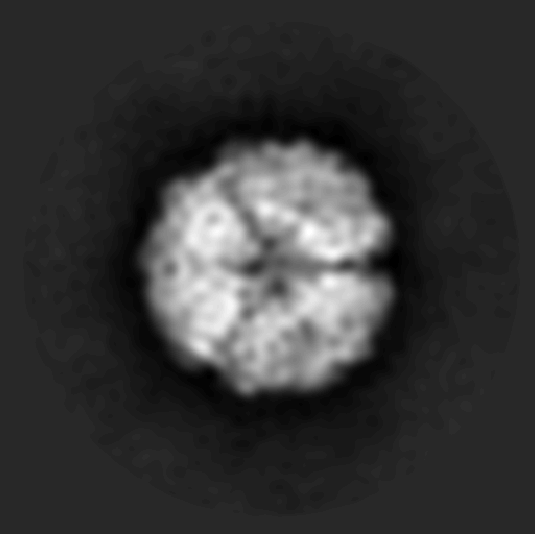}
		\includegraphics[width=0.15\textwidth]{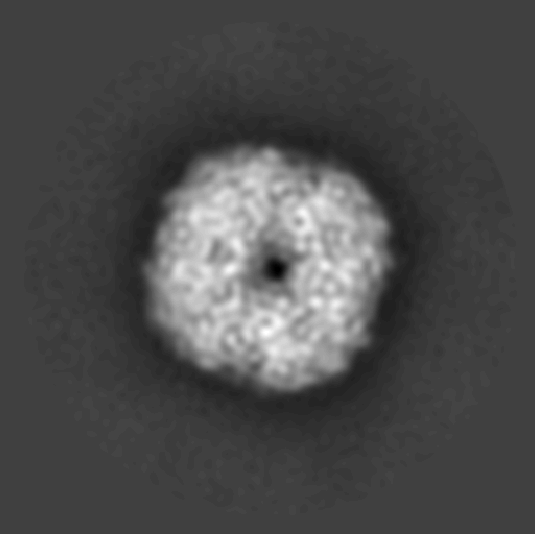}\\
		\includegraphics[width=0.15\textwidth]{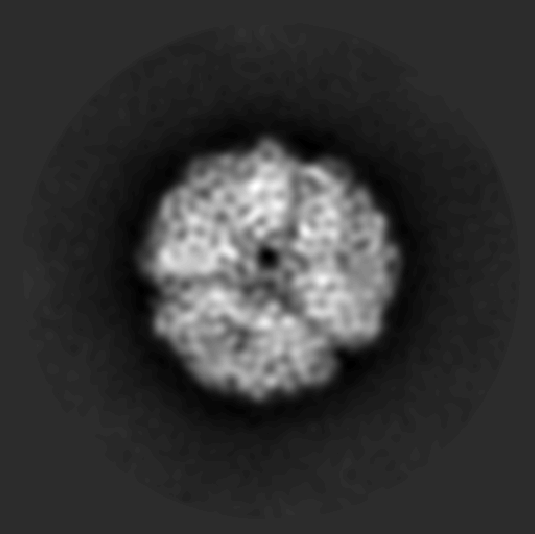}
		\includegraphics[width=0.15\textwidth]{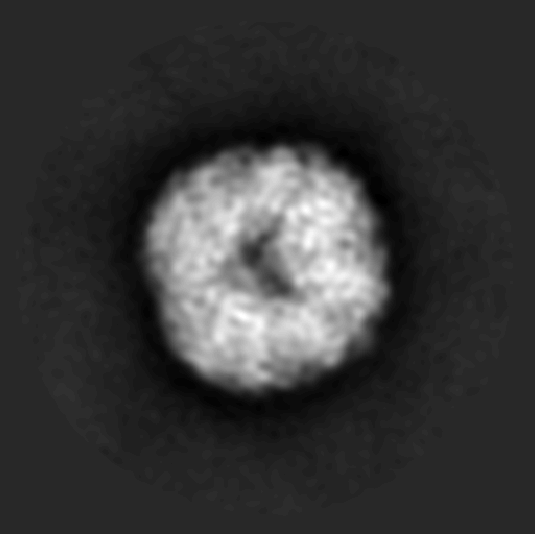}
		\includegraphics[width=0.15\textwidth]{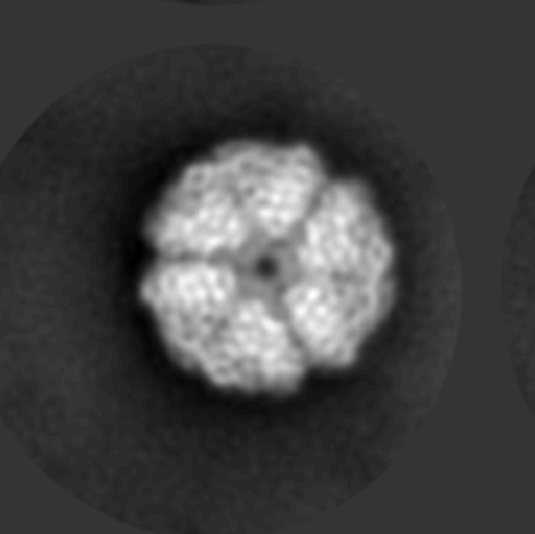}
		\includegraphics[width=0.15\textwidth]{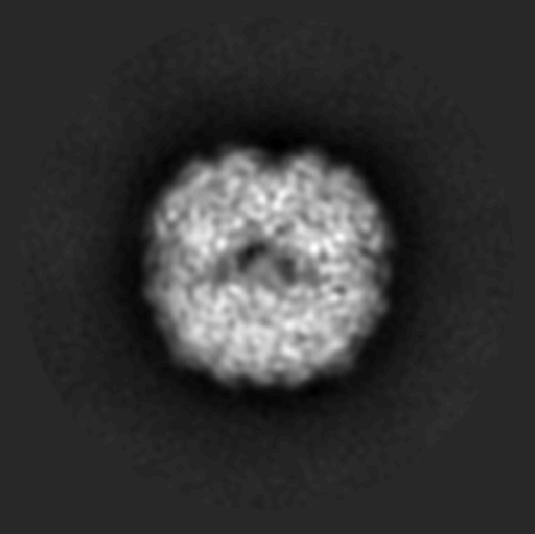}\\
		\includegraphics[width=0.15\textwidth]{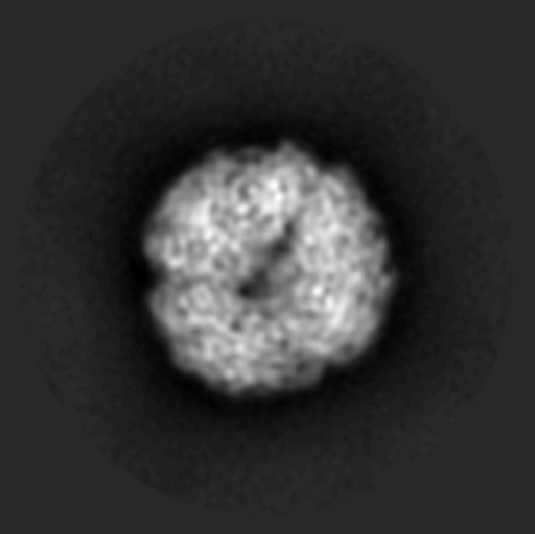}
		\includegraphics[width=0.15\textwidth]{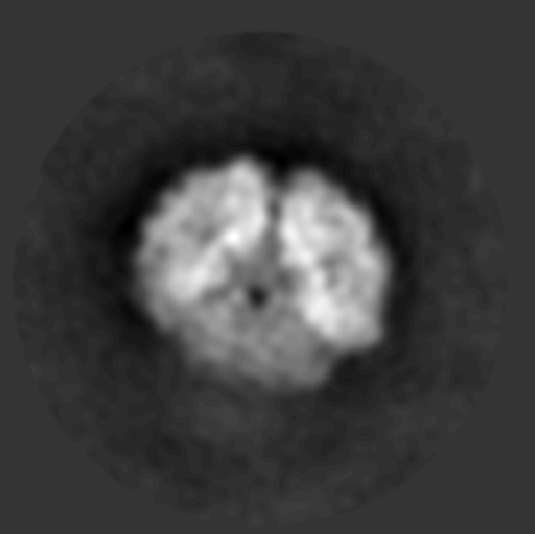}
		\includegraphics[width=0.15\textwidth]{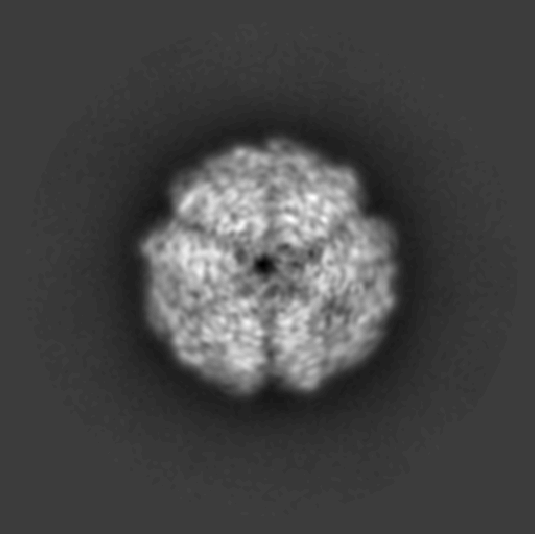}
		\includegraphics[width=0.15\textwidth]{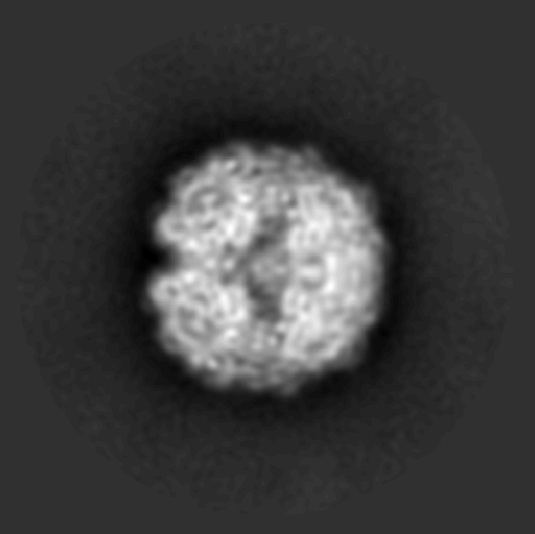}
	\end{center}
	\caption{16 of the class averages used to generate an ab-initio model for EMPIAR-10389.}
	\label{fig:classes10389}
\end{figure}

\begin{figure}
	\begin{center}
		\includegraphics[width=0.45\textwidth]{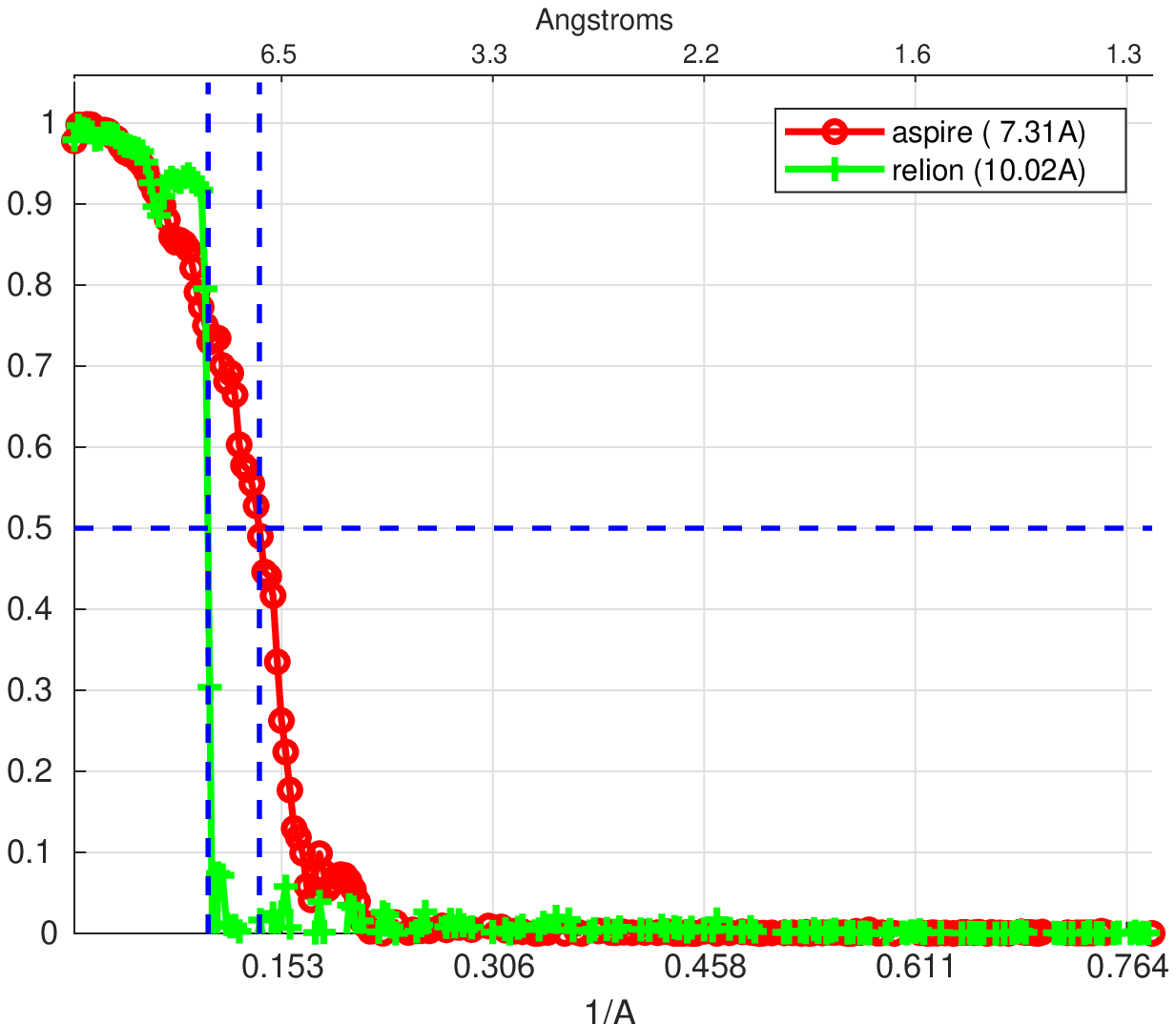}
	\end{center}
	\caption{Fourier shell correlation curves for ab-initio models for EMPIAR-10389.}
	\label{fig:abinitio10389_fsc}
\end{figure}

\begin{figure}
	\begin{center}
		\subfloat[Reference]{
			\includegraphics[width=0.3\textwidth]{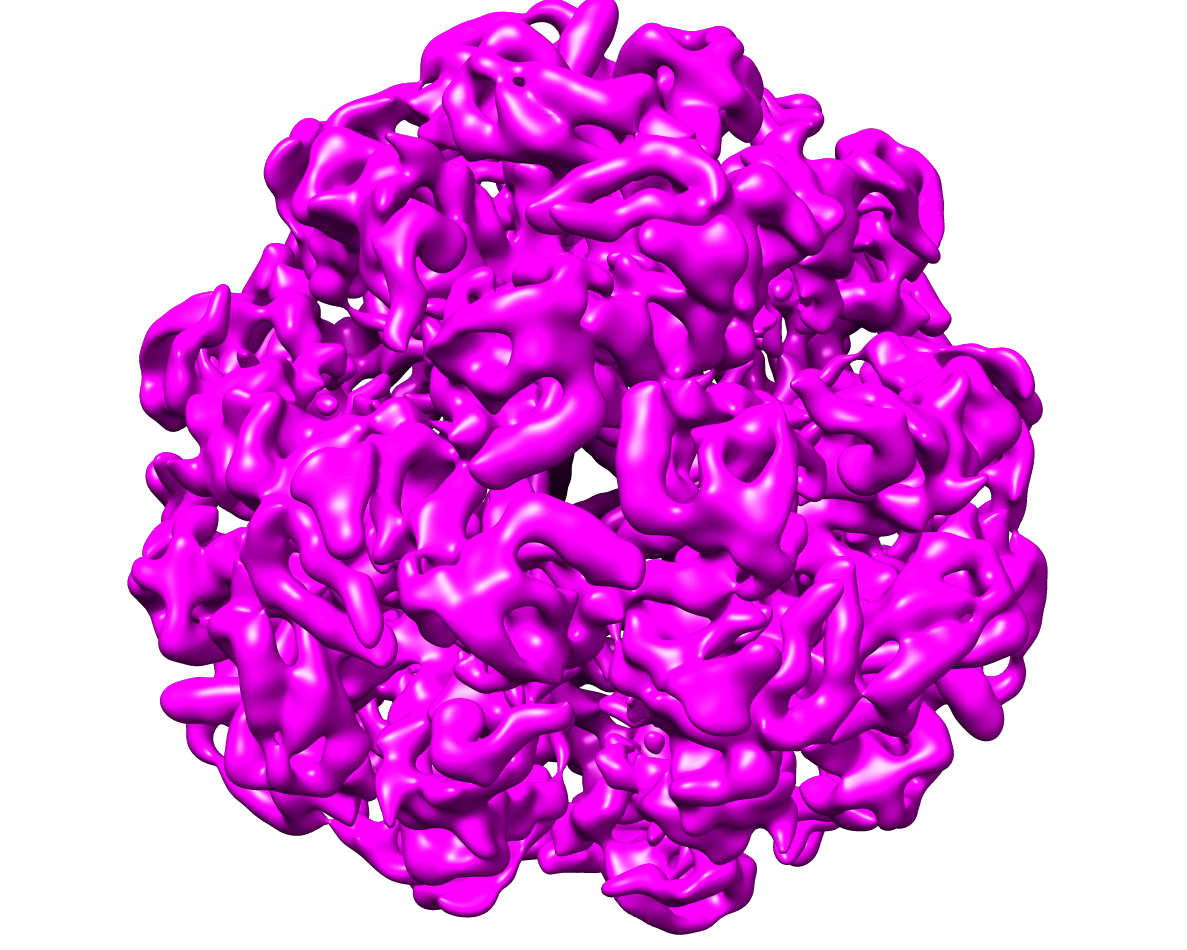} \label{fig:10389_ref}}
		\hfill
		\subfloat[ASPIRE]{
			\includegraphics[width=0.3\textwidth]{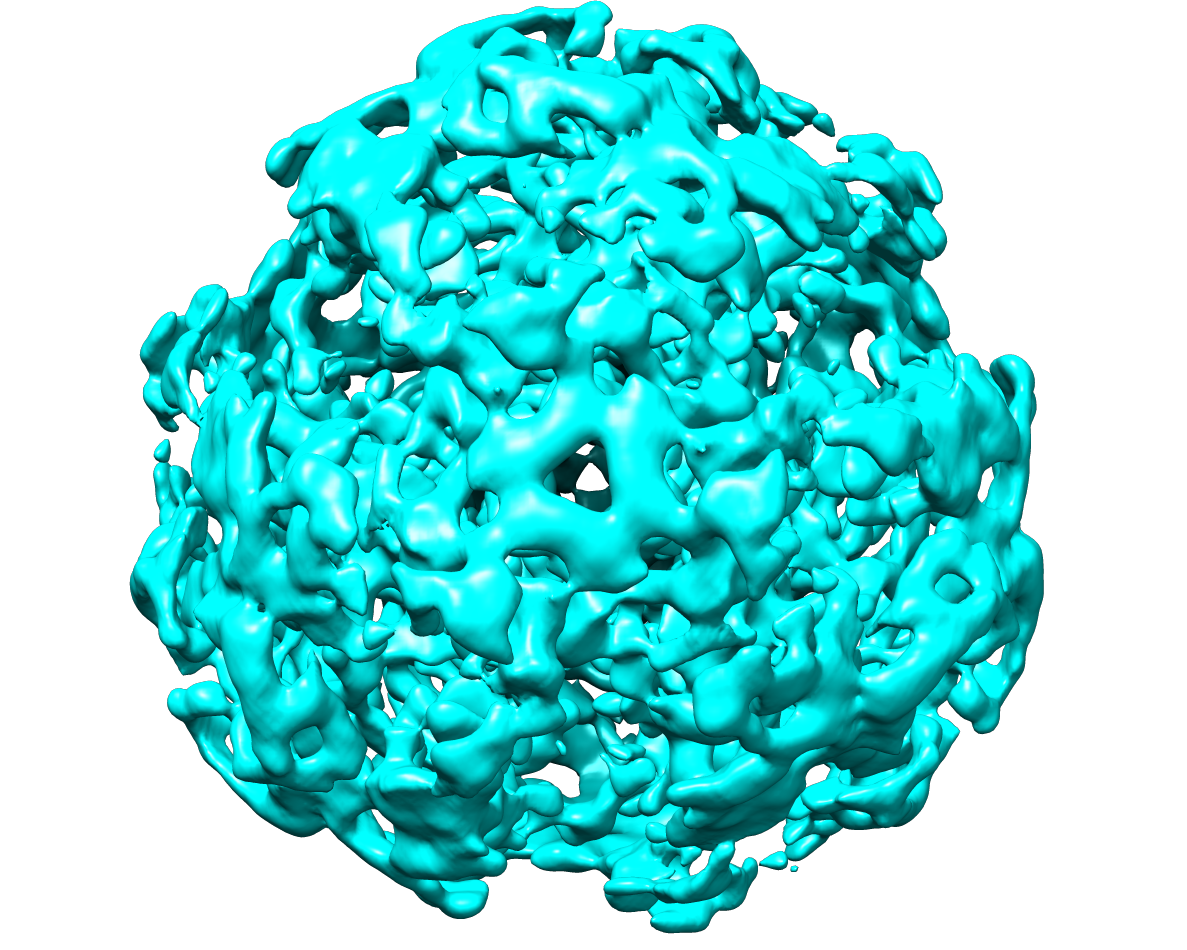} \label{fig:10389_aspire}}
		\hfill
		\subfloat[Relion]{
			\includegraphics[width=0.3\textwidth]{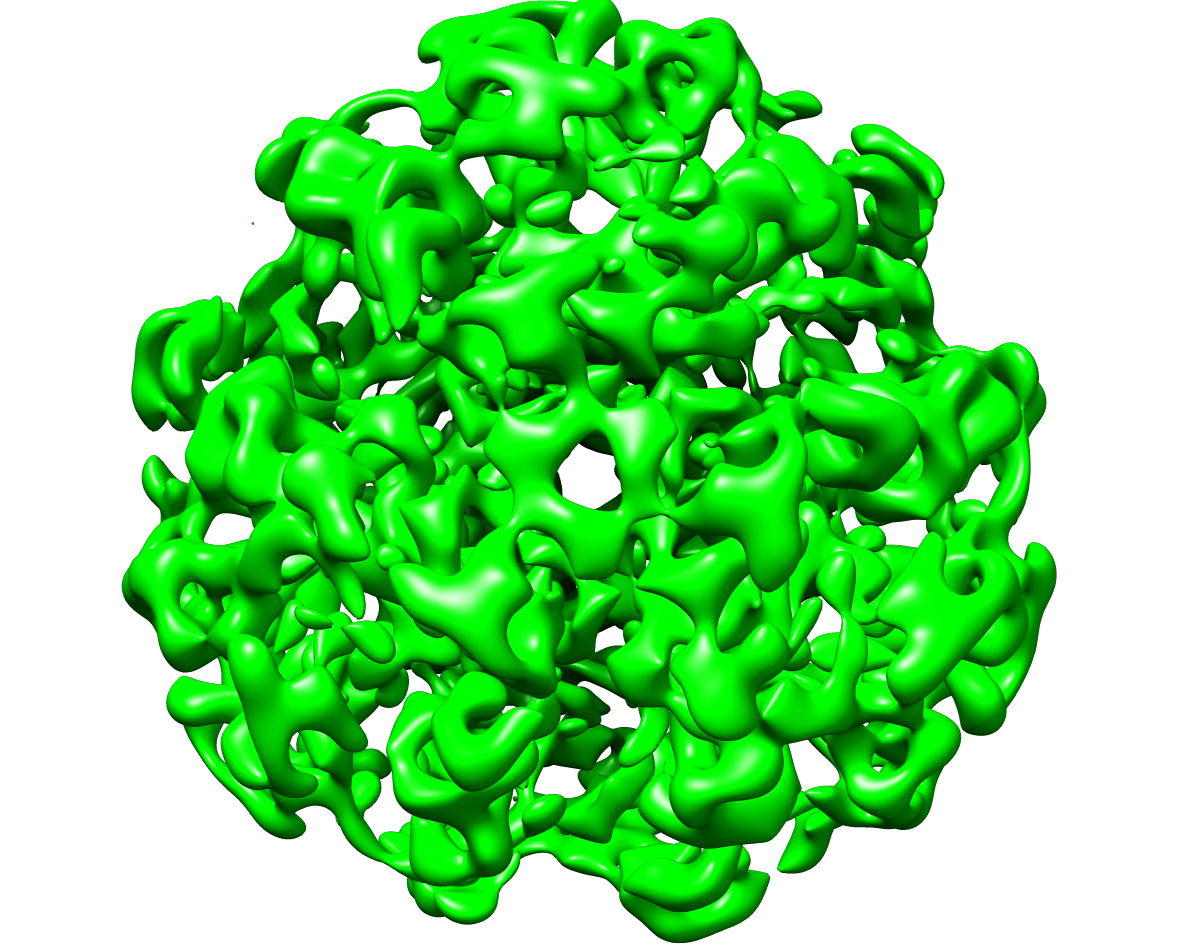} \label{fig:10389_relion}}
		\hfill
	\end{center}
	\caption{Ab-initio models for EMPIAR-10389.}
	\label{fig:abinitio10389}
\end{figure}

Next, we tested our algorithm on the EMPIAR-10389 data set, which has $\mathbb{T}$ symmetry. The EMPIAR-10389 data set consists of 4,313 dose-weighted micrographs with pixel size of 0.639~\AA/pixel. Automatic particle picking was done using the KLT picker~\cite{kltpicker}, resulting in 164,183 particles of size 512 $\times$ 512 pixels. (We note that for the EMPIAR-10272 data set discussed above, we used Relion's particle picker, as it gave superior results.) The particles were then imported into Relion~\cite{relion3}, and were subjected to 5 rounds of 2D classification, until 63,057 particles remained in 30 classes. These classes were used as the input to our algorithm, and are shown in Fig.~\ref{fig:classes10389}. 
The resolution of the resulting ab-initio model compared to the ground-truth density map EMD-10835~\cite{EMD10835} is  7.31~\AA. Using Relion's initial model algorithm on the particles of these 30 classes resulted in a resolution of 10~\AA~(also compared to the ground-truth density map EMD-10835). The Fourier shell correlation curves for the initial models generated by our algorithm and by Relion are shown in Fig.~\ref{fig:abinitio10389_fsc}. As before, we show in Fig.~\ref{fig:abinitio10389} two-dimensional views of the ground-truth volume EMD-10835, the volume reconstructed by our algorithm, and the volume reconstructed using Relion. 

%The resulting ab-initio model is shown in Figure~\ref{fig:abinitio10389_vol}. Its resolution, compared to the ground-truth density map EMD-10835~\cite{EMD10835} is 6.22~\AA, as shown in Figure~\ref{fig:abinitio10389_fsc}. This ab-initio model was used as the reference for 3D classification of the 63,057 particles  into four classes, out of which 58,640 particle were retained. High resolution refinement of this particle set resulted in a density map whose resolution is 2.5~\AA.

% Filter volref_10389.mrc to 2.4
% Colors: REF (1,0,1) ASPIRE (0,1,1) RELION (0,1,0)

% ===============================================================
% Chpter 9: Future work
% ===============================================================

\section{Future work}\label{Future work}
In this work, we proposed a method for estimating the orientations corresponding to a given set of projection-images of a molecule with tetrahedral or octahedral symmetry. The method relies on the observation that the elements of the tetrahedral and octahedral symmetry groups may be represented as rotation matrices with exactly one nonzero entry in each row and each column which is equal to either 1 or -1.

A future extension of this work would be to generalize it to molecules with icosahedral symmetry denoted by $\mathbb{I}$.
Since the elements of the icosahedral symmetry group cannot be represented as rotation matrices with exactly one nonzero entry in each row and each column which is equal to either 1 or -1, the method suggested in this work is not applicable to this symmetry.   

\section*{Acknowledgments}
This research was supported by the European Research Council (ERC) under the European Union's Horizon 2020 research and innovation programme (grant
agreement 723991 - CRYOMATH) and by the NIH/NIGMS Award R01GM136780-01.

% ===============================================================
% Appendices
% ===============================================================

\begin{appendices}
\section{Symmetry group elements}
\subsection{Tetrahedral group $\mathbb{T}$} \label{tab: T symmetry group elements representations}
%\begin{table}[H]
\begin{longtable}{ c c c c c c}
 element & matrix & axis & angle & single-entry sum & one-line notation \\ 
 $g^{(1)}$ & $
\begin{pmatrix}
1 & 0 & 0\\
0 & 1 & 0\\
0 & 0 & 1
\end{pmatrix}$ & any & 0 &  $e_{11}+e_{22}+e_{33}$ & $\sigma_1 = \begin{pmatrix}
1 & 2 & 3
\end{pmatrix}$   \\
$g^{(2)}$ & $
\begin{pmatrix}
0 & 0 & 1\\
1 & 0 & 0\\
0 & 1 & 0
\end{pmatrix}$ & [1,1,1] & $\nicefrac{2\pi}{3}$ &  $e_{13}+e_{21}+e_{32}$ & $\sigma_2 = \begin{pmatrix}
3 & 1 & 2
\end{pmatrix}$   \\
$g^{(3)}$ & $
\begin{pmatrix}
0 & 1 & 0\\
0 & 0 & 1\\
1 & 0 & 0
\end{pmatrix}$ & [1,1,1] & $\nicefrac{4\pi}{3}$ &  $e_{12}+e_{23}+e_{31}$ & $\sigma_3 = \begin{pmatrix}
2 & 3 & 1
\end{pmatrix}$   \\
$g^{(4)}$ & $
\begin{pmatrix}
0 & 0 & -1\\
1 & 0 & 0\\
0 & -1 & 0
\end{pmatrix}$ & [-1,-1,1] & $\nicefrac{2\pi}{3}$ &  $-e_{13}+e_{21}-e_{32}$ & $\sigma_4 = \begin{pmatrix}
-3 & 1 & -2
\end{pmatrix}$   \\
$g^{(5)}$ & $
\begin{pmatrix}
0 & 1 & 0\\
0 & 0 & -1\\
-1 & 0 & 0
\end{pmatrix}$ & [-1,-1,1] & $\nicefrac{4\pi}{3}$ &  $e_{12}-e_{23}-e_{31}$ & $\sigma_5 = \begin{pmatrix}
2 & -3 & -1
\end{pmatrix}$   \\
$g^{(6)}$ & $
\begin{pmatrix}
0 & 0 & -1\\
-1 & 0 & 0\\
0 & 1 & 0
\end{pmatrix}$ & [1,-1,-1] & $\nicefrac{2\pi}{3}$ &  $-e_{13}-e_{21}+e_{32}$ & $\sigma_6 = \begin{pmatrix}
-3 & -1 & 2
\end{pmatrix}$   \\
$g^{(7)}$ & $
\begin{pmatrix}
0 & -1 & 0\\
0 & 0 & 1\\
-1 & 0 & 0
\end{pmatrix}$ & [1,-1,-1] & $\nicefrac{4\pi}{3}$ &  $-e_{12}+e_{23}-e_{31}$ & $\sigma_7 = \begin{pmatrix}
-2 & 3 & -1
\end{pmatrix}$   \\
$g^{(8)}$ & $
\begin{pmatrix}
0 & 0 & 1\\
-1 & 0 & 0\\
0 & -1 & 0
\end{pmatrix}$ & [-1,1,-1] & $\nicefrac{2\pi}{3}$ &  $e_{13}-e_{21}-e_{32}$ & $\sigma_8 = \begin{pmatrix}
3 & -1 & -2
\end{pmatrix}$   \\
$g^{(9)}$ & $
\begin{pmatrix}
0 & -1 & 0\\
0 & 0 & -1\\
1 & 0 & 0
\end{pmatrix}$ & [-1,1,-1] & $\nicefrac{4\pi}{3}$ &  $-e_{12}-e_{23}+e_{31}$ & $\sigma_9 = \begin{pmatrix}
-2 & -3 & 1
\end{pmatrix}$   \\
$g^{(10)}$ & $
\begin{pmatrix}
1 & 0 & 0\\
0 & -1 & 0\\
0 & 0 & -1
\end{pmatrix}$ & [1,0,0] & $\pi$ &  $e_{11}-e_{22}-e_{33}$ & $\sigma_{10} = \begin{pmatrix}
1 & -2 & -3
\end{pmatrix}$   \\
$g^{(11)}$ & $
\begin{pmatrix}
-1 & 0 & 0\\
0 & 1 & 0\\
0 & 0 & -1
\end{pmatrix}$ & [0,1,0] & $\pi$ &  $-e_{11}+e_{22}-e_{33}$ & $\sigma_{11} = \begin{pmatrix}
-1 & 2 & -3
\end{pmatrix}$   \\
$g^{(12)}$ & $
\begin{pmatrix}
-1 & 0 & 0\\
0 & -1 & 0\\
0 & 0 & 1
\end{pmatrix}$ & [0,0,1] & $\pi$ &  $-e_{11}-e_{22}+e_{33}$ & $\sigma_{12} = \begin{pmatrix}
-1 & -2 & 3
\end{pmatrix}$
\\
\\
\\
\end{longtable}
%\end{table}
\subsection{Octahedral group $\mathbb{O}$} \label{tab: O symmetry group elements representations}
\begin{longtable}{ c c c c c c}
 element & matrix & axis & angle & single-entry sum & one-line notation \\
 $g^{(1)}$ & $
\begin{pmatrix}
1 & 0 & 0\\
0 & 1 & 0\\
0 & 0 & 1
\end{pmatrix}$ & any & 0 &  $e_{11}+e_{22}+e_{33}$ & $\sigma_{15} = \begin{pmatrix}
1 & 2 & 3
\end{pmatrix}$   \\
$g^{(2)}$ & $
\begin{pmatrix}
0 & -1 & 0\\
1 & 0 & 0\\
0 & 0 & 1
\end{pmatrix}$ & [0,0,1] & $\nicefrac{\pi}{2}$ &  $-e_{12}+e_{21}+e_{33}$ & $\sigma_{1} = \begin{pmatrix}
-2 & 1 & 3
\end{pmatrix}$   \\
$g^{(3)}$ & $
\begin{pmatrix}
0 & 1 & 0\\
-1 & 0 & 0\\
0 & 0 & 1
\end{pmatrix}$ & [0,0,1] & $\nicefrac{3\pi}{2}$ &  $e_{12}-e_{21}+e_{33}$ & $\sigma_{2} = \begin{pmatrix}
2 & -1 & 3
\end{pmatrix}$   \\
$g^{(4)}$ & $
\begin{pmatrix}
1 & 0 & 0\\
0 & 0 & -1\\
0 & 1 & 0
\end{pmatrix}$ & [1,0,0] & $\nicefrac{\pi}{2}$ &  $e_{11}-e_{23}+e_{32}$ & $\sigma_{3} = \begin{pmatrix}
1 & -3 & 2
\end{pmatrix}$   \\
$g^{(5)}$ & $
\begin{pmatrix}
1 & 0 & 0\\
0 & 0 & 1\\
0 & -1 & 0
\end{pmatrix}$ & [1,0,0] & $\nicefrac{3\pi}{2}$ &  $e_{11}+e_{23}-e_{32}$ & $\sigma_{4} = \begin{pmatrix}
1 & 3 & -2
\end{pmatrix}$   \\
$g^{(6)}$ & $
\begin{pmatrix}
0 & -1 & 0\\
0 & 0 & -1\\
1 & 0 & 0
\end{pmatrix}$ & [1,-1,1] & $\nicefrac{2\pi}{3}$ &  $-e_{12}-e_{23}+e_{31}$ & $\sigma_{5} = \begin{pmatrix}
-2 & -3 & 1
\end{pmatrix}$   \\
$g^{(7)}$ & $
\begin{pmatrix}
0 & 0 & 1\\
-1 & 0 & 0\\
0 & -1 & 0
\end{pmatrix}$ & [1,-1,1] & $\nicefrac{4\pi}{3}$ &  $e_{13}-e_{21}-e_{32}$ & $\sigma_{6} = \begin{pmatrix}
3 & -1 & -2
\end{pmatrix}$   \\
$g^{(8)}$ & $
\begin{pmatrix}
0 & -1 & 0\\
0 & 0 & 1\\
-1 & 0 & 0
\end{pmatrix}$ & [-1,1,1] & $\nicefrac{2\pi}{3}$ &  $-e_{12}+e_{23}-e_{31}$ & $\sigma_{7} = \begin{pmatrix}
-2 & 3 & -1
\end{pmatrix}$   \\
$g^{(9)}$ & $
\begin{pmatrix}
0 & 0 & -1\\
-1 & 0 & 0\\
0 & 1 & 0
\end{pmatrix}$ & [-1,1,1] & $\nicefrac{4\pi}{3}$ &  $-e_{13}-e_{21}+e_{32}$ & $\sigma_{8} = \begin{pmatrix}
-3 & -1 & 2
\end{pmatrix}$   \\
$g^{(10)}$ & $
\begin{pmatrix}
0 & 1 & 0\\
0 & 0 & 1\\
1 & 0 & 0
\end{pmatrix}$ & [-1,-1,-1] & $\nicefrac{2\pi}{3}$ &  $e_{12}+e_{23}+e_{31}$ & $\sigma_{9} = \begin{pmatrix}
2 & 3 & 1
\end{pmatrix}$   \\
$g^{(11)}$ & $
\begin{pmatrix}
0 & 0 & 1\\
1 & 0 & 0\\
0 & 1 & 0
\end{pmatrix}$ & [-1,-1,-1] & $\nicefrac{4\pi}{3}$ &  $e_{13}+e_{21}+e_{32}$ & $\sigma_{10} = \begin{pmatrix}
3 & 1 & 2
\end{pmatrix}$   \\
$g^{(12)}$ & $
\begin{pmatrix}
0 & 0 & 1\\
0 & 1 & 0\\
-1 & 0 & 0
\end{pmatrix}$ & [0,1,0] & $\nicefrac{\pi}{2}$ &  $e_{13}+e_{22}-e_{31}$ & $\sigma_{11} = \begin{pmatrix}
3 & 2 & -1
\end{pmatrix}$   \\
$g^{(13)}$ & $
\begin{pmatrix}
0 & 0 & -1\\
0 & 1 & 0\\
1 & 0 & 0
\end{pmatrix}$ & [0,1,0] & $\nicefrac{3\pi}{2}$ &  $-e_{13}+e_{22}+e_{31}$ & $\sigma_{12} = \begin{pmatrix}
-3 & 2 & 1
\end{pmatrix}$   \\
$g^{(14)}$ & $
\begin{pmatrix}
0 & 1 & 0\\
0 & 0 & -1\\
-1 & 0 & 0
\end{pmatrix}$ & [1,1,-1] & $\nicefrac{2\pi}{3}$ &  $e_{12}-e_{23}-e_{31}$ & $\sigma_{13} = \begin{pmatrix}
2 & -3 & -1
\end{pmatrix}$   \\
$g^{(15)}$ & $
\begin{pmatrix}
0 & 0 & -1\\
1 & 0 & 0\\
0 & -1 & 0
\end{pmatrix}$ & [1,1,-1] & $\nicefrac{4\pi}{3}$ &  $-e_{13}+e_{21}-e_{32}$ & $\sigma_{14} = \begin{pmatrix}
-3 & 1 & -2
\end{pmatrix}$   \\
$g^{(16)}$ & $
\begin{pmatrix}
-1 & 0 & 0\\
0 & -1 & 0\\
0 & 0 & 1
\end{pmatrix}$ & [0,0,1] & $\pi$ &  $-e_{11}-e_{22}+e_{33}$ & $\sigma_{16} = \begin{pmatrix}
-1 & -2 & 3
\end{pmatrix}$ \\
$g^{(17)}$ & $
\begin{pmatrix}
-1 & 0 & 0\\
0 & 0 & -1\\
0 & -1 & 0
\end{pmatrix}$ & [0,1,-1] & $\pi$ &  $-e_{11}-e_{23}-e_{32}$ & $\sigma_{17} = \begin{pmatrix}
-1 & -3 & -2
\end{pmatrix}$ \\
$g^{(18)}$ & $
\begin{pmatrix}
1 & 0 & 0\\
0 & -1 & 0\\
0 & 0 & -1
\end{pmatrix}$ & [1,0,0] & $\pi$ &  $e_{11}-e_{22}-e_{33}$ & $\sigma_{18} = \begin{pmatrix}
1 & -2 & -3
\end{pmatrix}$   \\
$g^{(19)}$ & $
\begin{pmatrix}
0 & -1 & 0\\
-1 & 0 & 0\\
0 & 0 & -1
\end{pmatrix}$ & [1,-1,0] & $\pi$ &  $-e_{12}-e_{21}-e_{33}$ & $\sigma_{19} = \begin{pmatrix}
-2 & -1 & -3
\end{pmatrix}$ \\
$g^{(20)}$ & $
\begin{pmatrix}
-1 & 0 & 0\\
0 & 1 & 0\\
0 & 0 & -1
\end{pmatrix}$ & [0,1,0] & $\pi$ &  $-e_{11}+e_{22}-e_{33}$ & $\sigma_{20} = \begin{pmatrix}
-1 & 2 & -3
\end{pmatrix}$   \\
$g^{(21)}$ & $
\begin{pmatrix}
0 & 1 & 0\\
1 & 0 & 0\\
0 & 0 & -1
\end{pmatrix}$ & [1,1,0] & $\pi$ &  $e_{12}+e_{21}-e_{33}$ & $\sigma_{21} = \begin{pmatrix}
2 & 1 & -3
\end{pmatrix}$ \\
$g^{(22)}$ & $
\begin{pmatrix}
-1 & 0 & 0\\
0 & 0 & 1\\
0 & 1 & 0
\end{pmatrix}$ & [0,1,1] & $\pi$ &  $-e_{11}+e_{23}+e_{32}$ & $\sigma_{22} = \begin{pmatrix}
-1 & 3 & 2
\end{pmatrix}$ \\
$g^{(23)}$ & $
\begin{pmatrix}
0 & 0 & 1\\
0 & -1 & 0\\
1 & 0 & 0
\end{pmatrix}$ & [1,0,1] & $\pi$ &  $e_{13}-e_{22}+e_{31}$ & $\sigma_{23} = \begin{pmatrix}
3 & -2 & 1
\end{pmatrix}$ \\
$g^{(24)}$ & $
\begin{pmatrix}
0 & 0 & -1\\
0 & -1 & 0\\
-1 & 0 & 0
\end{pmatrix}$ & [1,0,-1] & $\pi$ &  $-e_{13}-e_{22}-e_{31}$ & $\sigma_{24} = \begin{pmatrix}
-3 & -2 & -1
\end{pmatrix}$ 
\\
\\
\\
\end{longtable}

\section{Proof of Lemma~\ref{lema:group properties}} \label{Proof of Lemma sym prop}

\begin{proof}
First, note that for the matrices $e_{ij}$ and $e_{kl}$ defined in Definition~\ref{def:single entry matrix}, it holds that

\begin{equation} \label{eq: single entry mul}
 e_{ij}e_{kl} = \left \{
\begin{array}{rl}
  \ e_{il} & \text{if } k=j,\\ 
  -e_{il} & \text{if } k=-j,\\ 
  0_{3\times3} & \text{else. }  
\end{array}
\right .
\end{equation}
In addition, for any single entry matrix defined in Definition~\ref{def:single entry matrix}, it follows by a direct calculation that
$ e_{ij}^T = e_{ji}$. By expressing $g_1$ and $g_2$ using Lemma~\ref{lema:group rep}, we have
$$g_1 = e_{1\sigma_1(1)}+ e_{2\sigma_1(2)}+ e_{3\sigma_1(3)}, \quad 
g_2 = e_{1\sigma_2(1)}+ e_{2\sigma_2(2)}+ e_{3\sigma_2(3)}. $$
For~\eqref{eq:property 1}, we have that for $m=1,2,3$
\[
\begin{split}
g_1^Te_{mm}g_2 &= (e_{1\sigma_1(1)}+ e_{2\sigma_1(2)}+ e_{3\sigma_1(3)})^Te_{mm}
(e_{1\sigma_2(1)}+ e_{2\sigma_2(2)}+ e_{3\sigma_2(3)})  \\ & 
= (e_{\sigma_1(1)1}+ e_{\sigma_1(2)2}+ e_{\sigma_1(3)3})e_{mm}
(e_{1\sigma_2(1)}+ e_{2\sigma_2(2)}+ e_{3\sigma_2(3)}) \\ & 
=e_{\sigma_1(m)\sigma_2(m)},
\end{split}
\]
where the last equality follows by~\eqref{eq: single entry mul}.
For~\eqref{eq:property 2}, we use \eqref{eq:property 1} and obtain
$$
g_1^Te_{mm}g_1 = 
e_{\sigma_1(m)\sigma_1(m)}.
$$
For~\eqref{eq:property 3},
\[
\begin{split}
\{g_1^Te_{mr}g_2\}_{m,r=1}^3 &= 
\{
(e_{1\sigma_1(1)}+ e_{2\sigma_1(2)}+ e_{3\sigma_1(3)})^T
e_{mr} 
(e_{1\sigma_2(1)}+ e_{2\sigma_2(2)}+ e_{3\sigma_2(3)})
\}_{m,r=1}^3  \\ & 
=
\{
(e_{\sigma_1(1)1}+ e_{\sigma_1(2)2}+ e_{\sigma_1(3)3})
e_{mr}
(e_{1\sigma_2(1)}+ e_{2\sigma_2(2)}+ e_{3\sigma_2(3)})
\}_{m,r=1}^3 \\ & 
= \{e_{\sigma_1(m)\sigma_2(r)}\}_{m,r=1}^3,
\end{split} 
\]
where the last equality follows by~\eqref{eq: single entry mul}, and thus
$$
\{\pm g_1^Te_{mr}g_2\}_{m,r=1}^3   
= \{\pm e_{\sigma_1(m)\sigma_2(r)}\}_{m,r=1}^3 
= \{\pm e_{mr}\}_{m,r=1}^3.
$$
\end{proof}

\section{Constructing $SO_G(3)$} \label{Constructing $SO_G(3)$}

We denote by $SO_G(3) \subset SO(3)$ the finite subset of rotations for the symmetry group $G$ on which we search for the optimum of the score function $\pi_{ij}$ of~\eqref{eq:score cl scl}.
A naive choice for $SO_G(3)$ would be an almost equally spaced grid of rotations from $SO(3)$, denoted as $\widetilde{SO}(3)$ and defined below. However, the symmetry of $G$ allows us to significantly reduce the number of rotations in this naive set while maintaining the same accuracy of our algorithm.
Note that for any $Q_r, Q_s \in SO(3)$ and $g \in G$,
it holds that
$\{ Q_r^T g^{(k)}{Q_s}\}_{k=1}^n = \{ Q_r^T g^{(k)}{g Q_s}\}_{k=1}^n$,
and so the set of local coordinates
$\{(\alpha^{k,1}_{Q_r,Q_s}, \alpha _{Q_r,Q_s}^{k,2})\}_{k \in [n]}$
is equal to the set of local coordinates
$\{(\alpha^{k,1}_{Q_r,g Q_s}, \alpha _{Q_r,g Q_s}^{k,2})\}_{k \in [n]}$.
Thus, 
keeping both $Q_s$ and $g Q_s$ in $SO_{G}(3)$ is redundant. 
% It is unlikely that for each gQs we will find Ql out of a finite set such that gQs=Ql. I think that the index r is confusing.
Consequently, our objective is to find all pairs of rotations $Q_s, Q_l \in \widetilde{SO}(3)$ for which there exists $g^{(k)} \in G \setminus I$ such that $Q_s = g^{(k)}Q_l$, and filter either $Q_s$ or $Q_l$ from $\widetilde{SO}(3)$.
The resulting set would be $SO_G(3)$.

Since $\widetilde{SO}(3)$ is finite, an exact equality between $Q_s$ and $g^{(k)}Q_l$ is unlikely.
Therefore, the proximity between $Q_s$ and $g^{(k)} Q_l$ is determined up to pre-defined thresholds, based on their representation using viewing direction and in-plane rotation (see~\cite{viewang}) as follows. 
The viewing directions of $Q_s$ and $g^{(k)}Q_l$ are given by their third columns ${Q_s^3}$ and $(g^{(k)}Q_l)^3$, respectively. If $Q_s$ and $g^{(k)}Q_l$ are two rotations with the same viewing direction, i.e.,
$\langle {Q_s^3}, (g^{(k)}Q_l)^3 \rangle = 1$, then the rotation
matrix ${Q_s}^T g^{(k)}Q_l$ is an in-plane rotation matrix which has the form
\begin{equation} \label{inplane rot}
  \begin{pmatrix}
\cos( \theta_{sl}^{(k)}) & -\sin( \theta_{sl}^{(k)}) & 0\\
\sin( \theta_{sl}^{(k)}) &  \cos( \theta_{sl}^{(k)}) & 0\\
0 & 0 & 1
\end{pmatrix},
\end{equation}
where $\theta_{sl}^{(k)} \in [0,360 \degree)$
is the in-plane rotation angle (see~\cite{viewang}).
If  $\theta_{sl}^{(k)} =0$, then ${Q_s}^T g^{(k)}Q_l = I$, and so ${Q_s}= g^{(k)}Q_l$.
Hence, we define two thresholds; 
the viewing direction threshold $\epsilon_1$, and the in-plane rotation angle threshold $\epsilon_2$.
For the viewing direction threshold, we define $\epsilon_1 = 5 \degree$ along with the  condition
\begin{equation} \label{eq: proximity cond 1}
    \langle {Q_s^3}, (g^{(k)}Q_l)^3 \rangle > \cos (\epsilon_1).
\end{equation}
Satisfying condition~\eqref{eq: proximity cond 1} implies that the rotations $Q_r$ and $g^{(k)}Q_s$ have nearby viewing directions, 
and so it is reasonable to assume that the angle
\begin{equation} \label{eq: proximity angle} 
    \tilde \theta_{sl}^{(k)} = \arctan \left( \frac{({Q_s}^T g^{(k)}Q_l)_{2,1}}{({Q_s}^T g^{(k)}Q_l)_{1,1}} \right)
\end{equation}
approximates the in-plane rotation angle $\theta_{sl}^{(k)}$ of~\eqref{inplane rot}.
We therefore define $\epsilon_2 = 5 \degree$ along with the condition
\begin{equation} \label{eq: proximity cond 2} 
    \tilde \theta_{sl}^{(k)} < \epsilon_2.
\end{equation}
Once both conditions~\eqref{eq: proximity cond 1} and~\eqref{eq: proximity cond 2} hold, the proximity between $Q_s$ and $g^{(k)}Q_l$ is sufficient to remove either $Q_s$ or $Q_l$ from $\widetilde{SO}(3)$.

Of course, it is possible to replace the proximity measure we have used above with any other proximity measure. The advantage of the measure we use is its simple geometric interpretation, which allows to easily set and interpret its thresholds.
%We note that the choice of this proximity follows by its geometrical interpretation, which reflects the viewing directions. 

\bigskip

It remain to show how to construct the set $\widetilde{SO}(3)$, which is the input of the above pruning procedure. To that end, we let~$L$ be a positive integer, and let $\tau,\theta,\varphi$ denote Euler angles. We construct $\widetilde{SO}(3)$ by sampling the Euler angles in equally spaced increments as follows. First, we sample $\tau \in \{0,\ldots,\frac{\pi}{2}\}$ at $\lfloor \frac{L}{4}\rfloor$ points. Then, for each $\tau$, we sample $\theta \in \{0,\ldots, \pi\}$ at  $\lfloor \frac{L}{2}\sin(\tau)\rfloor$ points. Finally, for each pair $(\tau,\theta)$, we sample  $\varphi \in \{0,\ldots, 2\pi\}$ at $\lfloor \frac{L}{2}\sin(\tau)\sin(\theta)\rfloor$ points. For each $(\tau,\theta,\varphi)$ on this grid, we compute a corresponding rotation matrix~$R$ by

\begin{equation*}
    R = R_z(\tau)R_y(\theta)R_x(\varphi),
\end{equation*}
where
\begin{align*}
R_z(\tau) &= \begin{pmatrix}
\cos\tau & -\sin\tau & 0\\
\sin\tau & \cos\tau & 0\\
0 & 0 & 1\\
\end{pmatrix},\\ 
R_y(\theta) &= \begin{pmatrix}
\cos\theta & 0 & \sin\theta  \\
0 & 1 & 0\\
-\sin\theta & 0 & \cos\theta \\
\end{pmatrix},\\ 
R_x(\varphi)&= \begin{pmatrix}
1 & 0 & 0\\
0 & \cos\varphi & -\sin\varphi \\
0 & \sin\varphi & \cos\varphi \\
\end{pmatrix}.
\end{align*}

\section{$N_{SO(3)}(\mathbb T) = \mathbb O$ and $N_{SO(3)}(\mathbb O) = \mathbb O$} \label{prop: group normalizers}

\begin{proof}
A classification of the closed subgroups of $SO(3)$ is given in~\cite{golubitsky1988singularities}, stating that every closed subgroup of $SO(3)$ is conjugate to one of $SO(3)$, $O(2)$, $SO(2)$, $D_n(n \geq 2)$, $C_n(n \geq 2)$, $\mathbb T$, $\mathbb O$, $\mathbb I$ (the icosahedral symmetry), $\mathds{1}$ (the trivial group). 
Moreover, $\mathbb T$ and $\mathbb O$ are closed subgroups of $SO(3)$.
Since for topological groups the normalizer of a closed subgroup is closed (Lemma \ref{lemma: normalizer is close} below)
and since $SO(3)$ is indeed a topological group, the normalizers of the closed subgroups $\mathbb T$ and $\mathbb O$ in $SO(3)$, i.e.  $N_{SO(3)}(\mathbb T)$ and $N_{SO(3)}(\mathbb O) $, are also closed subgroups, thus conjugate to one of the closed subgroups of $SO(3)$. 

By definition of the normalizer, $G \subseteq N_{SO(3)}(G)$, which precludes $O(2)$, $SO(2)$, $D_n(n \geq 2)$, $C_n(n \geq 2)$ and~$\mathds{1}$ from being the normalizers of $\mathbb T$ or $\mathbb O$, since each has at most one symmetry axis of order larger than 2, while both $\mathbb T$ and $\mathbb O$ have more than one such axis. 
In addition, $SO(3)$ and $\mathbb I$ are simple groups~\cite{StillWell2008,Artin2011}, 
and so have no non-trivial normal subgroups.
By definition of the normalizer, $G $ is a normal subgroup of $ N_{SO(3)}(G)$.
Thus, since $SO(3)$ and $\mathbb I$ have no non-trivial normal subgroups, neither $\mathbb T$ nor $\mathbb O$ are normal subgroups of $\mathbb I$ or $SO(3)$,
which precludes $SO(3)$ and $\mathbb I$ from being the normalizers of $\mathbb T$ or $\mathbb O$.
Since $\mathbb T$ is normal in $\mathbb O$~\cite{Artin2011}, we have that $\mathbb{O} \subseteq N_{SO(3)}(\mathbb{T})$ and thus it must hold that $N_{SO(3)}(\mathbb{T})=\mathbb{O}$ and $N_{SO(3)}(\mathbb O) =\mathbb O$. 
\end{proof}

\begin{lemma}\label{lemma: normalizer is close}
Suppose $\tilde H$ is a topological group. Then, the normalizer of a closed subgroup $\tilde G$ of~$\tilde H$ 
$$N_{\tilde H}(\tilde G) = 
\{\tilde h \in \tilde H: \tilde h^{-1}\tilde G \tilde h = \tilde G\}$$
is a closed subgroup.
\end{lemma}
\begin{proof}
Fix $\tilde g \in \tilde G$ and define the map $f_{\tilde g}: \tilde H \rightarrow \tilde H $ by $f_{\tilde g}(\tilde h) = \tilde h^{-1}\tilde g\tilde h$.
Since $\tilde H$ is a topological group, $f_{\tilde g}$ is continuous as the composition of multiplication and inversion maps.
Thus, the preimage of the closed subgroup $\tilde G$ under $f_{\tilde g}$,
defined by 
$f^{-1}_{\tilde g}(\tilde G) =
\{\tilde h \in \tilde H: f_{\tilde g}(\tilde h) \in \tilde G\} =
\{\tilde h \in \tilde H: \tilde h^{-1}\tilde g\tilde h \in \tilde G\}$, is closed.
As any intersection of closed sets is closed, the intersection 
$$\bigcap_{\tilde g \in \tilde G} f^{-1}_{\tilde g}(\tilde G) =
\{\tilde h \in \tilde H: \tilde h^{-1}\tilde g\tilde h \in \tilde G \  \forall \tilde g \in \tilde G\} = N_{\tilde H}(\tilde G)$$
is closed. 

\end{proof}

\end{appendices}

% ===============================================================
% Bibliography
% ===============================================================

\bibliographystyle{plain}
\bibliography{TO}

\begin{thebibliography}{10}

\bibitem{aspire}
{ASPIRE} - algorithms for single particle reconstruction.
\newblock \url{http://spr.math.princeton.edu/}.

\bibitem{Artin2011}
M.~Artin.
\newblock {\em Algebra}.
\newblock Pearson, 2nd edition, 2010.

\bibitem{xmipp}
J.~M. de~la Rosa-Trev\'{i}n, J.~Ot\'{o}n, R.~Marabini, A.~Zald\'{i}var,
  J.~Vargas, J.~M. Carazo, and C.~O.~S Sorzano.
\newblock {Xmipp} 3.0: an improved software suite for image processing in
  electron microscopy.
\newblock {\em Journal of structural biology}, 184(2):321--328, 2013.

\bibitem{kltpicker}
A.~Eldar, B.~Landa, and Y.~Shkolnisky.
\newblock {KLT} picker: Particle picking using data-driven optimal templates.
\newblock {\em Journal of structural biology}, 210(2):107473, 2020.

\bibitem{Frank2006}
{Frank, J.}
\newblock {\em Three-Dimensional Electron Microscopy of Macromolecular
  Assemblies: Visualization of Biological Molecules in Their Native State}.
\newblock Oxford, 2006.

\bibitem{golubitsky1988singularities}
M.~Golubitsky, I.~Stewart, and D.~G. Schaeffer.
\newblock {\em Singularities and groups in bifurcation theory: Volume II},
  volume~69 of {\em Applied Mathematical Sciences}.
\newblock Springer, 1988.

\bibitem{greenberg2017common}
I.~Greenberg and Y.~Shkolnisky.
\newblock Common lines modeling for reference free ab-initio reconstruction in
  cryo-{EM}.
\newblock {\em Journal of structural biology}, 200(2):106--117, 2017.

\bibitem{iudin2016empiar}
A.~Iudin, P.~K. Korir, J.~Salavert-Torres, G.~J. Kleywegt, and A.~Patwardhan.
\newblock {EMPIAR}: a public archive for raw electron microscopy image data.
\newblock {\em Nature methods}, 13(5):387--388, 2016.

\bibitem{Natr2001a}
F.~Natterer.
\newblock {\em The Mathematics of Computerized Tomography}.
\newblock Classics in Applied Mathematics. SIAM, 2001.

\bibitem{EMD4905}
K.~Naydenova, M.~J. Peet, and C.~J. Russo.
\newblock Multifunctional graphene supports for electron cryomicroscopy.
\newblock {\em Proceedings of the National Academy of Sciences of the United
  States of America}, 116(24):11718--11724, June 2019.

\bibitem{pragier2016graph}
G.~Pragier, I.~Greenberg, X.~Cheng, and Y.~Shkolnisky.
\newblock A graph partitioning approach to simultaneous angular reconstitution.
\newblock {\em IEEE transactions on computational imaging}, 2(3):323--334,
  2016.

\bibitem{pragier2019common}
G.~Pragier and Y.~Shkolnisky.
\newblock A common lines approach for ab initio modeling of cyclically
  symmetric molecules.
\newblock {\em Inverse Problems}, 35, 2019.

\bibitem{cryosparc}
A.~Punjani, J.~L. Rubinstein, D.~J. Fleet, and M.~A. Brubaker.
\newblock {cryoSPARC}: algorithms for rapid unsupervised cryo-{EM} structure
  determination.
\newblock {\em Nature Methods}, 14:290--296, 2017.

\bibitem{EMD10835}
R.~D. Righetto, L.~Anton, R.~Adaixo, R.~P. Jakob, J.~Zivanov, M.-A. Mahi,
  P.~Ringler, T.~Schwede, T.~Maier, and H.~Stahlberg.
\newblock High-resolution cryo-em structure of urease from the pathogen
  yersinia enterocolitica.
\newblock {\em Nature communications}, 11(1):5101, October 2020.

\bibitem{ROHOU2015216}
A.~Rohou and N.~Grigorieff.
\newblock {CTFFIND4}: Fast and accurate defocus estimation from electron
  micrographs.
\newblock {\em Journal of Structural Biology}, 192(2):216--221, 2015.

\bibitem{syncD2}
E.~Rosen and Y.~Shkolnisky.
\newblock Common lines ab initio reconstruction of {$D_{2}$}-symmetric
  molecules in cryo-electron microscopy.
\newblock {\em SIAM Journal on Imaging Sciences}, 13(4):1898--1944, 2020.

\bibitem{asymsync}
Y.~Shkolnisky and A.~Singer.
\newblock Viewing direction estimation in {Cryo-EM} using synchronization.
\newblock {\em SIAM Journal on Imaging Sciences}, 5(3):1088--1110, 2012.

\bibitem{viewang}
A.~Singer.
\newblock Viewing angle classification of cryo-electron microscopy images using
  eigenvectors.
\newblock {\em SIAM Journal on Imaging Sciences}, pages 723--759, 2011.

\bibitem{voting}
A.~Singer, R.~R. Coifman, F.~J. Sigworth, D.~W. Chester, and Y.~Shkolnisky.
\newblock Detecting consistent common lines in cryo-{EM} by voting.
\newblock {\em Journal of Structural Biology}, 169(3):312--322, 2010.

\bibitem{springer2012}
A.~Singer and Y.~Shkolnisky.
\newblock {\em Modeling Nanoscale Imaging in Electron Microscopy}, chapter
  Center of Mass operators for CryoEM - Theory and implementation, pages
  147--177.
\newblock Nanostructure Science and Technology. Springer, New York, 2012.

\bibitem{StillWell2008}
J.~StillWell.
\newblock {\em Naive Lie Theory}.
\newblock Undergraduate Texts in Mathematics. Springer, 2008.

\bibitem{EMAN}
G.~Tang, L.~Peng, P.~R. Baldwin, D.~S. Mann, W.~Jiang, I.~Rees, and S.~J.
  Ludtke.
\newblock {EMAN2}: an extensible image processing suite for electron
  microscopy.
\newblock {\em Journal of Structural Biology}, 157(1):38--46, 2007.

\bibitem{VanHeel1987}
M.~Van~Heel.
\newblock Angular reconstitution: a posteriori assignment of projection
  directions for {3D} reconstruction.
\newblock {\em Ultramicroscopy}, 21(2):111--123, 1987.

\bibitem{vanHeel_Schatz}
M.~Van~Heel and M.~Schatz.
\newblock Fourier shell correlation threshold criteria.
\newblock {\em J. Struct. Biol.}, 151(3):250--262, 2005.

\bibitem{Zheng2017}
S.~Zheng, E.~Palovcak, J.~P. Armache, K.~Verba, Y.~Cheng, and D.~Agard.
\newblock {MotionCor2}: anisotropic correction of beam-induced motion for
  improved cryo-electron microscopy.
\newblock {\em Nature methods}, 14(4):331--332, 2017.

\bibitem{relion3}
J.~Zivanov, T.~Nakane, B.~O. Forsberg, D.~Kimanius, W.~J. Hagen, E.~Lindahl,
  and S.~H. Scheres.
\newblock New tools for automated high-resolution cryo-{EM} structure
  determination in {RELION-3}.
\newblock {\em Elife}, 7:e42166, 2018.

\end{thebibliography}

\thispagestyle{empty}

\end{document}